\documentclass[10pt,amsfonts,amssymb,amsmath,ps,prb,longbibliography,nobibnotes,twocolumn,groupedaddress]{revtex4-1}
\usepackage[english]{babel}
\usepackage[utf8]{inputenc}
\usepackage[colorinlistoftodos, color=green!40, prependcaption]{todonotes}
\usepackage{amsthm}
\usepackage{mathtools}
\usepackage{blkarray}
\usepackage{amsmath}
\usepackage{physics}
\usepackage{xcolor}
\usepackage{graphicx}
\usepackage{subcaption}
\usepackage[left=23mm,right=13mm,top=35mm,columnsep=15pt]{geometry} 
\usepackage{adjustbox}
\usepackage{placeins}
\usepackage[T1]{fontenc}
\usepackage{lmodern}
\usepackage{lipsum}
\usepackage{csquotes}
\usepackage{tikz}
\usetikzlibrary{decorations.pathreplacing,calc}
\usepackage{qcircuit}
\usepackage{array}
\usepackage{multirow}
\usepackage{rotating}
\usepackage{diagbox}
\usepackage{ragged2e}
\usepackage{makecell}
\usepackage[unicode=true, colorlinks=true]{hyperref} 
\usepackage{cleveref}
\newcommand*{\citen}[1]{%
  \begingroup
    \romannumeral-`\x 
    \setcitestyle{numbers}%
    \cite{#1}%
  \endgroup
}

\begin{document}
\title{A Grid-Based Quantum Algorithm for the Time-Dependent Simulation of Infrared Spectra}

\author{Xiaoning Feng}
    \email[Email: ]{xiaoning.feng@chem.ox.ac.uk}
    \affiliation{Physical and Theoretical Chemistry Laboratory, University of Oxford, Oxford OX1 3QZ, United Kingdom}
\author{Arman Nejad}
    \affiliation{Physical and Theoretical Chemistry Laboratory, University of Oxford, Oxford OX1 3QZ, United Kingdom}
\author{David P. Tew}
    \email[Email: ]{david.tew@chem.ox.ac.uk}
    \affiliation{Physical and Theoretical Chemistry Laboratory, University of Oxford, Oxford OX1 3QZ, United Kingdom}

\date{\today}

\begin{abstract}
We develop a time-dependent, grid-based framework for simulating infrared spectra that is specifically designed for quantum computers. The proposed circuit employs a probabilistic strategy for applying the non-unitary dipole operator and an Split Operator-Quantum Fourier Transform time evolution scheme. Using a vibrational model of the water molecule as a test system, our classical emulation results demonstrate accurate determination of fundamental and overtone band positions and intensities via Fourier-transformed dipole-dipole autocorrelation functions. We also identify the optimal time parameters that minimise gate depths while maintaining high fidelity. For further resource reduction, we validate the feasibility of utilising harmonic oscillator approximations in state preparation and dipole operator truncations. With its scalability to higher-dimensional normal mode spaces, this wavefunction-based approach establishes a robust foundation for studying IR spectra on future quantum hardware.
\end{abstract}

\maketitle

\section{Introduction}

The simulation of infrared spectra has long been a significant topic in molecular characterisation, as it is crucial for interpreting molecular structures and predicting vibrational behaviours. 
It proceeds by first constructing Potential Energy Surface (PES) and Dipole Moment Surface (DMS) from electronic-structure calculations and then solving nuclear-vibrational dynamics on these surfaces.
The former typically relies on Density Functional Theory\cite{Kim2022, KAZA2022, Yanji2020} or post-Hartree–Fock methods\cite{Brito2025, Khire2024, KNAANIE2014, Bartlett1994}.
Established nuclear-vibrational treatments include Vibrational Perturbation Theory (VPT2 or higher-order variants)\cite{Sibert_JCP_1988, Barone_JCP_2005, Gong_JCP_2018} that offer great advantages for anharmonic corrections, Vibrational Configuration Interaction (VCI)\cite{Bowman_IRPC_2003, SIBAEV2016, Benjamin2024, Schneider_JCP_2024_214118} with high accuracy at substantial cost when many modes are included, Ab Initio Molecular Dynamics (AIMD)\cite{Shima2022, Gastegger2017, Fischer2016, Gaigeot2003} and Path Integral techniques\cite{Althorpe2024, Shepherd2021, Althorpe2021, Liu2016} which sample the PES/DMS on-the-fly along discretised trajectories or beads.
These mature classical methods have wide applicability in medium-sized molecules. However, as molecular complexity increases, their scalability becomes a critical bottleneck. Accurately capturing fine spectral details for very large molecules often requires prohibitively high computational demands due to the exponential scaling of traditional algorithms.

Quantum computing algorithms surpass classical approaches by exploiting the powerful scaling of qubits.\cite{Wu2021, Madsen2022, Deng2023} The unique properties of qubits, such as superposition and entanglement, offer the potential to overcome inherent limitations of classical computational techniques,\cite{Daley2022} particularly the exponential scaling of the required computational resources in many-body quantum simulations. Vibrational problems, which are central to understanding phenomena like energy transfer and spectral properties, are identified as more tractable on quantum hardware compared to electronic structure problems.\cite{Sawaya2021} In spite that significant progress has been made in applying quantum computing to tackling molecular vibrational problems,\cite{Sawaya2021, Pauline2020, McArdle2019, Alexander2019, HU2018, MacDonell2023, Richerme2023, Komarova2020, trenev2023} there is limited focus on IR-specific applications. Our work bridges this critical gap by introducing a grid-based quantum framework capable of deriving IR spectra from time-evolved molecular wavefunctions.

Among the representation schemes essential for vibrational dynamics modelling, grid-based methods, combined with efficient strategies such as Smolyak quadrature\cite{LAUVERGNAT2014,avila2011} or tensor-train decompositions,\cite{Maxim2016} distinguish themselves by enabling flexible discretisation of spatially localised vibrational modes. Several recent studies have demonstrated the synergies of integrating grid-based representations with quantum computing.\cite{Chan2023, Childs2022, Pauline2023} In stark contrast to the exponential growth in complexity encountered by classical simulations, quantum computing's ability to scale polynomially with system size\cite{Lloyd1996} positions the grid-based representation as a very compatible choice in the quantum domain.

The Split Operator Fourier Transform (SO-FT) method, which originates from the Trotter-Suzuki decomposition,\cite{SUZUKI1990, Ostmeyer2023, Dhand2014} effectively handles the time-propagation of discrete wavefunctions and has been widely adopted in time-dependent simulations for its computational efficacy.\cite{Zhu2024, Semenova2021, Sathya2021, Roulet2021, Bandrauk1993, BANDRAUK1991, Hermann1988} On quantum platforms, the Quantum Fourier Transform (QFT) further enhances this method with exponential speed-ups over classical Fourier methods.\cite{Chen2023, Dixit2022, Camps2020} The combination of the SO-QFT together with grid-based methods therefore emerges as a highly promising strategy for quantum implementations to target classically intractable problems.

Quantum simulation of vibrational wavefunctions broadly falls into two paradigms: analogue approaches that emulate vibrational Hamiltonians on bosonic platforms such as trapped ions or photonic systems,\cite{Wang_2020,Chen_2021} and digital gate-based approaches that encode vibrational degrees of freedom in qubit registers and simulate the Hamiltonians via universal quantum circuits.\cite{Sawaya_2020} Within the digital paradigm, two complementary formulations can be distinguished. Eigenstate-based methods, including quantum phase estimation and variational quantum eigensolver strategies,\cite{Ove2026,Erik2021} aim to determine stationary vibrational energy levels. In contrast, real-time propagation schemes enable direct evaluation of dynamical observables, providing access to spectroscopic information without explicit eigenstate construction. The SO-QFT scheme adopted in this work belongs to the latter class.


Building on the SO-QFT scheme, we develop a novel time-dependent quantum algorithm for computing infrared spectra. In this framework, the vibrational wavefunction is represented on a multi-dimensional grid and propagated in time. The infrared spectrum is obtained from the Fourier transform of the dipole–dipole autocorrelation function. To map the proposed algorithm onto quantum registers, we design a customised set of quantum circuits that optimise qubit usage and minimise gate depth. A significant challenge arises in implementing the non-unitary dipole operator, which we resolve by integrating a probabilistic encoding scheme with the use of one ancillary qubit. We employ classical emulation to demonstrate the feasibility of this algorithm and successfully reproduce the characteristic fundamental bands of water and higher-energy overtones and combination bands with high fidelity.

This paper is structured as follows: Section~\ref{Background Theory} provides the theoretical and methodological foundation underlying the workflow. Section~\ref{Quantum Framework} outlines the complete quantum circuit design, from the initial state preparation, through the implementation of the non-unitary dipole operator and the SO-QFT time evolution scheme, to the measurement strategies utilised for obtaining the infrared spectrum. Based on classical emulation results of the water molecule, Section~\ref{Characteristic Peak Analysis} assesses the performance of this algorithm in its general feasibility, the sensitivity to varying simulation time parameters, and the quantitative impact of replacing exact components with approximate counterparts. Section~\ref{Discussion} discusses the scope of the framework from a broader perspective. Finally, Section~\ref{Conclusion} concludes the paper.

\section{Background Theory}
\label{Background Theory}


\subsection{Vibrational Hamiltonian and Dipole Surface}
\label{subsec:Hamiltonian}

In this work, we employ the pure vibrational Hamiltonian $H=K+V$ in dimensionless normal coordinates. The kinetic energy operator takes the diagonal structure
\begin{align}
    \frac{K}{hc} = - \sum_i \frac{\omega_i}{2} \frac{\partial^2}{\partial Q_i^2}
    \label{kinetic}
    ,
\end{align}
where $\omega_i$ is the harmonic wavenumber of mode $i$ which runs over all $3N_\mathrm{atom}-6$ vibrational indices. 
Vibrational angular momentum and pseudopotential terms are neglected in our framework. These terms are of the order of the rotational constants and generally small compared to other terms.\cite{avila2011}


The potential energy surface (PES) and dipole moment surface (DMS) are represented as Taylor expansions about the equilibrium geometry in sum-of-products form, which we truncate at fourth and third order, respectively:
\begin{align}
    \frac{V}{hc} & =
    \sum_i \frac{\omega_i}{2} Q_i^2 + \sum_{ijk} k_{ijk} Q_i Q_j Q_k
    \nonumber \\
    &+ \sum_{ijkl} k_{ijkl} Q_i Q_j Q_k Q_l
    \label{vhat}
    ,
    \\
    \mu^{(\alpha)} &= \mu_e^{(\alpha)} + \sum_{i} \mu_i^{(\alpha)} Q_i + \sum_{ij} \mu_{ij}^{(\alpha)} Q_i Q_j
    \nonumber \\
    &+ \sum_{ijk} \mu_{ijk}^{(\alpha)} Q_i Q_j Q_k
    \label{dipole function}
    .
\end{align}
Here, $\alpha$ is the axis of the body-fixed Eckart frame and $k_{ijk...}$ and $\mu_{ijk...}$ correspond to the derivatives of the potential and dipole moment with respect to dimensionless normal coordinates. In the present work, we do not include the constant term $\mu_e^{(\alpha)}$, as it does not affect the vibrational band positions and intensities.


\subsection{Time Evolution}
\label{subsec:trotter}


We adopt the second-order SO-QFT method to evolve the wavefunction in time. This scheme partitions the time evolution operator into kinetic and potential energy components and alternately applies them in the position and momentum spaces. However, this decomposition suffers from inherent Trotter errors due to the non-commutativity of the potential and kinetic operators. A second-order formulation mitigates this error to $\mathcal{O}(dt^{3})$ through a symmetrised decomposition,\cite{SUZUKI1991,Hatano2005,jones2019} where $dt$ denotes the spacing between adjacent discrete time points during evolution.

The high-order polynomial structure of $V$ presents a considerable gate depth demands in the quantum circuit. To address this, we select a more efficient scheme where the kinetic energy operator takes on the half-time step split rather than the potential operator:  
\begin{align}
     e^{-iHdt/\hbar}
    &\approx
    \underbrace{e^{-iKdt/2\hbar}}_{U_K}
    \cdot
    \underbrace{e^{-iVdt/\hbar}}_{U_V}
    \cdot
    \underbrace{e^{-iKdt/2\hbar}}_{U_K}
\end{align}
with time evolution operators $U_K$ and $U_V$ implemented in the momentum and position spaces, respectively. This adjustment allows us to minimise gate depth, as each time step only requires a single potential operation. Although two additional QFT operations are needed at the initial and final time points, the overall gate count is still considerably lower than that of the more conventional splitting pattern, where the potential is split across half-time steps. We have verified through classical emulations of both splitting patterns that the resultant infrared spectra are quantitatively indistinguishable. For water, band centre differences were found to be below $10^{-5}$\,cm$^{-1}$.

\subsection{Simulation of the Infrared Spectrum}

In our time-dependent approach, we compute the infrared spectrum from the dipole-dipole autocorrelation function\cite{Balint1990}
\begin{align}
     A_{\mu^{(\alpha)}} (t) = \bra{\Psi_{\mu^{(\alpha)}}(t_0)} e^{-iHt/\hbar} \ket{\Psi_{\mu^{(\alpha)}}(t_0)}
     \label{eq:autocorr}
     ,
\end{align}
where $\ket{\Psi_{\mu^{(\alpha)}}(t_0)} \equiv \mu^{(\alpha)}\ket{\Psi(t_0)}$.
The infrared cross section is related to the Fourier transform of the dipole-dipole autocorrelation and given by\cite{Balint1990, Vendrell2007}
\begin{align}
    \sigma(E) &= \frac{E}{3c\varepsilon_0 \hbar^2}\sum_\alpha \operatorname{\mathbb{R}e}\left\{\int_0^\infty e^{i(E+E_0)t/\hbar} A_{\mu^{(\alpha)}} (t) \mathrm{d}t \right\}
    \label{eq:sigma(E)}
    .
\end{align}
Here, $E_0$ is the vibrational ground state energy and $E$ is the energy relative to the ground state.

While the simulation of $\sigma(E)$ in principle requires an infinite time evolution, in practice it must be truncated at a finite time $T$. This truncation leads to Gibbs-type oscillations in the frequency domain, leading to locally negative features in  the Fourier-transformed spectra. To mitigate this effect, a damping function $D(t)$ is applied to the autocorrelation function prior to the Fourier transform. Several functional forms have been proposed for $D(t)$, including $\cos({\pi t/{2T}})$\cite{Vendrell2007,BECK2000} and $\cos({\pi t/{2T}})^{2}$\cite{PELAEZ2017}. In the present work, we adopt the latter.

 

From the simulated spectrum, we determine band positions by calculating the centroid, i.e. the intensity-weighted wavenumber average, via
\begin{align}
    \tilde{\nu}_c = \frac{\sum_i \tilde{\nu}_i \sigma_i}{\sum_i\sigma_i},
\end{align}
where $\tilde{\nu}_i=E_i/hc$ and $\sigma_i$ are the wavenumber and IR cross section at the $i^{th}$ data point, respectively.
For each band, the analysis window is empirically chosen by inspection of the spectrum so as to capture the full extent of the target peak while minimising contamination from neighbouring bands.
By integrating over the band, we finally obtain the infrared intensity
\begin{align}
    I_\mathrm{band} &= \int_\mathrm{band} N_A\sigma(E)\mathrm{d}E
    \label{eq:I_mol}
    .
\end{align}

\section{Quantum Framework}
\label{Quantum Framework}

\begin{figure*}[!htbp]
    \centering
    \includegraphics[scale=0.14]{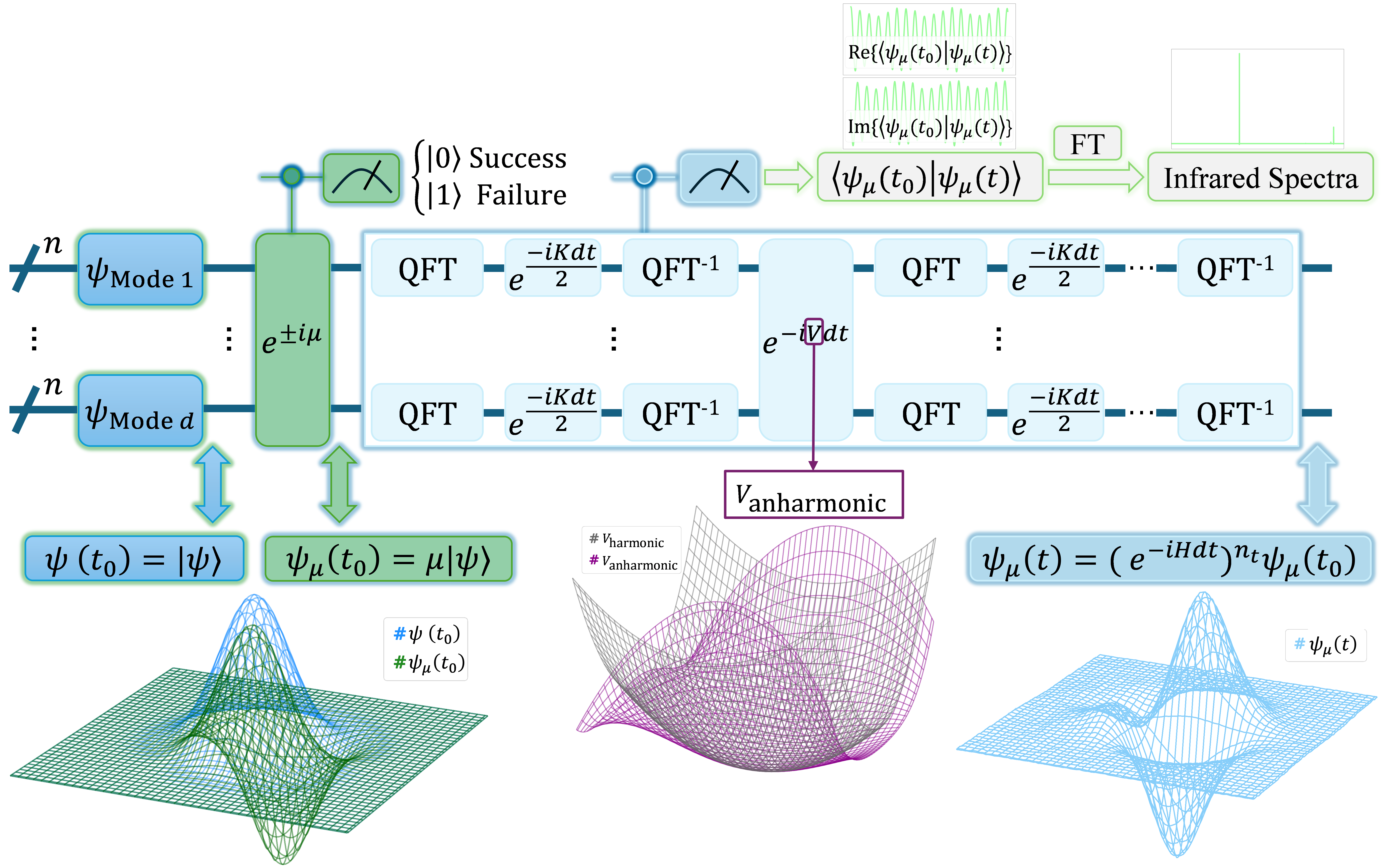}
    \caption{\justifying Schematic overview of the proposed quantum framework for simulating infrared spectra.}
    \label{schemefig}
\end{figure*}

We aim to develop scalable quantum circuit architectures that are directly applicable to future noise-free, fault-tolerant quantum hardwares. Our quantum implementation is structured into four key components: approximating and preparing the initial state, applying the non-unitary dipole operator, evolving the system via SO-QFT time operators constructed from polynomial Hamiltonian terms, and finally measuring the dipole-dipole autocorrelation to retrieve the infrared spectrum, as illustrated in Figure~\ref{schemefig}. This framework builds on our previous SO-QFT approach for vibronic dynamics\cite{Feng_arxiv_2025} and adapts it to infrared spectroscopy by incorporating explicit treatment of non-unitary dipole operators and higher-order Hamiltonian terms.

\subsection{Initial State Preparation}
\label{ISP}

In the context of quantum computing, the grid-based representation enables an intuitive mapping of spatial coordinates onto qubit states $\ket{x}$, with $x$ indexing the grid point in binary form. The qubit superposition allows the initial state to be compactly stored and manipulated in quantum registers, using only a logarithmic number of qubits relative to the system size. For a given wavefunction defined over $2^n$ grid points, the amplitude $c_x$ at the grid point indexed by $x$ is encoded as the probability of the corresponding qubit state $\ket{x}$, leading to $\ket{\psi}=\sum_{x=0}^{2^n-1}c_x\ket{x}$. This formalism makes it feasible to exploit quantum parallelism for following state preparation.

The discretisation of grid-based spaces facilitates the use of adaptive grid densities to either optimise state fidelity or reduce computational resources. Due to the absence of singularities in the potential, vibrational wavefunctions are smooth and a low grid density will suffice to represent the relevant normal modes. For each normal mode, we allocate $2^4 = 16$ grids within a spatial range of [-5, 5] in dimensionless unit, corresponding to $n=4$ qubits per quantum register. Quantum registers here function as collections of qubits representing different normal modes. In terms of water, this context leads to a total 12 qubits distributed across 3 registers.

Here, we consider two possibilities for the initial state: the harmonic ground state, used as a cost-saving approximation, and the anharmonic ground state that corresponds to the  Hamiltonian introduced in Section~\ref{Background Theory}.
We prepare the anharmonic initial state via imaginary time evolution (ITE) method, using the harmonic ground state as initial guess.\cite{LEHTOVAARA2007,McClean2015} By evolving the system in imaginary time, excited-state components exponentially decay, thus projecting the wavefunction towards the ground state. 

While implementation of ITE is straightforward on classical computers, it becomes much more tedious and expensive on quantum devices. Various quantum algorithms, such as variational,\cite{McArdle20192, Gomes2020} probabilistic,\cite{Leadbeater2024, Xie2024} or fully quantum methods\cite{Motta2019, Yeter2020} exist that provide diverse circuits for this purpose. However, the presence of high-order polynomial terms in the potential makes the quantum implementation of ITE significantly more computationally demanding. 
To reduce the overall quantum resource cost, we therefore explore using harmonic oscillator wavefunctions as approximate initial states. Provided that the subsequent real-time evolution is governed by a potential operator that fully retains anharmonic contributions, this approach is expected to yield reliable vibrational spectra.  We quantify the accuracy of this approximation for water in Section~\ref{Characteristic Peak Analysis}.
One key advantage of using these harmonic wavefunctions is the ability to prepare initial states for various normal modes in parallel. Such approximation greatly simplifies the computational complexity as each wavefunction is generated independently in its own register. This architecture makes scaling to higher-dimensional systems highly efficient by merely adding more registers without increasing gate depth for state preparation.

Owing to their smoothness, the use of Gaussian functions eliminates the need for dense grids. Several efficient methods exist for preparing Gaussian states on quantum computers, such as the uniformly controlled rotational method,\cite{Mottenen2005, Moosa2023} the Matrix Product State technique,\cite{Iaconis2024} the Mid-Circuit Measurement and Reuse scheme,\cite{Rattew2021} the Kitaev-Webb algorithm,\cite{bauer2021} etc. Among the various approaches, utilising uniformly controlled rotations, i.e. $\mathbf{R_y}$ and $\mathbf{CNOT}$ gates, allows for constructing the real valued harmonic wavefunction for each normal mode with a gate depth of only 29. The validity of the corresponding circuits has been demonstrated in our previous work on the vibronic dynamics of photoexcited pyrazine,\cite{Feng_arxiv_2025} in which the rotation angles of the relevant gates were directly calculated from the harmonic wavefunction amplitudes. 

\subsection{Elementary Evolution Operators}
\label{elementary}

In our simulation, both the application of the dipole operator and the time evolution under the Hamiltonian require building unitaries of the form $e^{if}$, where $f$ is a polynomial function of the normal mode coordinates $\mathbf{Q}$.
Since the wavefunction is represented on a grid, each discrete $Q$ is defined over $[-L/2,L/2]$ and is encoded as $Q_i=-L/2+x_idQ$ for quantum implementation, where $x_i\in\{0,1,\cdots,2^n-1\}$ labels the computational basis states represented by $n$ qubits and $dQ$ denotes the grid spacing. 
Under this mapping, phase functions that depend polynomially on $Q$ are expressed as sums of terms with different polynomial orders in $x$. For example, a cubic polynomial in $Q$ will generate contributions up to cubic order in $x$ upon expansion. Accordingly, a target unitary can be decomposed into a product of elementary $U$-operators, which are defined as:
\begin{align}
    U_{0}    &= e^{ib                }, \\
    U_{i}    &= e^{ib x_i              }, \\
    U_{ij}   &= e^{ib x_i x_j        }, \\
    U_{ijk}  &= e^{ib x_i x_j x_k    }, \\
    U_{ijkl} &= e^{ib x_i x_j x_k x_l},
\end{align}
where $b$ is a constant phase factor.
Note that the indices appearing in the subscript of each unitary ($i,j,k,...$) are not necessarily distinct.
It is common to encounter coinciding indices, in which case identical indices indicate the same normal mode coordinates and thus the same quantum register. For example, when $i=j$, $U_{ij}$ reduces to a specific class whose exponent involves quadratic-order polynomials.

Since all such phase operators are diagonal in the computational basis and mutually commute, different unitaries acting on the same normal mode registers and sharing the same polynomial order can be consolidated into a single effective unitary. This allows us to count only the grouped terms when estimating the total circuit depth.
Before assembling these effective terms in sequence to form the specific composite operations required by our simulation, we analyse their individual circuit structure and analyse the associated gate depths in this section.

Our prior work\cite{Feng_arxiv_2025} has detailed the circuits for $U_{0}$, $U_{i}$, and $U_{ij}$ using $\mathbf{U_1}$ gates. Here, we focus on the third- and fourth-order terms $U_{ijk}$ and $U_{ijkl}$, which are necessary to capture anharmonic interactions in our model. In the following discussions, qubits within a register of $x_i$ are denoted as $q_{i^\prime}$ accordingly. With $n$ qubits assigned to a specific $x_i$ register, the number of the possible $i^\prime$ values becomes $n$, i.e., $q_{i^\prime} \in \{q_0,q_1,\cdots,q_{n-1}\}$. In general, the polynomial order in the exponents of these unitaries determines the corresponding gate complexities, with $U_{0}$, $U_{i}$, $U_{ij}$, $U_{ijk}$, and $U_{ijkl}$ requiring 4, $n$, $n^2$, $\mathcal{O}(n^3)$, $\mathcal{O}(n^4)$ gates, respectively. 


\subsubsection{Gates for cubic polynomials}

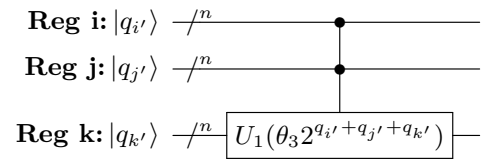
\begin{figure}[htbp]
\centering
\scalebox{1.1}{
\Qcircuit @C=1em @R=1.5em {
\lstick{\textbf{Reg i:}\ket{q_{i^\prime}}} &\qw{/}^{n} & \ctrl{1}& \qw \\
\lstick{\textbf{Reg j:}\ket{q_{j^\prime}}} &\qw{/}^{n}& \ctrl{1} &\qw  \\
\lstick{\textbf{Reg k:}\ket{q_{k^\prime}}} &\qw{/}^{n}& \gate{U_1(\theta_3 2^{q_{i^\prime}+q_{j^\prime}+q_{k^\prime}})} &\qw  \\
}}
\caption{\justifying Quantum circuit for implementing $U_{ijk}$, the time evolution operators of third-order polynomial terms.}
\label{fig:third}
\end{figure}

We first illustrate the general strategy for building circuits that implement $U_{ijk}$ in Figure~\ref{fig:third}, accompanied by an angle definition of $\theta_3=c$. In this structure, we assign the two registers, $x_{i}$ and $x_{j}$, as controlling registers, while the third $x_{k}$ one serves as the target register accommodating $\mathbf{U_1}$ gates.

\begin{figure*}[htbp]
\flushright
\subfloat[When $i=j=k$, the operation only involves one register.\label{thirdexample1cir}]{
\vspace{2.5em} 
\scalebox{0.6}{
\Qcircuit @C=1em @R=1.5em {
\makebox[4em][l]{\textbf{Reg i, Reg j, Reg k refer to the same Reg:}}\\
\lstick{\ket{q_0}}&\gate{U_1(\theta_32^0)} & \ctrl{1} & \ctrl{2} & \ctrl{3} 
&\gate{U_1(\theta_32^1)} &\ctrl{1} &\ctrl{1} &\ctrl{1} 
&\gate{U_1(\theta_32^2)} & \ctrl{1} & \ctrl{2} & \ctrl{2} 
&\gate{U_1(\theta_32^3)} & \ctrl{1} & \ctrl{2} & \ctrl{3}&\qw \\
\lstick{\ket{q_1}}&\qw & \gate{U_1(\theta_32^1)}&\qw &\qw 
&\ctrl{-1}& \gate{U_1(\theta_32^2)}&\ctrl{1} & \ctrl{2} 
&\qw & \gate{U_1(\theta_32^3)}&\qw &\qw
&\qw & \gate{U_1(\theta_32^4)}&\qw &\qw&\qw\\
\lstick{\ket{q_2}}&\qw & \qw & \gate{U_1(\theta_32^2)} &\qw 
& \qw &\qw &\gate{U_1(\theta_32^3)} &\qw
&\ctrl{-2} & \ctrl{-1}&\gate{U_1(\theta_32^4)}&\ctrl{1}
& \qw &\qw &\gate{U_1(\theta_32^5)} &\qw&\qw\\
\lstick{\ket{q_3}}&\qw & \qw &\qw & \gate{U_1(\theta_32^3)} 
& \qw &\qw&\qw &\gate{U_1(\theta_32^4)}
&\qw & \qw &\qw &\gate{U_1(\theta_32^5)}
&\ctrl{-3} & \ctrl{-2} &\ctrl{-1} &\gate{U_1(\theta_32^6)}&\qw\\
}
\begin{tikzpicture}[overlay, remember picture]
    \draw[decorate, decoration={brace, amplitude=8pt, mirror}, yshift=-2ex] 
        (-27.8,-4.2) -- (-21.5,-4.2) node[midway, yshift=-2.5em, align=center]{$q_{i^\prime} = q_0$, $q_{j^\prime} = q_0$, \\ $q_{k^\prime} = \{q_0,q_1,q_2,q_3\}$};
    \draw[decorate, decoration={brace, amplitude=8pt, mirror}, yshift=-2ex] 
        (-20.8,-4.2) -- (-14.5,-4.2) node[midway, yshift=-2.5em, align=center]{$q_{i^\prime} = q_0$, $q_{j^\prime} = q_1$, \\ $q_{k^\prime} = \{q_0,q_1,q_2,q_3\}$};
    \draw[decorate, decoration={brace, amplitude=8pt, mirror}, yshift=-2ex] 
        (-13.8,-4.2) -- (-7.5,-4.2) node[midway, yshift=-2.5em, align=center]{$q_{i^\prime} = q_0$, $q_{j^\prime} = q_2$, \\ $q_{k^\prime} = \{q_0,q_1,q_2,q_3\}$};
    \draw[decorate, decoration={brace, amplitude=8pt, mirror}, yshift=-2ex] 
        (-6.8,-4.2) -- (-0.5,-4.2) node[midway, yshift=-2.5em, align=center]{$q_{i^\prime} = q_0$, $q_{j^\prime} = q_3$, \\ $q_{k^\prime} = \{q_0,q_1,q_2,q_3\}$};
\end{tikzpicture}}}
\\
\vspace{1em}
\subfloat[When $i=j\neq k$, the operation involves two register.\label{thirdexample2cir}]{
\vspace{2.5em} 
\scalebox{0.6}{
\Qcircuit @C=1em @R=1.5em {
\makebox[4em][l]{\textbf{Reg i, Reg j refer to the same Reg:}}\\
\lstick{\ket{q_0}}&\ctrl{5}&\ctrl{6}&\ctrl{7}&\ctrl{8}   
&\ctrl{5}&\ctrl{6}&\ctrl{7}&\ctrl{8}
&\ctrl{5}&\ctrl{6}&\ctrl{7}&\ctrl{8}
&\ctrl{5}&\ctrl{6}&\ctrl{7}&\ctrl{8}&\qw\\
\lstick{\ket{q_1}}&\qw&\qw&\qw&\qw   
&\ctrl{4}&\ctrl{5}&\ctrl{6}&\ctrl{7}
&\qw&\qw&\qw&\qw 
&\qw&\qw&\qw&\qw &\qw\\
\lstick{\ket{q_2}}&\qw&\qw&\qw&\qw   
&\qw&\qw&\qw&\qw 
&\ctrl{3}&\ctrl{4}&\ctrl{5}&\ctrl{6}
&\qw&\qw&\qw&\qw &\qw \\
\lstick{\ket{q_3}}&\qw&\qw&\qw&\qw   
&\qw&\qw&\qw&\qw 
&\qw&\qw&\qw&\qw 
&\ctrl{2}&\ctrl{3}&\ctrl{4}&\ctrl{5}&\qw \\
\makebox[4em][l]{\textbf{Reg k:}}\\
\lstick{\ket{q_0}}&\gate{U_1(\theta_32^0)} &\qw & \qw &\qw   
&\gate{U_1(\theta_32^1)} &\qw & \qw &\qw
&\gate{U_1(\theta_32^2)} &\qw & \qw &\qw
&\gate{U_1(\theta_32^3)} &\qw & \qw &\qw&  \qw \\
\lstick{\ket{q_1}}&\qw &\gate{U_1(\theta_32^1)} & \qw &\qw   
&\qw &\gate{U_1(\theta_32^2)} & \qw &\qw   
&\qw &\gate{U_1(\theta_32^3)} & \qw &\qw   
&\qw &\gate{U_1(\theta_32^4)} & \qw &\qw   &\qw \\
\lstick{\ket{q_2}}&\qw & \qw &\gate{U_1(\theta_32^2)} &\qw   
&\qw & \qw &\gate{U_1(\theta_32^3)} &\qw   
&\qw & \qw &\gate{U_1(\theta_32^4)} &\qw   
&\qw & \qw &\gate{U_1(\theta_32^5)} &\qw   &\qw \\
\lstick{\ket{q_3}}&\qw & \qw &\qw &\gate{U_1(\theta_32^3)}   
&\qw & \qw &\qw &\gate{U_1(\theta_32^4)}   
&\qw & \qw &\qw &\gate{U_1(\theta_32^5)}  
&\qw & \qw &\qw &\gate{U_1(\theta_32^6)}  &\qw \\
}
\begin{tikzpicture}[overlay, remember picture]
    \draw[decorate, decoration={brace, amplitude=8pt, mirror}, yshift=-2ex] 
        (-27.8,-7) -- (-21.5,-7) node[midway, yshift=-2em, align=center]{$q_{i^\prime} = q_0$, $q_{j^\prime} = q_0$, \\ $q_{k^\prime} = \{q_0,q_1,q_2,q_3\}$};
    \draw[decorate, decoration={brace, amplitude=8pt, mirror}, yshift=-2ex] 
        (-20.8,-7) -- (-14.5,-7) node[midway, yshift=-2em, align=center]{$q_{i^\prime} = q_0$, $q_{j^\prime} = q_1$, \\ $q_{k^\prime} = \{q_0,q_1,q_2,q_3\}$};
    \draw[decorate, decoration={brace, amplitude=8pt, mirror}, yshift=-2ex] 
        (-13.8,-7) -- (-7.5,-7) node[midway, yshift=-2em, align=center]{$q_{i^\prime} = q_0$, $q_{j^\prime} = q_2$, \\ $q_{k^\prime} = \{q_0,q_1,q_2,q_3\}$};
    \draw[decorate, decoration={brace, amplitude=8pt, mirror}, yshift=-2ex] 
        (-6.8,-7) -- (-0.5,-7) node[midway, yshift=-2em, align=center]{$q_{i^\prime} = q_0$, $q_{j^\prime} = q_3$, \\ $q_{k^\prime} = \{q_0,q_1,q_2,q_3\}$};
\end{tikzpicture}}
}
\caption{\justifying Example fragment quantum circuit of $U_{ijk}$ operators when fixing $q_{i^\prime}=q_0$ and each register has 4 qubits.}
\label{fig:third examples}
\end{figure*}
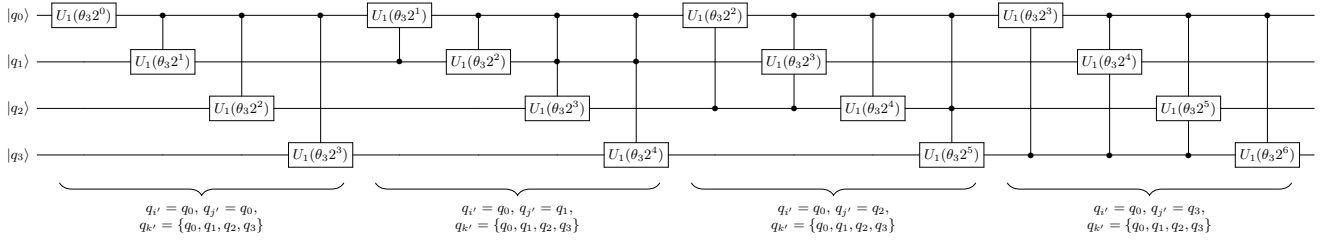
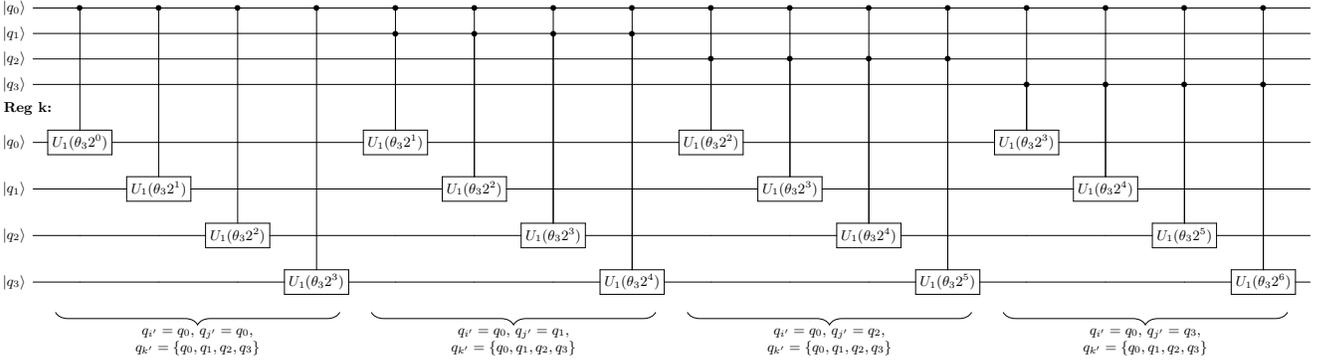

This schematic circuit corresponds to the case $i\neq j \neq k$, where all indices differ and three distinct normal modes are involved. Its fully expanded representation can be depicted by traversing all control-target configurations across the three registers, which manifests as $n^3$ possible combinations of the involved qubits $\{q_{i^\prime},q_{j^\prime},q_{k^\prime}\}$ and consequently $n^3$ controlled-controlled gates.

In cases where two or all three indices in $\{i,j,k\}$ coincide, the overlap of registers reduces the number of control nodes. For the $i=j=k$ case, all controlling and target qubits ($q_{i^\prime}$, $q_{j^\prime}$ and $q_{k^\prime}$) reside within the same $n$-qubit register. Similarly, in the $i=j\neq k$ case, the circuit complexity is reduced as $x_{i}$ and $x_{j}$ refer to the same normal mode, allowing $q_{i^\prime}$ and $q_{j^\prime}$ within the same register to coincide under certain configurations. When allocating $n=4$ qubits for each register, we present two example fragment circuits in Figure~\ref{fig:third examples} to illustrate $i=j=k$ and $i=j\neq k$ cases. These examples fix the controlling qubit $q_{i^\prime} = q_0$ and iteratively traverse all possible controlling $q_{j^\prime}$ and target $q_{k^\prime}$ values. The proposed architecture is readily adaptable to other values of $q_{i^\prime}$ (i.e. $\{q_1,q_2,q_3\}$), by following the consistent logic.

Scaling to the total circuit depth involves summing contributions from individual fragment circuits. Given that there are 4 possible values of $q_{i^\prime}$ under the $n=4$ setting, the full circuit depth for either $i=j=k$ or $i=j\neq k$ cases is 4 times the 16 gate count of the fragment circuits depicted in Figure~\ref{fig:third examples}. Extending this approach to $n$-qubit registers also yields a generalised gate count of $n^3$. 

We further decompose these multi-controlled structures into circuits operating at a more elementary level. Since a $\mathbf{C^2U}$ can be implemented using 5 two-qubit gates,\cite{nielsen2010} we accordingly convert the naive gate estimate of $n^3$ for $U_{ijk}$, to $5n^3-12n^2+8n$, $5n^3-4n^2$ and $5n^3$ for the cases $i=j=k$, $i=j\neq k$ and $i\neq j\neq k$, respectively, as listed in Table~\ref{Fragment Gate}.

\subsubsection{Gates for quartic polynomials}

\begin{figure}[htbp]
\centering
\scalebox{1.1}{
\Qcircuit @C=1em @R=1.5em {
\lstick{\textbf{Reg i:}\ket{q_{i^\prime}}} &\qw{/}^{n}& \ctrl{1} &\qw \\
\lstick{\textbf{Reg j:}\ket{q_{j^\prime}}} &\qw{/}^{n}& \ctrl{1} &\qw  \\
\lstick{\textbf{Reg k:}\ket{q_{k^\prime}}} &\qw{/}^{n}& \ctrl{1} &\qw  \\
\lstick{\textbf{Reg l:}\ket{q_{l^\prime}}} &\qw{/}^{n}&\gate{U_1(\theta_4 2^{q_{i^\prime}+q_{j^\prime}+q_{k^\prime}+q_{l^\prime}})} &\qw  \\
}}
\caption{\justifying Quantum circuit for implementing $U_{ijkl}$, the time evolution operators of fourth-order polynomial terms.}
\label{fig:fourth}
\end{figure}
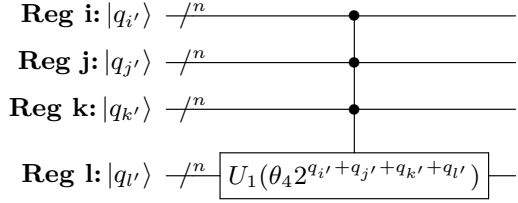

The fourth-order polynomial terms generate time evolution operators of the form $U_{ijkl}$, which requires circuits with up to 3-controlled operations. By defining $\theta_4 = c$, Figure~\ref{fig:fourth} schematically illustrates the circuit of $U_{ijkl}$ for the generic case $i\neq j\neq k\neq l$ (i.e. $x_i$, $x_j$, $x_k$ and $x_l$ indicate four different registers). This configuration results in $n^4$ $\mathbf{C^3U_1}$ gates.

As suggested in\cite{nielsen2010}, any multi-controlled gate $\mathbf{C^n{U}}$ (with $\mathbf{n}>2$) can be constructed from $2(\mathbf{n}-1)$ $\mathbf{C^2{U}}$ plus a two-qubit operations, using a minimal number of $\mathbf{n}-1$ ancillary qubits initialised and returned to the state $\ket{0}$. Specifically, a $\mathbf{C^3U}$ gate requires 2 ancillary qubits and decomposes into 4 $\mathbf{C^2U}$ gates and 1 two-qubit gate. Given that each $\mathbf{C^2U}$ decomposes into 5 two-qubit gates, the total cost amounts to 21 two-qubit gates. The circuit shown in Figure~\ref{fig:fourth} therefore translates to $21n^4$ two-qubit gates with 2 ancillary qubits. Other scenarios of $\{i,j,k,l\}$, where some of the indices coincide, exhibit reduced circuit depth due to overlaps in control qubits. The corresponding fragment circuits are outlined in Appendix~\ref{App evolve}. We herein classify the gate depth of $U_{ijkl}$ into various $\{i,j,k,l\}$ cases in Table~\ref{Fragment Gate}.

\begin{table}[htbp]
    \centering 
    \renewcommand{\arraystretch}{1.3}
    \caption{\justifying General gate depth calculations of evolution operators $U_{i}$, $U_{ij}$, $U_{ijk}$ and $U_{ijkl}$.} 
    \label{Fragment Gate} 
    \begin{tabular}{|c<{\centering\arraybackslash}|c<{\centering\arraybackslash}|c<{\centering\arraybackslash}|c<{\centering\arraybackslash}|} 
        \hline
        \multicolumn{2}{|c|}{Unitary} & Gate Depth & $n=4$ \\ 
        \hline
        \multicolumn{2}{|c|}{$U_{0}$} & $4$ & 4 \\
        \hline
        \multicolumn{2}{|c|}{$U_{i}$} & $n$ & 4 \\
        \hline
        \multirow{2}{*}{$U_{ij}$}& $i=j$ & $n^2$ & 16\\
        \cline{2-4}
        & $i\neq j$ & $n^2$ & 16\\
        \hline
        \multirow{3}{*}{$U_{ijk}$}& $i=j=k$ & $5n^3-12n^2+8n$ &160\\ 
        \cline{2-4}
        & $i=j\neq k$ & $5n^3-4n^2$ & 256\\
        \cline{2-4}
        &$i\neq j\neq k$ & $5n^3$ & 320\\
        \hline
        \multirow{5}{*}{$U_{ijkl}$}& $i=j=k=l$  & $21n^4-96n^3+148n^2-72n$ & 1312\\
        \cline{2-4}
        & $i=j=k\neq l$ & $21n^4-48n^3+28n^2$ & 2752 \\   
        \cline{2-4}
        & $i=j\neq k=l$  & $21n^4-32n^3+12n^2$ & 3520 \\ 
        \cline{2-4}
        & $i=j\neq k\neq l$ &  $21n^4-16n^3$ & 4352\\
        \cline{2-4}
        & $i\neq j\neq k\neq l$ &  $21n^4$ & 5376\\
        \hline
    \end{tabular}
\end{table}


\subsection{Non-Unitary Dipole Operator}
\label{dipole operator}

The quantum simulation begins with the application of the dipole operator onto the initial state. As discussed in Section~\ref{ISP}, a harmonic approximation to the initial state is feasible and effective. Similarly, approximations in the dipole operator are also explored to reduce computational complexity. 
In agreement with standard practice in vibrational spectroscopy, we find that truncating the dipole operator to first order is sufficient to reproduce the fundamental bands (see section~\ref{Characteristic Peak Analysis}). Strictly speaking, a purely linear dipole leads to vanishing intensity for overtone transitions. However, in the context of our simulations, the anharmonic Hamiltonian produces state mixing that yields non-zero overtone signals even when the dipole is linear, although accurate calculations of their absolute intensities does require inclusion of higher-order terms.
No matter how high or low the polynomial truncation is taken, difficulties in quantum implementation arise due to the inherently non-unitary nature of the dipole operator, which inevitably requires another approximation strategy that balances computational feasibility with physical fidelity for its quantum circuit constructions. 

The non-unitary challenge presented from the dipole operator is similar to that of imaginary time evolution in requiring a non-unitary operation, where probabilistic quantum circuit designs become highly applicable. Drawing inspiration from the probabilistic imaginary time evolution illustrated in~\cite{Kosugi2022}, we build on their circuit architecture for applications of our non-unitary dipole operator. The realisation of $\mu^{(\alpha)}$ requires an ancillary qubit to mediate probabilistic control, together with a phase operator
\begin{align}
    \Theta = \arccos\left({\frac{\tilde{\mu}^{(\alpha)}+\sqrt{1-(\tilde{\mu}^{(\alpha)})^2}}{\sqrt{2}}}\right).
    \label{exactdip}
\end{align}
To define this phase angle, we rescale the physical dipole operator by a fixed constant $\beta$ with the same physical units and introduce the dimensionless quantity $\tilde{\mu}=\mu/\beta$. For Eq.~\ref{exactdip} to be well defined, $\beta$ is chosen to satisfy $\|\tilde\mu\|_\infty \leq 1$, ensuring $\Theta$ is real-valued.

\begin{figure}[htbp]
\flushright
\scalebox{1}{
\Qcircuit @C=1em @R=1.5em {
\makebox[4em][l]{\textbf{Ancilla:}}\\
\lstick{\ket{0}} &\gate{H} & \gate{W} & \ctrlo{2}& \ctrl{2}& \gate{R_z(-2\theta_0)}& \gate{W^{\dagger}}&\qw \\
\makebox[4em][l]{\textbf{System:}}\\
\lstick{\ket{\Psi(t_0)}} &\qw{/}^{3n}&\qw & \gate{e^{i\theta_1}}& \gate{e^{-i\theta_1}} &\qw &\qw &\qw \\
}}
\caption{\justifying Quantum circuit for implementing the non-unitary dipole moment operator $\mu^{(\alpha)}$. The single-qubit gate $\mathbf{W}$ is defined as $\mathbf{W} = \frac{1}{\sqrt{2}}\begin{pmatrix}
    1 & -i   \\
    1  & i
    \end{pmatrix}$.}
\label{fig:non-unitary dipole}
\end{figure}
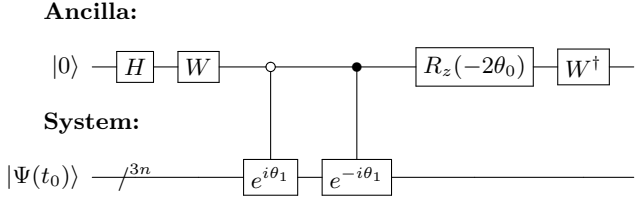

Since the exact calculation of $\Theta$ is analytically intractable, we expand it as a Taylor series in powers of $\tilde{\mu}^{(\alpha)}$:
\begin{equation}
    \Theta \approx \frac{\pi}{4}-\tilde{\mu}^{(\alpha)}.
    \label{taylordipole}
\end{equation}
This approximation is a small-parameter expansion and thus requires $\|\tilde\mu\|_\infty \ll 1$, so that the neglected $\mathcal{O}(\tilde{\mu}^2)$ terms are negligible. Consequently, the choice of $\beta$ is not unique, as long as it is sufficiently large to ensure that the overall magnitude of $\tilde\mu$ lies within the regime of validity of the Taylor truncation.
Accordingly, the practical circuit employs two explicit phases, $\theta_0=\frac{\pi}{4}$ and $\theta_1=-\tilde{\mu}^{(\alpha)}$, as depicted in Figure~\ref{fig:non-unitary dipole}.

This circuit converts the $\ket{\Psi(t_0)}\otimes\ket{0}$ to the $$\tilde{\mu}^{(\alpha)}\ket{\Psi(t_0)}\otimes\ket{0}+\sqrt{1-(\tilde{\mu}^{(\alpha)})^2}\ket{\Psi(t_0)}\otimes\ket{1},$$ which indicates that the successful application of $\tilde\mu^{(\alpha)}$ is contingent on measuring the ancilla qubit in the $\ket{0}$ state, with failure cases (ancilla in $\ket{1}$) rendering the results invalid. Despite the inherent probabilistic nature, this approach offers exceptional adaptability in tackling non-unitary operations for quantum hardware implementation.

We evaluate the inner product between the exact dipole-operated state and the approximate counterpart generated by the circuits in Figure~\ref{fig:non-unitary dipole}, obtaining an overlap greater than 0.99999. 
Furthermore, across simulations with various time parameters, the fundamental bands obtained using the approximate preparation, where first-order dipole operators are implemented via the approximate preparation circuit and applied to the approximate harmonic ground state (see Table~\ref{approxdipoledata} in Appendix~\ref{app:watertable}), show excellent agreement with those obtained from the corresponding exact preparation (see the last row of Table~\ref{vary time} in Section~\ref{time parameters}).
Such close correspondence demonstrates the robustness of Eq.~\ref{taylordipole}, enabling us to bypass the mathematical complexity related to $\Theta$ and simplify the circuit with high fidelity.

By exploiting the success probability of the ancilla qubit, we implement the non-unitary dipole moment operator via a block-encoding strategy. Concretely, the operation $e^{\pm i\theta_1}$ used in Figure~\ref{fig:non-unitary dipole} is decomposed into a sequence of unitary blocks specified as: 
\begin{equation}
    e^{\pm i\theta_1}=e^{\pm i\tilde{\mu}^{(\alpha)}}=\prod_i U_{\mu,i}^{(\alpha)} \prod_{ij} U_{\mu,ij}^{(\alpha)} \prod_{ijk} U_{\mu,ijk}^{(\alpha)} \cdots,
\end{equation}
where $\tilde\mu^{(\alpha)}$ accommodates polynomial terms of different orders, depending on the level of truncation selected. The subscripts of $U_\mu$ blocks label the polynomial orders of their exponents in $Q$ and each of these unitaries is further decomposed into elementary phase operators according to Section~\ref{elementary}:
\begin{align}
    U_{\mu,i}^{(\alpha)} 
    &= e^{\pm i\tilde{\mu}_i^{(\alpha)} Q_{i}} =e^{\pm i\tilde{\mu}_i^{(\alpha)} (-L/2+xdQ_{i})}  \nonumber \\
    &=\prod_{\mathcal{S}_{i}}  U_{\mathcal{S}_{i}}, \\
    U_{\mu,ij}^{(\alpha)} 
    &= e^{\pm i\tilde{\mu}_{ij}^{(\alpha)} \hspace{-0.8em}\prod\limits_{s\in\{i,j\}}\hspace{-0.8em}Q_{s}} =e^{\pm i\tilde{\mu}_{ij}^{(\alpha)} \hspace{-0.8em}\prod\limits_{s\in\{i,j\}}\hspace{-0.7em}(-L/2+xdQ_{s})} \nonumber  \\
    &=\prod_{\mathcal{S}_{ij}}  U_{\mathcal{S}_{ij}}, \\
    U_{\mu,ijk}^{(\alpha)} 
    &= e^{\pm i\tilde{\mu}_{ijk}^{(\alpha)} \hspace{-1em}\prod\limits_{s\in\{i,j,k\}}\hspace{-1em}Q_{s}} =e^{\pm i\tilde{\mu}_{ijk}^{(\alpha)} \hspace{-1em}\prod\limits_{s\in\{i,j,k\}}\hspace{-1em}(-L/2+xdQ_{s})}\nonumber   \\
    &=\prod_{\mathcal{S}_{ijk}} U_{\mathcal{S}_{ijk}}.
\end{align}
Here $\mathcal{S}_{i}\in\{0,i\}$, $\mathcal{S}_{ij}\in\{0,i,j,ij\}$ and $\mathcal{S}_{ijk}\in\{0,i,j,k,ij,ik,jk,ijk\}$. For notational clarity we denote the elementary operators as $U_{\mathcal{S}}$ and omit their difference in block-specific constant prefactors.

After grouping phase operators with identical subscripts originating from different $U_\mu$ blocks, the circuit depth is determined by the number of remaining effective elementary operators.
The ancilla-specific gates in Figure~\ref{fig:non-unitary dipole}, such as $\mathbf{H}$, $\mathbf{W}$, $\mathbf{R_z}$ and $\mathbf{W^{\dagger}}$, are aligned in parallel with neighbouring initial state preparation and time evolution segments, thereby avoiding any contribution to overall circuit depth.

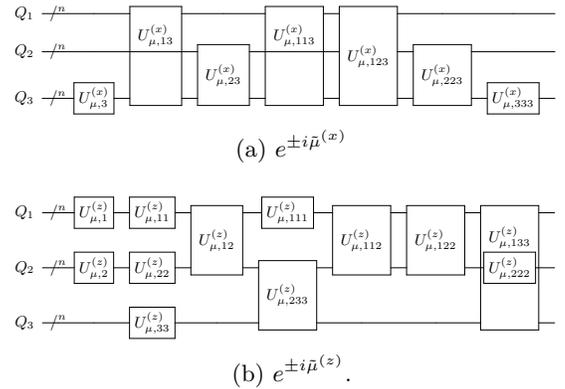
\begin{figure}[htbp]
\centering
\subfloat[$e^{\pm i\tilde{\mu}^{(x)}}$]{
\scalebox{0.65}{
\Qcircuit @C=1em @R=1.5em {
\lstick{Q_1}&\qw{/}^{n} &\qw & \multigate{2}{\raisebox{7ex}{$U_{\mu,13}^{(x)}$}} &\qw & \multigate{2}{\raisebox{7ex}{$U_{\mu,113}^{(x)}$}} & \multigate{2}{U_{\mu,123}^{(x)}} &\qw &\qw &\qw \\
\lstick{Q_2}&\qw{/}^{n} &\qw & \qw & \multigate{1}{U_{\mu,23}^{(x)}} & \qw & \ghost{U_{\mu,123}^{(x)}} & \multigate{1}{U_{\mu,223}^{(x)}} &\qw &\qw \\
\lstick{Q_3}&\qw{/}^{n}&\gate{U_{\mu,3}^{(x)}} & \ghost{\raisebox{7ex}{$U_{\mu,13}^{(x)}$}} & \ghost{U_{\mu,23}^{(x)}} & \ghost{\raisebox{7ex}{$U_{\mu,113}^{(x}$}} & \ghost{U_{\mu,123}^{(x)}} & \ghost{U_{\mu,223}^{(x)}} &\gate{U_{\mu,333}^{(x)}}&\qw \\
}}}
\\
\vspace{1em}
\subfloat[$e^{\pm i\tilde{\mu}^{(z)}}$.]{
\scalebox{0.65}{
\Qcircuit @C=1em @R=1.5em {
\lstick{Q_1}&\qw{/}^{n}&\gate{U_{\mu,1}^{(z)}} & \gate{U_{\mu,11}^{(z)}} & \multigate{1}{U_{\mu,12}^{(z)}} & \gate{U_{\mu,111}^{(z)}} & \multigate{1}{U_{\mu,112}^{(z)}} & \multigate{1}{U_{\mu,122}^{(z)}} & \multigate{2}{\raisebox{10ex}{$U_{\mu,133}^{(z)}$}} &\qw\\
\lstick{Q_2}&\qw{/}^{n}&\gate{U_{\mu,2}^{(z)}} & \gate{U_{\mu,22}^{(z)}} & \ghost{U_{\mu,12}^{(z)}} & \multigate{1}{U_{\mu,233}^{(z)}} & \ghost{U_{\mu,112}^{(z)}} & \ghost{U_{\mu,122}^{(z)}} &\gate{U_{\mu,222}^{(z)}} &\qw\\
\lstick{Q_3}&\qw{/}^{n}&\qw & \gate{U_{\mu,33}^{(z)}}&\qw & \ghost{U_{\mu,233}^{(z)}} &\qw &\qw & \ghost{\raisebox{10ex}{$U_{\mu,133}^{(z)}$}} &\qw\\
}}
}
\caption{\justifying Unitary blocks required for implementing $e^{\pm i\tilde{\mu}^{(\alpha)}}$ on the water system.}
\label{dipolesequence}
\end{figure}

\begin{table}[htbp]
    \centering 
    \caption{\justifying Elementary phase operators required for applying $e^{\pm i\tilde{\mu}^{(\alpha)}}$ of water, with the associated gate depths.} 
    \label{Dipole Count} 
    \renewcommand{\arraystretch}{1.3}
    \begin{tabular}{|c<{\centering\arraybackslash}|c<{\centering\arraybackslash}|c<{\centering\arraybackslash}|} 
        \hline
        $e^{\pm i\tilde{\mu}^{(x)}}$& Gate Depth & $n=4$ \\
        \hline
        $U_{0}$ & 4 & 4\\
        \hline
        $ \{U_1,U_2,U_3\}$ & $3n$ & 12\\
        \hline
        \makecell[c]{\rule{0pt}{2.2ex}$ \{U_{11},U_{22},U_{33},$\\$U_{12},U_{13},U_{23}\}$} & $6n^2$ & 96 \\
        \hline
        $ \{U_{113},U_{123},U_{223},U_{333}\}$ & $20n^3-20n^2+8n$ & 992 \\
        \hline
        Total& $20n^3-14n^2+11n+4$ & 1104\\
        \hline
        \hline
        $e^{\pm i\mu^{(z)}}$& Gate Depth & $n=4$ \\
        \hline
        $U_{0}$ & 4 & 4\\
        \hline
        $ \{U_1,U_2,U_3\}$ & $3n$ & 12\\
        \hline
        \makecell[c]{\rule{0pt}{2.2ex}$ \{U_{11},U_{22},U_{33},$\\$U_{12},U_{13},U_{23}\}$} & $6n^2$ & 96 \\
        \hline
        \makecell[c]{\rule{0pt}{2.2ex}$ \{U_{111},U_{112},U_{122},$\\$U_{133},U_{222},U_{233}\}$} & $30n^3-40n^2+16n$ & 1344 \\
        \hline
        Total& $30n^3-34n^2+19n+4$ & 1456 \\
        \hline
    \end{tabular}
\end{table}

\begin{figure*}[htbp]
\centering
\scalebox{0.9}{
\Qcircuit @C=1em @R=1.5em {
\lstick{Q_1}&\qw{/}^{n} &\gate{U_{V,11}} &\gate{U_{V,111}}  &\multigate{1}{U_{V,112}} & \multigate{1}{U_{V,122}} & \multigate{2}{\raisebox{7ex}{$U_{V,133}$}} &\gate{U_{V,1111}}& \multigate{2}{\raisebox{7ex}{$U_{V,1133}$}} &\multigate{1}{U_{V,1122}} & \multigate{1}{U_{V,1112}} &\multigate{1}{U_{V,1222}} &\multigate{2}{U_{V,1233}} &\qw\\
\lstick{Q_2}&\qw{/}^{n} &\gate{U_{V,22}} & \multigate{1}{U_{V,233}} & \ghost{U_{V,112}} & \ghost{U_{V,122}} &\gate{U_{V,222}} & \multigate{1}{U_{V,2233}}&\gate{U_{V,2222}} & \ghost{U_{V,1122}} & \ghost{U_{V,1112}} & \ghost{U_{V,1222}} & \ghost{U_{V,1233}} &\qw\\
\lstick{Q_3}&\qw{/}^{n} &\gate{U_{V,33}} & \ghost{U_{V,233}} &\qw &\qw & \ghost{U_{V,133}} & \ghost{U_{V,2233}} & \ghost{U_{V,1133}} &\gate{U_{V,3333}} &\qw &\qw & \ghost{U_{V,1233}} &\qw\\
}}
\caption{\justifying Unitary blocks required for implementing $U_V=e^{-iVdt/\hbar}$ on the water system.}
\label{Vsequence}
\end{figure*}
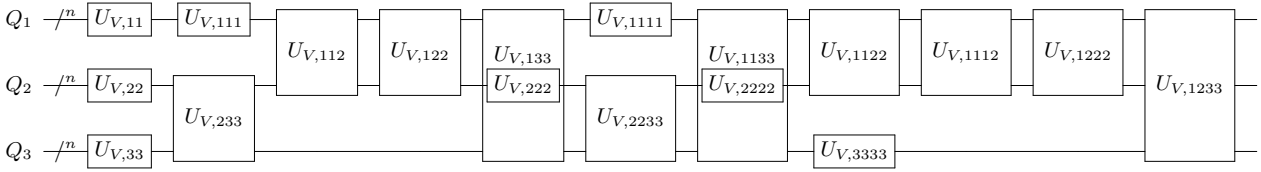

\begin{table*}[btp]
    \centering 
    \renewcommand{\arraystretch}{1.3}
    \caption{\justifying Gate depth for applying $U_V$ of water.} 
    \label{UV Count} 
    \begin{tabular}{|c<{\centering\arraybackslash}|c<{\centering\arraybackslash}|c<{\centering\arraybackslash}|} 
        \hline
        Elementary Phase Operators Required for $U_V$ & Gate Depth & $n=4$ \\
        \hline
        $U_{0}$ & 4 & 4\\
        \hline
        $ \{U_1,U_2,U_3,$\} & $3n$ & 12\\
        \hline
        $ \{U_{11},U_{22},U_{33},U_{12},U_{13},U_{23}\}$ & $6n^2$ & 96 \\
        \hline
        $\{U_{111},U_{112},U_{113},U_{122},U_{123},U_{133},U_{222},U_{223},U_{233},U_{333}\}$ & $50n^3-60n^2+24n$& 2336 \\
        \hline
        $\{U_{1111},U_{1112},U_{1122},U_{1133},U_{1222},U_{1233},U_{2222},U_{2233},U_{3333}\}$ & $189n^4-496n^3+536n^2-216n$ &  24352 \\
        \hline
        Total $U_V$& $189n^4-446n^3+482n^2-189n+4$ & 26800\\
        \hline
    \end{tabular}
\end{table*}

As an illustrative example, we consider the water molecule oriented in the $xz$-plane with dipole operators $\mu^{(x)}$ and $\mu^{(z)}$, both truncated at third order. With $n$ qubits per normal mode register, the required unitary blocks are explicitly illustrated in Figure~\ref{dipolesequence}, and the decomposition of $e^{\pm i\tilde{\mu}^{(\alpha)}}$ is summarised in Table~\ref{Dipole Count} with corresponding depth estimates.
Using the basic gate counts from Table~\ref{Fragment Gate}, the circuit depth for implementing the $e^{\pm i\tilde{\mu}^{(\alpha)}}$ scales as $20n^3-14n^2+11n+4$ gates for $\alpha=x$ and $30n^3-34n^2+19n+4$ gates for $\alpha=z$ (e.g. 1,104 and 1,456 gates when $n=4$). As both the positive and negative exponentials must be executed under ancilla control, the total cost for each dipole operator is twice these values. Comparatively, a first-order truncation in $\mu^{(\alpha)}$ offers a low-depth alternative, demanding only $U_{\mu,i}^{(\alpha)}$ operations with a total cost of $2n+8$ and $4n+8$ gates for $\alpha=x$ and $\alpha=z$, respectively.


\subsection{SO-QFT Time Evolution}
\label{Time Evolve Theory}

The preliminary operation of dipole moment functions establishes a dipole operated ground state as the starting state for the following time evolution, which we perform using the second-order SO-QFT method.
The potential contribution to the time-evolution operator produces exponentials whose phases are polynomials in the normal mode coordinate $Q$; specifically, second-, third- and fourth-order terms arise.
We therefore express the potential propagator as
\begin{equation}
    e^{-iVdt/\hbar} = \prod_{ij} U_{V,ij} \prod_{ijk}U_{V,ijk} \prod_{ijkl}U_{V,ijkl} \cdots. 
\end{equation}

Mapping $Q$ to the computational basis states $x$ decomposes these unitary blocks into elementary phase operators defined in Section~\ref{elementary}:
\begin{align}
    U_{V,ii} &=e^{-ib_{ii}Q_{i}^2} =e^{- ib_{ii}(-L/2+xdQ_{i})^2} \nonumber \\
    &=\prod_{\mathcal{S}_{ii}}U_{\mathcal{S}_{ii}}, \; \text{with} \; b_{ii}=2\pi c\omega_idt\\
    U_{V,ijk}&=e^{-ib_{ijk}\hspace{-1.2em}\prod\limits_{s\in\{i,j,k\}}\hspace{-1.2em}Q_{s}}=e^{-ib_{ijk}\hspace{-1.2em}\prod\limits_{s\in\{i,j,k\}}\hspace{-1.1em}(-L/2+xdQ_{s})} \nonumber \\
    &=\prod_{\mathcal{S}_{ijk}}U_{\mathcal{S}_{ijk}},\; \text{with} \; b_{ijk}=2\pi ck_{ijk}dt\\
    U_{V,ijkl}&=e^{-ib_{ijkl}\hspace{-1.4em}\prod\limits_{s\in\{i,j,k,l\}}\hspace{-1.4em}Q_{s}}=e^{- ib_{ijkl}\hspace{-1.4em}\prod\limits_{s\in\{i,j,k,l\}}\hspace{-1.3em}(-L/2+xdQ_{s})} \nonumber \\
    &=\prod_{\mathcal{S}_{ijkl}}U_{\mathcal{S}_{ijkl}}, \; \text{with} \; b_{ijkl}=2\pi ck_{ijkl}dt
\end{align}
where $\mathcal{S}_{ii}\in\{0,i,ii\}$, $\mathcal{S}_{ijk}\in\{0,i,j,k,ij,ik,jk,ijk\}$ and $\mathcal{S}_{ijkl}\in\{0,i,j,k,ij,ik,jk,ijk,ijl,ikl,jkl\}$ enumerate all elementary operator types generated by the respective polynomial structures. Grouping terms with identical subscripts into a single elementary $U$-operator reduces the circuit depth to the count of remaining effective elementary operators.

To enable extraction of the autocorrelation function during time evolution, the $U_{V}$, $U_{K}$ and QFT blocks are globally controlled by a common ancillary qubit (see Section~\ref{Auto}). This prevents parallel execution of their internal operations, leading to a circuit depth that scales linearly with the number of normal modes.

Using the water force field as an example, we depict the unitary blocks that implement its $U_V$ in Figure~\ref{Vsequence}, together with the associated registers; their decomposition and the resulting single-step gate count are reported in Table~\ref{UV Count}. 
The kinetic component $U_K$ is local and diagonal in the momentum space with a quadratic form, and thus scaling as $n^2+n+4$ gates for an $n$-qubit register. 
In Table~\ref{water total} we present the gate contributions from all potential and kinetic components with the QFT processes. The total number of gates required for one time step of time evolution on the water system is $189n^4-446n^3+491n^2-177n+28$ (i.e., 27,016 for $n=4$). Extending this to the complete time evolution segment, the cumulative circuit depth becomes $n_t$ multiples of the single time step gate count, with $n_t$ being the time steps propagated. 

\begin{table}[htbp]
    \centering
    \caption{\justifying Gate depth calculations of the time evolution section on the water system within single time step.}
    \label{water total}
    \renewcommand{\arraystretch}{1.3}
    \begin{tabular}{|c<{\centering\arraybackslash}|c<{\centering\arraybackslash}|c<{\centering\arraybackslash}|}
         \hline
         Evolution Blocks & Gate Depth & $n=4$ \\
         \hline
         $U_V$  & \makecell[c]{\rule{0pt}{2.2ex}$189n^4-446n^3+482n^2$\\$-189n+4$} & 26800 \\
         \hline
        $U_K$  & $(n^2+n+4)\times3$ registers & 72 \\
        \hline
        QFT (or QFT$^{-1}$)  & $(n^2/2+n)\times3$ registers & 36 \\
        \hline
        $U_K$QFT$^{-1}$$U_V$QFT$U_K$  & \makecell[c]{\rule{0pt}{2.2ex}$189n^4-446n^3+491n^2$\\$-177n+28$} & 27016 \\
        \hline
    \end{tabular}
\end{table}

\subsection{Simulation of IR Spectra}
\label{Auto}

\begin{figure}[hbtp]
\centering
\begin{subfigure}{0.23\textwidth}
    \centering
    \scalebox{0.9}{
    \Qcircuit @C=1em @R=1.5em {
    \lstick{\ket{0}} & \gate{H} & \ctrl{1}   & \gate{H}  & \meter  \\
    \lstick{\ket{\psi}} & \qw {/}^{3n} & \gate{U} & \qw & \qw  
    }}
    \captionsetup{font=scriptsize} \caption{$\operatorname{\mathbb{R}e}\{A_\mu (t)\}=P_\text{a}(0)-P_\text{a}(1)$}
    \label{fig:a}
\end{subfigure}
\hspace{0.01\textwidth}
\begin{subfigure}{0.23\textwidth}
    \centering
    \scalebox{0.9}{
    \Qcircuit @C=1em @R=1.5em {
    \lstick{\ket{0}} & \gate{H} & \ctrl{1} & \gate{S}  & \gate{H}  & \meter  \\
    \lstick{\ket{\psi}} & \qw {/}^{3n} & \gate{U} & \qw & \qw  & \qw
    }}
    \captionsetup{font=scriptsize} \caption{$\operatorname{\mathbb{I}m}\{A_\mu (t)\}=P_\text{a}(1)-P_\text{a}(0)$}
    \label{fig:b}
\end{subfigure}
\caption{\justifying Hadamard tests for measuring (a) $\operatorname{\mathbb{R}e}\{A_\mu (t)\}$ and (b) $\operatorname{\mathbb{I}m}\{A_\mu (t)\}$ at a single time point using an ancillary qubit. Here $P_\text{a}(0)$ and $P_\text{a}(1)$ represent the probability of measuring the ancilla in state $\ket{0}$ and $\ket{1}$.}
\label{fig:had}
\end{figure}
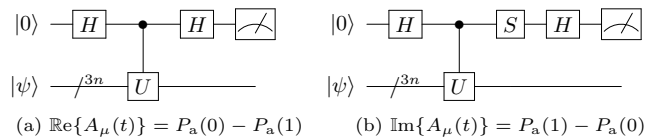

We eventually yield the IR spectra through Fourier Transform on the time-dependent dipole-dipole autocorrelation, $A_\mu (t)$, which is measured at a series of time points throughout the time evolution process. The Hadamard test is a well-established quantum algorithmic technique for estimating inner products and expectation values.\cite{Aharonov2009, Mitarai2019, Abhijith2022} It works by preparing an ancillary qubit in a superposition state and applying controlled operations conditioned on this qubit. As shown in Figure~\ref{fig:had}, repeated measurements of the ancilla enables calculations of real and imaginary parts of $A_\mu (t)$ from probability differences.

\section{Characteristic Peak Analysis}
\label{Characteristic Peak Analysis}

We have developed a time-dependent SO-QFT framework for modeling infrared spectra within a grid-based representation. We now use the water molecule as a model system to evaluate the overall performance of the proposed algorithm from three perspectives: (i) its fidelity relative to variational reference results, (ii) the optimal choice of time-evolution parameters, and (iii) the robustness of approximation schemes introduced to reduce quantum circuit depth and gate complexity in initial-state preparation and dipole operator implementation. Specifically, we focus on band positions and absolute infrared intensities as key spectroscopic metrics.
The potential energy and dipole moment coefficients (Eqs.~\ref{vhat}-\ref{dipole function}) are taken from Ref.~\citen{Culot1992} (SD-CI/TZ${+}$2P). 

Uncertainties in the centroid positions are expressed as 95\% confidence intervals (CI), estimated from the weighted standard deviation and the $t$-distribution. 
Variations in window width across bands lead to variations in the spectral sample counts entering the analysis, and hence the width of the CI. The Fourier Transform of the autocorrelation yields a wavenumber resolution that is inversely proportional to the total propagation time $T$. Thus, even for a fixed band window, increasing $T$ results in finer wavenumber resolution, more data points within that window, and correspondingly narrower CIs.

\subsection{Fidelity Assessment}
\label{feasibility}

We first assess the feasibility of our SO-QFT approach by comparing the simulated characteristic bands against fully converged variational results from time-independent calculations. As detailed in Section~\ref{Background Theory}, we prepared the ground state via ITE and then time-evolve the dipole-operated ground state. To obtain a well-resolved autocorrelation function, we simulated for a total propagation time of 19750\,fs over 300,000 time steps. The results for the zero-point vibrational energy (ZPVE), fundamental and multi-quantum transitions (overtone and combination bands) are summarised in Table~\ref{validation tab}.

\begin{table}[htbp]
    \centering
    \caption{\justifying Comparison of SO-QFT calculations for water band positions (cm$^{-1}$) and integrated intensities (km\,mol$^{-1}$) against variational reference calculations. Integrated intensities are given in parentheses.}
    \label{validation tab} 
    \begin{ruledtabular}
    \begin{tabular}{llcccccc}
    & & & & \multicolumn{2}{c}{PyVCI} & &  \\
    \cline{5-6} 
    & & \multicolumn{2}{c}{SO-QFT} & I & II & \multicolumn{2}{c}{Ref.~\citen{Culot1992}} \\
    \hline
    ZPVE            & $a_1$ & 4776.2           &        & 4776.4 & 4765.4 & 4765 &        \\ \\
    $\nu_1$         & $a_1$ & 3809.1 $\pm$ 0.5 &  (4.6) & 3808.4 & 3808.4 & 3808 &  (4.7) \\
    $\nu_2$         & $a_1$ & 1624.3 $\pm$ 0.2 & (76.7) & 1623.9 & 1637.6 & 1637 & (76.8) \\
    $\nu_3$         & $b_2$ & 3876.3 $\pm$ 0.3 & (37.0) & 3875.6 & 3889.6 & 3889 & (39.9) \\ \\
    $\nu_1{+}\nu_2$ & $a_1$ & 5384.7 $\pm$ 1.7 &  (0.1) & 5384.2 & 5397.6 & 5397 &  (0.1) \\
    $\nu_1{+}\nu_3$ & $b_2$ & 7643.8 $\pm$ 1.5 &  (2.2) & 7642.4 & 7657.3 & 7656 &  (2.0) \\
    $\nu_2{+}\nu_3$ & $b_2$ & 5428.2 $\pm$ 0.5 &  (7.8) & 5427.5 & 5481.7 & 5431\rlap{\footnote{VPT2 calculations give 5475\,cm$^{-1}$, consistent with column II.}} &  (4.5) \\ \\
    $2\nu_1$        & $a_1$ & 7595.0 $\pm$ 2.1 &  (0.5) & 7594.7 & 7595.6 & 7595 &  (0.3)\\
    $2\nu_2$        & $a_1$ & 3207.2 $\pm$ 2.7 &  (0.7) & 3207.7 & 3235.2 & 3235 &  (0.6)\\
    $2\nu_3$        & $a_1$ & 7727.9 $\pm$ 3.6 &  (0.3) & 7723.9 & 7753.7 & 7752 &  (0.004)\\
    \end{tabular}
    \end{ruledtabular}
\end{table}

Because our simulations employ a simplified vibrational Hamiltonian that omits the Coriolis and pseudopotential terms, a direct comparison with the full-Watson results of Ref.~\cite{Culot1992} is not appropriate. We therefore generated two sets of auxiliary variational results using the PyVCI program:\cite{SIBAEV2016} one with the same working Hamiltonian adopted in this work (column I), and one with the full Watson Hamiltonian (column II). The agreement between column II and Ref.~\cite{Culot1992} supports the use of column I as a benchmark for assessing the SO-QFT performance under the simplified Hamiltonian. As shown in Table~\ref{validation tab}, the deviations in SO-QFT band positions are generally below 1 cm$^{-1}$, with only one transition exhibiting a deviation of 4 cm$^{-1}$. This level of consistency reflects the high numerical fidelity of our time-dependent approach.

\subsection{Selection of Time Parameters}
\label{time parameters}

An important consideration in simulating infrared spectra in a time-dependent ansatz is the length of the time evolution. A longer propagation period $T$ enables the simulation to capture more subtle oscillatory components, which translates into better resolved spectral features.
Additionally, a smaller time-step size $dt$, obtained by dividing $T$ into a larger number of time steps $n_t$ (with $n_t=T/dt$), improves numerical accuracy by reducing the Trotter error. However, as presented in Section~\ref{Time Evolve Theory}, the gate depth of the quantum circuits is proportional to the number of evolved time steps $n_t$. To ensure an efficient implementation on real quantum computers, we seek to find the optimal time settings that minimise circuit depth while maintaining high fidelity in characterising IR bands.


\begin{table*}[htbp]
    \centering
    \caption{\justifying Impact of various time parameters and approximations of $\mu\left|\Psi\right>$ on the fundamental and overtone bands of water simulated with SO-QFT.
    $\tilde{\nu}_c$ and $I_\mathrm{IR}$ are the centroid band positions (in cm$^{-1}$) and integrated band intensities (in km\,mol$^{-1}$), respectively.}
    \label{vary time}
    \begin{ruledtabular}
    \begin{tabular}{lllrrr cc cc cc cc}
    & & & & & & \multicolumn{2}{c}{$\nu_1$} & \multicolumn{2}{c}{$\nu_2$} & \multicolumn{2}{c}{$\nu_3$} & \multicolumn{2}{c}{$2\nu_1$} \\
    \cline{7-8} \cline{9-10} \cline{11-12} \cline{13-14}
    & $\mu$ & $\left|\Psi\right>$ & \multicolumn{1}{c}{$T$ (fs)} & \multicolumn{1}{c}{$n_t$} & \multicolumn{1}{c}{$dt$ (fs)} &
    \multicolumn{1}{c}{$\tilde{\nu}_\mathrm{c}$} & \multicolumn{1}{c}{$I_\mathrm{IR}$} &
    \multicolumn{1}{c}{$\tilde{\nu}_\mathrm{c}$} & \multicolumn{1}{c}{$I_\mathrm{IR}$} &
    \multicolumn{1}{c}{$\tilde{\nu}_\mathrm{c}$} & \multicolumn{1}{c}{$I_\mathrm{IR}$} &
    \multicolumn{1}{c}{$\tilde{\nu}_\mathrm{c}$} & \multicolumn{1}{c}{$I_\mathrm{IR}$} \\
    \hline
    A  & full   & exact &  1975 &  30,000 & 0.066 & 3808.1 $\pm$ 14.5 & (4.5) & 1624.5 $\pm$ 1.9 & (76.7) & 3875.8 $\pm$ 7.4 & (37.0) & 7593.2 $\pm$ 17.8 & (0.5)\\
       & full   & exact &  3950 &  60,000 & 0.066 & 3809.1 $\pm$  4.1 & (4.6) & 1624.4 $\pm$ 1.2 & (76.7) & 3876.3 $\pm$ 3.1 & (37.0) & 7594.9 $\pm$ 6.0 & (0.5)\\
       & full   & exact &  7900 & 120,000 & 0.066 & 3809.1 $\pm$  0.8 & (4.6) & 1624.3 $\pm$ 0.7 & (76.7) & 3876.4 $\pm$ 0.5 & (37.0) & 7594.8 $\pm$ 3.9 & (0.5)\\
       & full   & exact & 13165 & 200,000 & 0.066 & 3809.1 $\pm$  0.7 & (4.6) & 1624.3 $\pm$ 0.2 & (76.7) & 3876.4 $\pm$ 0.4 & (37.0) & 7595.0 $\pm$ 2.7 & (0.5)\\ \hline
    B  & full   & exact &  3950 &  20,000 & 0.198 & 3815.2 $\pm$  3.9 & (4.6) & 1627.8 $\pm$ 1.8 & (76.9) & 3882.7 $\pm$ 2.7 & (37.1) & 7600.3 $\pm$ 10.8 & (0.6) \\
       & full   & exact &  3950 &  40,000 & 0.099 & 3810.2 $\pm$  4.0 & (4.6) & 1624.9 $\pm$ 0.8 & (76.7) & 3877.5 $\pm$ 2.9 & (37.0) & 7595.7 $\pm$ 6.9 & (0.5)\\
       & full   & exact &  3950 &  60,000 & 0.066 & 3809.1 $\pm$  4.1 & (4.6) & 1624.4 $\pm$ 1.2 & (76.7) & 3876.3 $\pm$ 3.1 & (37.0) & 7594.9 $\pm$ 6.0 & (0.5)\\
       & full   & exact &  3950 &  80,000 & 0.049 & 3808.7 $\pm$  3.9 & (4.5) & 1624.2 $\pm$ 1.3 & (76.7) & 3875.9 $\pm$ 3.0 & (37.0) & 7594.6 $\pm$ 5.6 & (0.5)\\ \hline
    C  & full   & exact &  3950 &  60,000 & 0.066 & 3809.1 $\pm$  4.1 & (4.6) & 1624.4 $\pm$ 1.2 & (76.7) & 3876.3 $\pm$ 3.1 & (37.0) & 7594.9 $\pm$  6.0 & (0.5) \\
       & linear & exact &  3950 &  60,000 & 0.066 & 3809.1 $\pm$  4.0 & (5.8) & 1624.4 $\pm$ 1.2 & (75.7) & 3876.3 $\pm$ 3.1 & (40.8) & 7593.0 $\pm$ 10.4 & (0.1) \\
       & full   & harm. &  3950 &  60,000 & 0.066 & 3809.1 $\pm$  4.1 & (4.4) & 1624.4 $\pm$ 1.2 & (72.5) & 3876.3 $\pm$ 3.1 & (34.5) & 7595.5 $\pm$  3.6 & (1.8) \\
       & linear & harm. &  3950 &  60,000 & 0.066 & 3809.1 $\pm$  4.0 & (4.7) & 1624.4 $\pm$ 1.2 & (70.9) & 3876.3 $\pm$ 3.1 & (36.6) & 7595.4 $\pm$  4.1 & (1.2) \\
    \end{tabular}
    \end{ruledtabular}
\end{table*}


Table~\ref{vary time} (A and B) summarises the sensitivity of band positions and integrated band intensities to variations in the total simulation time $T$ and the time-step size $dt$, obtained by systematically altering $T$ at a constant $dt$ (A), and varying $dt$ for a fixed $T$ (B). We observe convergence behaviours of fundamental band positions as a function of the time parameters, stabilising with extended simulation durations or finer time resolutions. Among the various parameters tested, we identify $T\approx 3950$\,fs, discretised by 60,000 steps (i.e. $dt\approx 0.066$\,fs), as an optimal setting that achieves quantitatively accurate band positions. While further increases in $T$ or $n_t$ enhance performance, the associated growth in resource demands makes such adjustments suboptimal for practical implementations. According to Section~\ref{Time Evolve Theory}, with $n_t=60,000$ time steps, the circuit depth of the time evolution section becomes $1.6\times 10^9$ gates. 

The time parameters optimised for the fundamental bands were subsequently employed to compute higher-order overtone transitions (see Appendix~\ref{App overtone}). As expected, a longer propagation time $T$ yields improved resolution of weaker multi-quantum features. These results further demonstrate the capability of our time-dependent SO-QFT framework in reproducing vibrational spectra in the near-infrared region.


\subsection{Approximation Scheme Validation}
\label{approxpsidipole}

Our framework is specifically tailored for compatibility with quantum computing paradigms, where we tend to leverage various approximations in the dipole operated ground state $\mu\ket{\Psi}$ to minimise quantum resource requirements without sacrificing the fidelity.
We now analyse the consequences of truncating the dipole operator (Eq.~\ref{dipole function}) after the linear terms and replacing the exact anharmonic ground state with the harmonic oscillator ground state, i.e. a Gaussian distribution. Although the linear dipole approximation is frequently invoked in time-independent classical treatments, its role in balancing spectral accuracy and quantum circuit depth within a time-dependent framework merits explicit examination.


Using the optimised time parameters identified above, Table~\ref{vary time} (C) reports results for the various approximation combinations considered, demonstrating that simplifying the dipole operator or the initial state has negligible impact on fundamental band positions. This remarkable robustness persists across a range of time parameters, as confirmed by the complete datasets provided in Appendix~\ref{app:watertable}. In contrast, fundamental band intensities are slightly more sensitive to these approximations. This behaviour is expected, as the band centres are determined by the Hamiltonian governing the time evolution and therefore remain unaffected, whereas the intensities depend explicitly on the dipole operator and the vibrational wavefunction.

Multi-quantum transitions, such as overtones and combination bands, on the other hand show greater variation across approximation levels, as evidenced in Table~\ref{vary time}. A comprehensive set of overtones and propagation-time scans under different approximation schemes is provided in the Appendix~\ref{App overtone}. If the linear dipole approximation is used in conjunction with a purely harmonic Hamiltonian, these transitions are dipole-forbidden and therefore would carry zero intensity. In the present work, however, the Hamiltonian employed is anharmonic. The resulting state mixing relaxes the strict harmonic selection rules, so that non-zero intensities emerge even with a linear dipole operator.

Overall, these approximations provide a cost-efficient strategy for capturing qualitative spectral structure under constrained quantum resource budgets. Nonetheless, dipole truncation may distort intensity borrowing mechanisms and alter the relative intensity ratios of resonance-coupled bands, while the harmonic initial-state approximation changes vibrational wavefunction overlaps and thus affects transition amplitudes. Quantitative overtone and combination intensities therefore require the inclusion of higher-order dipole contributions and the exact anharmonic initial state.

\section{Discussion}
\label{Discussion}

The value in a quantum implementation is the ability to simulate in high-dimensionality without introducing severe approximations in the quantum dynamics. The present work has so far focused on the quantum-algorithmic problem of obtaining infrared spectra using the normal mode Hamiltonian with a polynomial expansion of the PES and DMS. Larger molecules, however, 
are rarely semi-rigid and exhibit conformational flexibility through bond rotations, and many vibrational eigenstates are delocalised across rotamers. Classical methods therefore often switch from a normal mode coordinate representation to an internal coordinate representation, which concentrates the limited classical resource on the relevant low-energy regions of Hilbert space. This is not necessary in quantum algorithms since we do not truncate Hilbert space, except through the discretisation and range of the coordinate grid. However, different coordinate choices have significant implications for the practicality of the SO-FQT approach, in terms of obtaining the potential energy, in preparing good initial states, and in performing the time evolution, which we now discuss.

\subsection{Electronic Structure Input}
Our algorithm assumes access to analytic PES and DMS representations. For realistic molecules, however, constructing such models remains computationally intensive. For example, direct evaluation of electronic energies and dipoles on dense grids to obtain a global PES becomes computationally prohibitive beyond small systems. Practical treatments therefore rely on parameterised representations, such as polynomial force fields, n-mode expansions, or other fitted models derived from a finite set of electronic-structure calculations.\cite{Guntram2004,Bowman2009,Ove2016,Tew2014,Tew2016,Guntram2024} 

Although this strategy avoids dense-grid sampling, the scaling of the parameterisation step with system size is not yet fully resolved and depends on the strategy employed, i.e. what coordinates and functional form are used to represent the PES. The number of expansion coefficients required for an accurate representation grows rapidly with dimensionality and expansion order, and each coefficient requires electronic-structure data whose cost itself increases steeply with molecular size.\cite{Tew2012} These upstream costs are intrinsic to Hamiltonian construction and arise independently of the subsequent propagation algorithm.

\subsection{Coordinate Choices and Trade-Offs}
The choice of coordinates determines how computational complexity is distributed between kinetic and potential operators. In the rectilinear coordinates adopted here, the constant volume element and the kinetic operator diagonal in momentum space enable an efficient SO–QFT implementation. The trade-off then shifts to the potential term. For example, a torsional motion involves large displacements in many normal coordinates, and the potential can no longer be represented accurately through low-order Taylor expansion.

Many molecules have deep conformational wells, where the low-energy vibrational states are localised to one conformer. Here, spectroscopic accuracy can be obtained through a normal mode representation with higher-order force fields (e.g., sextic expansions). The zero point vibrational state is also well-approximated as a product of decoupled oscillators. In this regime, the scheme applied in this paper remains polynomial and systematically controllable. Propagators generated by polynomial potentials admit an explicit circuit implementation, with circuit depths scaling as $\mathcal{O}(n^k)$ for a $k^\text{th}$-order expansion on an $n$-qubit register. 

By contrast, for floppy molecules exhibiting large-amplitude motion, rectilinear representations typically require non-polynomial potentials, e.g., Morse functions, sums of Gaussians, splines or tabulated surfaces. Their propagators need to be synthesised as general unitaries via oracle-based constructions and thus demand large resource requirements. By using curvilinear coordinates, a more compact description of the potential can be obtained. Similarly, the initial state is well approximated as a product of oscillators in each coordinate. However, this scheme comes at the expense of introducing coordinate-dependent kinetic operators with non-trivial metric factors and volume elements. In such cases, the clean kinetic-potential split breaks down, which substantially increases the circuit depth and ancilla overheads. 

For complex vibrational systems, the central design question is therefore whether to simplify the kinetic structure or the potential representation. Within the SO-QFT scheme, preserving a diagonal kinetic operator tends to be the more robust choice, with potential complexity addressed through controlled expansion.

\subsection{Scope of SO–QFT Approach}
The SO–QFT algorithm developed here is not intended to resolve the construction of PES and DMS. Rather, it targets a distinct downstream task: given a Hamiltonian and dipole operator in a specified representation, it provides a time-dependent route to obtaining infrared spectra. Its principal strengths lie in exploiting diagonal kinetic structure and in favourable qubit-space scaling that allows access to Hilbert spaces beyond the reach of classical approaches. That said, its practical advantage ultimately depends on the availability of compact PES representations and on the extent to which strongly curvilinear couplings can be avoided or efficiently incorporated.

\section{Conclusion}
\label{Conclusion}

In this work, we develop and evaluate a time-dependent wavefunction-based methodology for simulating infrared spectra, accompanied by the design of a corresponding quantum circuit framework tailored for future quantum implementation. A series of uniformly controlled rotations first prepare the initial vibrational state of each normal mode in the grid based representation. For the following application of the dipole operator, we employ a probabilistic circuit design incorporating an ancillary qubit to handle its non-unitary nature. The time evolution is carried out by trotterised SO-QFT scheme under a suitably selected Hamiltonian. We illustrate the quantum circuits that manage the exponentiation of Hamiltonian terms across various polynomial orders with associated gate depth evaluations. The dipole-dipole autocorrelation is extracted at each time step through Hadamard test, and subsequently subjected to Fourier transformation which generates the final IR spectrum. Band positions and intensities are extracted through centroid and numerical integration analysis over the simulated spectrum, which deliver excellent agreement with time-independent references when using water molecule as a model system. 

While the water system is essentially a solved problem, with classical variational calculations already reproducing its rovibrational features to spectroscopic accuracy,\cite{Tennyson2016, Huang_JCP_2008_044312} it serves primarily as a benchmark for validating the proposed grid-based SO-QFT framework.
The principal merit of our framework lies in its direct compatibility for implementation on quantum hardware and thus the scalability to larger and more complex molecular systems, where classical treatments become computationally prohibitive beyond a few degrees of freedom.

Within this framework, the required qubit resources grow linearly with the number of vibrational modes. For the specific water model, this translates to a total of 16 qubits (4 qubits for each of the 3 normal mode register, 1 ancillary qubit for encoding the dipole operator, 2 ancillary qubits for the decomposition of multi-controlled gates, 1 ancillary qubit for the Hadamard test). 
The circuit depth of state preparation remains independent of the total number of modes, while the time evolution scales polynomially with the qubit count per mode $n$ and linearly with both the number of modes $d$ and the number of propagation time steps $n_t$: $\mathcal{O}(dn^4n_t)$. Here the linear dependence on the mode count arises from the global ancilla control required in the Hadamard-test measurement. Overcoming this restriction will require control-free strategies\cite{Yang2024, Lu2021, Brien2021, Moosa2023} that can restore the concurrent execution of kinetic, potential, or QFT blocks operating on disjoint registers.

Efforts to optimise our algorithm focus on minimising the total circuit depth, from either the aspect of optimal time parameters or cost-effective approximation schemes in the dipole operated ground state. We first systematically investigate the influence of different time settings on the accuracy of the calculated water IR spectra. Our results reveal a convergence behaviour in the band positions as the total simulation time is extended and the time resolution becomes finer. Through this analysis, we identify an optimal configuration of 60,000 time steps across a total time of 3950~fs for the water system. Although longer simulation times, such as 13165~fs, capture more weak overtone features, 3950~fs is sufficient for interpreting all fundamental bands and the most significant overtones in water.

We then explore how approximating either the initial ground state or the dipole operator affects the band characterisation. These approximations introduce minimal deviations in fundamental and overtone band positions, demonstrating their suitability for practical routine use. Although the impact on overtone intensities is evident, the intensities of fundamental bands remain largely unaffected by these approximations, which ensures the overall fidelity of the spectra. As IR spectroscopy primarily emphasises fundamental bands (with overtones often playing a supplementary role), we conclude that the most cost-saving scheme with linear dipole and harmonic ground state strikes an appropriate balance between circuit efficiency and spectral accuracy. The corresponding circuit depth including the state preparation and time evolution of water over 60,000 times steps becomes $1.6\times 10^9$ gates. 

An important direction for future work is the extension of the present framework to floppy systems characterised by torsional or other large-amplitude motions. For such systems, hybrid strategies that combine rectilinear normal modes for small-amplitude vibrations with curvilinear coordinates for selected large-amplitude degrees of freedom may provide a practical and balanced coordinate representation.



\section{Data Availability}

All data supporting the findings of this study are available within the paper.

\section{Author Contributions}

Xiaoning Feng conceived the project, developed the algorithm, performed the simulations, carried out the data analysis, and wrote the original draft of the manuscript.
Arman Nejad provided time-independent reference data for comparison and revised the manuscript. 
David P. Tew supervised the research and provided scientific guidance on the research design and interpretation of the results.

\section{Acknowledgements}
This work was supported by the UKRI through the QCi3 Hub EP/Z53318X/1.

\section{Conflicts of Interest}
The authors have no conflicts to disclose.

\bibliography{references.bib}

\begin{thebibliography}{99}%
\makeatletter
\providecommand \@ifxundefined [1]{%
 \@ifx{#1\undefined}
}%
\providecommand \@ifnum [1]{%
 \ifnum #1\expandafter \@firstoftwo
 \else \expandafter \@secondoftwo
 \fi
}%
\providecommand \@ifx [1]{%
 \ifx #1\expandafter \@firstoftwo
 \else \expandafter \@secondoftwo
 \fi
}%
\providecommand \natexlab [1]{#1}%
\providecommand \enquote  [1]{``#1''}%
\providecommand \bibnamefont  [1]{#1}%
\providecommand \bibfnamefont [1]{#1}%
\providecommand \citenamefont [1]{#1}%
\providecommand \href@noop [0]{\@secondoftwo}%
\providecommand \href [0]{\begingroup \@sanitize@url \@href}%
\providecommand \@href[1]{\@@startlink{#1}\@@href}%
\providecommand \@@href[1]{\endgroup#1\@@endlink}%
\providecommand \@sanitize@url [0]{\catcode `\\12\catcode `\$12\catcode `\&12\catcode `\#12\catcode `\^12\catcode `\_12\catcode `\%12\relax}%
\providecommand \@@startlink[1]{}%
\providecommand \@@endlink[0]{}%
\providecommand \url  [0]{\begingroup\@sanitize@url \@url }%
\providecommand \@url [1]{\endgroup\@href {#1}{\urlprefix }}%
\providecommand \urlprefix  [0]{URL }%
\providecommand \Eprint [0]{\href }%
\providecommand \doibase [0]{http://dx.doi.org/}%
\providecommand \selectlanguage [0]{\@gobble}%
\providecommand \bibinfo  [0]{\@secondoftwo}%
\providecommand \bibfield  [0]{\@secondoftwo}%
\providecommand \translation [1]{[#1]}%
\providecommand \BibitemOpen [0]{}%
\providecommand \bibitemStop [0]{}%
\providecommand \bibitemNoStop [0]{.\EOS\space}%
\providecommand \EOS [0]{\spacefactor3000\relax}%
\providecommand \BibitemShut  [1]{\csname bibitem#1\endcsname}%
\let\auto@bib@innerbib\@empty
\bibitem [{\citenamefont {Kim}\ \emph {et~al.}(2022)\citenamefont {Kim}, \citenamefont {Yoon}, \citenamefont {Ryu}, \citenamefont {Jeong}, \citenamefont {il~Kim}, \citenamefont {Park}, \citenamefont {Kye}, \citenamefont {Hwang}, \citenamefont {Kim}, \citenamefont {cho},\ and\ \citenamefont {Jeong}}]{Kim2022}%
  \BibitemOpen
  \bibfield  {author} {\bibinfo {author} {\bibfnamefont {Honghyun}\ \bibnamefont {Kim}}, \bibinfo {author} {\bibfnamefont {Ung~Hwi}\ \bibnamefont {Yoon}}, \bibinfo {author} {\bibfnamefont {Tae~In}\ \bibnamefont {Ryu}}, \bibinfo {author} {\bibfnamefont {Hey~Jin}\ \bibnamefont {Jeong}}, \bibinfo {author} {\bibfnamefont {Sung}\ \bibnamefont {il~Kim}}, \bibinfo {author} {\bibfnamefont {Jinseon}\ \bibnamefont {Park}}, \bibinfo {author} {\bibfnamefont {Young~Sik}\ \bibnamefont {Kye}}, \bibinfo {author} {\bibfnamefont {Seung-Ryul}\ \bibnamefont {Hwang}}, \bibinfo {author} {\bibfnamefont {Dongwook}\ \bibnamefont {Kim}}, \bibinfo {author} {\bibfnamefont {Yoonjae}\ \bibnamefont {cho}}, \ and\ \bibinfo {author} {\bibfnamefont {Keunhong}\ \bibnamefont {Jeong}},\ }\bibfield  {title} {\enquote {\bibinfo {title} {Calculation of the infrared spectra of organophosphorus compounds and prediction of new types of nerve agents},}\ }\href {\doibase 10.1039/D2NJ00850E} {\bibfield  {journal} {\bibinfo  {journal} {New J. Chem.}\
  }\textbf {\bibinfo {volume} {46}},\ \bibinfo {pages} {8653--8661} (\bibinfo {year} {2022})}\BibitemShut {NoStop}%
\bibitem [{\citenamefont {Kazachenko}\ \emph {et~al.}(2022)\citenamefont {Kazachenko}, \citenamefont {Malyar}, \citenamefont {Ghatfaoui}, \citenamefont {Issaoui}, \citenamefont {Al-Dossary}, \citenamefont {Wojcik}, \citenamefont {Kazachenko}, \citenamefont {Miroshnikova},\ and\ \citenamefont {Berezhnaya}}]{KAZA2022}%
  \BibitemOpen
  \bibfield  {author} {\bibinfo {author} {\bibfnamefont {Aleksandr~S.}\ \bibnamefont {Kazachenko}}, \bibinfo {author} {\bibfnamefont {Yuriy~N.}\ \bibnamefont {Malyar}}, \bibinfo {author} {\bibfnamefont {Sofian}\ \bibnamefont {Ghatfaoui}}, \bibinfo {author} {\bibfnamefont {Noureddine}\ \bibnamefont {Issaoui}}, \bibinfo {author} {\bibfnamefont {Omar}\ \bibnamefont {Al-Dossary}}, \bibinfo {author} {\bibfnamefont {Marek~J.}\ \bibnamefont {Wojcik}}, \bibinfo {author} {\bibfnamefont {Anna~S.}\ \bibnamefont {Kazachenko}}, \bibinfo {author} {\bibfnamefont {Angelina~V.}\ \bibnamefont {Miroshnikova}}, \ and\ \bibinfo {author} {\bibfnamefont {Yaroslava~D.}\ \bibnamefont {Berezhnaya}},\ }\bibfield  {title} {\enquote {\bibinfo {title} {A density functional theory calculations of infrared spectra of galactomannan butyl ether},}\ }\href {\doibase https://doi.org/10.1016/j.molstruc.2021.131998} {\bibfield  {journal} {\bibinfo  {journal} {Journal of Molecular Structure}\ }\textbf {\bibinfo {volume} {1251}},\ \bibinfo {pages}
  {131998} (\bibinfo {year} {2022})}\BibitemShut {NoStop}%
\bibitem [{\citenamefont {Ji}\ \emph {et~al.}(2020)\citenamefont {Ji}, \citenamefont {Yang}, \citenamefont {Ji}, \citenamefont {Zhu}, \citenamefont {Ma}, \citenamefont {Chen}, \citenamefont {Jia}, \citenamefont {Tang},\ and\ \citenamefont {Cao}}]{Yanji2020}%
  \BibitemOpen
  \bibfield  {author} {\bibinfo {author} {\bibfnamefont {Yan}\ \bibnamefont {Ji}}, \bibinfo {author} {\bibfnamefont {Xiaoliang}\ \bibnamefont {Yang}}, \bibinfo {author} {\bibfnamefont {Zhi}\ \bibnamefont {Ji}}, \bibinfo {author} {\bibfnamefont {Linhui}\ \bibnamefont {Zhu}}, \bibinfo {author} {\bibfnamefont {Nana}\ \bibnamefont {Ma}}, \bibinfo {author} {\bibfnamefont {Dejun}\ \bibnamefont {Chen}}, \bibinfo {author} {\bibfnamefont {Xianbin}\ \bibnamefont {Jia}}, \bibinfo {author} {\bibfnamefont {Junming}\ \bibnamefont {Tang}}, \ and\ \bibinfo {author} {\bibfnamefont {Yilin}\ \bibnamefont {Cao}},\ }\bibfield  {title} {\enquote {\bibinfo {title} {Dft-calculated ir spectrum amide i, ii, and iii band contributions of n-methylacetamide fine components},}\ }\href {\doibase 10.1021/acsomega.9b04421} {\bibfield  {journal} {\bibinfo  {journal} {ACS Omega}\ }\textbf {\bibinfo {volume} {5}},\ \bibinfo {pages} {8572--8578} (\bibinfo {year} {2020})}\BibitemShut {NoStop}%
\bibitem [{\citenamefont {Brito}\ \emph {et~al.}(2025)\citenamefont {Brito}, \citenamefont {Aquino}, \citenamefont {dos Santos~Politi},\ and\ \citenamefont {Martins}}]{Brito2025}%
  \BibitemOpen
  \bibfield  {author} {\bibinfo {author} {\bibfnamefont {Airan F.~S.}\ \bibnamefont {Brito}}, \bibinfo {author} {\bibfnamefont {Adelia J.~A.}\ \bibnamefont {Aquino}}, \bibinfo {author} {\bibfnamefont {José~Roberto}\ \bibnamefont {dos Santos~Politi}}, \ and\ \bibinfo {author} {\bibfnamefont {João B.~L.}\ \bibnamefont {Martins}},\ }\bibfield  {title} {\enquote {\bibinfo {title} {Infrared spectroscopy of ethanethiol monomers and dimers at mp2 level: Characterizing the dimer formation and hydrogen bond},}\ }\href {\doibase https://doi.org/10.1002/jcc.27540} {\bibfield  {journal} {\bibinfo  {journal} {Journal of Computational Chemistry}\ }\textbf {\bibinfo {volume} {46}},\ \bibinfo {pages} {e27540} (\bibinfo {year} {2025})}\BibitemShut {NoStop}%
\bibitem [{\citenamefont {Khire}\ \emph {et~al.}(2024)\citenamefont {Khire}, \citenamefont {Nakajima},\ and\ \citenamefont {Gadre}}]{Khire2024}%
  \BibitemOpen
  \bibfield  {author} {\bibinfo {author} {\bibfnamefont {Subodh~S.}\ \bibnamefont {Khire}}, \bibinfo {author} {\bibfnamefont {Takahito}\ \bibnamefont {Nakajima}}, \ and\ \bibinfo {author} {\bibfnamefont {Shridhar~R.}\ \bibnamefont {Gadre}},\ }\bibfield  {title} {\enquote {\bibinfo {title} {Cluster-in-cluster approach for computing mp2-level vibrational infrared spectra of large molecular clusters},}\ }\href {\doibase 10.1021/acs.jpca.4c00952} {\bibfield  {journal} {\bibinfo  {journal} {The Journal of Physical Chemistry A}\ }\textbf {\bibinfo {volume} {128}},\ \bibinfo {pages} {3703--3710} (\bibinfo {year} {2024})}\BibitemShut {NoStop}%
\bibitem [{\citenamefont {Knaanie}\ \emph {et~al.}(2014)\citenamefont {Knaanie}, \citenamefont {Šebek}, \citenamefont {Kalinowski},\ and\ \citenamefont {{Benny Gerber}}}]{KNAANIE2014}%
  \BibitemOpen
  \bibfield  {author} {\bibinfo {author} {\bibfnamefont {Roie}\ \bibnamefont {Knaanie}}, \bibinfo {author} {\bibfnamefont {Jiří}\ \bibnamefont {Šebek}}, \bibinfo {author} {\bibfnamefont {Jaroslaw}\ \bibnamefont {Kalinowski}}, \ and\ \bibinfo {author} {\bibfnamefont {R.}~\bibnamefont {{Benny Gerber}}},\ }\bibfield  {title} {\enquote {\bibinfo {title} {Hybrid mp2/mp4 potential surfaces in vscf calculations of ir spectra: Applications for organic molecules},}\ }\href {\doibase https://doi.org/10.1016/j.saa.2013.06.035} {\bibfield  {journal} {\bibinfo  {journal} {Spectrochimica Acta Part A: Molecular and Biomolecular Spectroscopy}\ }\textbf {\bibinfo {volume} {119}},\ \bibinfo {pages} {2--11} (\bibinfo {year} {2014})}\BibitemShut {NoStop}%
\bibitem [{\citenamefont {Bartlett}\ and\ \citenamefont {Stanton}(1994)}]{Bartlett1994}%
  \BibitemOpen
  \bibfield  {author} {\bibinfo {author} {\bibfnamefont {Rodney~J.}\ \bibnamefont {Bartlett}}\ and\ \bibinfo {author} {\bibfnamefont {John~F.}\ \bibnamefont {Stanton}},\ }\enquote {\bibinfo {title} {Applications of post-hartree—fock methods: A tutorial},}\ in\ \href {\doibase https://doi.org/10.1002/9780470125823.ch2} {\emph {\bibinfo {booktitle} {Reviews in Computational Chemistry}}}\ (\bibinfo  {publisher} {John Wiley \& Sons, Ltd},\ \bibinfo {year} {1994})\ pp.\ \bibinfo {pages} {65--169}\BibitemShut {NoStop}%
\bibitem [{\citenamefont {{Sibert III}}(1988)}]{Sibert_JCP_1988}%
  \BibitemOpen
  \bibfield  {author} {\bibinfo {author} {\bibfnamefont {Edwin~L.}\ \bibnamefont {{Sibert III}}},\ }\bibfield  {title} {\enquote {\bibinfo {title} {Theoretical studies of vibrationally excited polyatomic molecules using canonical {Van Vleck} perturbation theory},}\ }\href {\doibase 10.1063/1.453797} {\bibfield  {journal} {\bibinfo  {journal} {J. Chem. Phys.}\ }\textbf {\bibinfo {volume} {88}},\ \bibinfo {pages} {4378--4390} (\bibinfo {year} {1988})}\BibitemShut {NoStop}%
\bibitem [{\citenamefont {Barone}(2005)}]{Barone_JCP_2005}%
  \BibitemOpen
  \bibfield  {author} {\bibinfo {author} {\bibfnamefont {Vincenzo}\ \bibnamefont {Barone}},\ }\bibfield  {title} {\enquote {\bibinfo {title} {Anharmonic vibrational properties by a fully automated second-order perturbative approach},}\ }\href {\doibase 10.1063/1.1824881} {\bibfield  {journal} {\bibinfo  {journal} {J. Chem. Phys.}\ }\textbf {\bibinfo {volume} {122}},\ \bibinfo {pages} {014108} (\bibinfo {year} {2005})}\BibitemShut {NoStop}%
\bibitem [{\citenamefont {Gong}\ \emph {et~al.}(2018)\citenamefont {Gong}, \citenamefont {Matthews}, \citenamefont {Changala},\ and\ \citenamefont {Stanton}}]{Gong_JCP_2018}%
  \BibitemOpen
  \bibfield  {author} {\bibinfo {author} {\bibfnamefont {Justin~Z.}\ \bibnamefont {Gong}}, \bibinfo {author} {\bibfnamefont {Devin~A.}\ \bibnamefont {Matthews}}, \bibinfo {author} {\bibfnamefont {P.~Bryan}\ \bibnamefont {Changala}}, \ and\ \bibinfo {author} {\bibfnamefont {John~F.}\ \bibnamefont {Stanton}},\ }\bibfield  {title} {\enquote {\bibinfo {title} {Fourth-order vibrational perturbation theory with the {W}atson {H}amiltonian: Report of working equations and preliminary results},}\ }\href {\doibase 10.1063/1.5040360} {\bibfield  {journal} {\bibinfo  {journal} {J. Chem. Phys.}\ }\textbf {\bibinfo {volume} {149}},\ \bibinfo {pages} {114102} (\bibinfo {year} {2018})}\BibitemShut {NoStop}%
\bibitem [{\citenamefont {Bowman}\ \emph {et~al.}(2003)\citenamefont {Bowman}, \citenamefont {Carter},\ and\ \citenamefont {Huang}}]{Bowman_IRPC_2003}%
  \BibitemOpen
  \bibfield  {author} {\bibinfo {author} {\bibfnamefont {Joel~M.}\ \bibnamefont {Bowman}}, \bibinfo {author} {\bibfnamefont {Stuart}\ \bibnamefont {Carter}}, \ and\ \bibinfo {author} {\bibfnamefont {Xinchuan}\ \bibnamefont {Huang}},\ }\bibfield  {title} {\enquote {\bibinfo {title} {{MULTIMODE}: A code to calculate rovibrational energies of polyatomic molecules},}\ }\href {\doibase 10.1080/0144235031000124163} {\bibfield  {journal} {\bibinfo  {journal} {Int. Rev. Phys. Chem.}\ }\textbf {\bibinfo {volume} {22}},\ \bibinfo {pages} {533--549} (\bibinfo {year} {2003})}\BibitemShut {NoStop}%
\bibitem [{\citenamefont {Sibaev}\ and\ \citenamefont {Crittenden}(2016)}]{SIBAEV2016}%
  \BibitemOpen
  \bibfield  {author} {\bibinfo {author} {\bibfnamefont {Marat}\ \bibnamefont {Sibaev}}\ and\ \bibinfo {author} {\bibfnamefont {Deborah~L.}\ \bibnamefont {Crittenden}},\ }\bibfield  {title} {\enquote {\bibinfo {title} {Pyvci: A flexible open-source code for calculating accurate molecular infrared spectra},}\ }\href {\doibase https://doi.org/10.1016/j.cpc.2016.02.026} {\bibfield  {journal} {\bibinfo  {journal} {Computer Physics Communications}\ }\textbf {\bibinfo {volume} {203}},\ \bibinfo {pages} {290--297} (\bibinfo {year} {2016})}\BibitemShut {NoStop}%
\bibitem [{\citenamefont {Schr{\"o}der}\ and\ \citenamefont {Rauhut}(2024{\natexlab{a}})}]{Benjamin2024}%
  \BibitemOpen
  \bibfield  {author} {\bibinfo {author} {\bibfnamefont {Benjamin}\ \bibnamefont {Schr{\"o}der}}\ and\ \bibinfo {author} {\bibfnamefont {Guntram}\ \bibnamefont {Rauhut}},\ }\bibfield  {title} {\enquote {\bibinfo {title} {From the automated calculation of potential energy surfaces to accurate infrared spectra},}\ }\href {\doibase 10.1021/acs.jpclett.4c00186} {\bibfield  {journal} {\bibinfo  {journal} {The Journal of Physical Chemistry Letters}\ }\textbf {\bibinfo {volume} {15}},\ \bibinfo {pages} {3159--3169} (\bibinfo {year} {2024}{\natexlab{a}})}\BibitemShut {NoStop}%
\bibitem [{\citenamefont {Schneider}\ and\ \citenamefont {Rauhut}(2024)}]{Schneider_JCP_2024_214118}%
  \BibitemOpen
  \bibfield  {author} {\bibinfo {author} {\bibfnamefont {Moritz}\ \bibnamefont {Schneider}}\ and\ \bibinfo {author} {\bibfnamefont {Guntram}\ \bibnamefont {Rauhut}},\ }\bibfield  {title} {\enquote {\bibinfo {title} {{VSCF/VCI} theory based on the {P}odolsky hamiltonian},}\ }\href {\doibase 10.1063/5.0213401} {\bibfield  {journal} {\bibinfo  {journal} {J. Chem. Phys.}\ }\textbf {\bibinfo {volume} {160}},\ \bibinfo {pages} {214118} (\bibinfo {year} {2024})}\BibitemShut {NoStop}%
\bibitem [{\citenamefont {Taherivardanjani}\ \emph {et~al.}(2022)\citenamefont {Taherivardanjani}, \citenamefont {Elfgen}, \citenamefont {Reckien}, \citenamefont {Suarez}, \citenamefont {Perlt},\ and\ \citenamefont {Kirchner}}]{Shima2022}%
  \BibitemOpen
  \bibfield  {author} {\bibinfo {author} {\bibfnamefont {Shima}\ \bibnamefont {Taherivardanjani}}, \bibinfo {author} {\bibfnamefont {Roman}\ \bibnamefont {Elfgen}}, \bibinfo {author} {\bibfnamefont {Werner}\ \bibnamefont {Reckien}}, \bibinfo {author} {\bibfnamefont {Estela}\ \bibnamefont {Suarez}}, \bibinfo {author} {\bibfnamefont {Eva}\ \bibnamefont {Perlt}}, \ and\ \bibinfo {author} {\bibfnamefont {Barbara}\ \bibnamefont {Kirchner}},\ }\bibfield  {title} {\enquote {\bibinfo {title} {Benchmarking the computational costs and quality of vibrational spectra from ab initio simulations},}\ }\href {\doibase https://doi.org/10.1002/adts.202100293} {\bibfield  {journal} {\bibinfo  {journal} {Advanced Theory and Simulations}\ }\textbf {\bibinfo {volume} {5}},\ \bibinfo {pages} {2100293} (\bibinfo {year} {2022})}\BibitemShut {NoStop}%
\bibitem [{\citenamefont {Gastegger}\ \emph {et~al.}(2017)\citenamefont {Gastegger}, \citenamefont {Behler},\ and\ \citenamefont {Marquetand}}]{Gastegger2017}%
  \BibitemOpen
  \bibfield  {author} {\bibinfo {author} {\bibfnamefont {Michael}\ \bibnamefont {Gastegger}}, \bibinfo {author} {\bibfnamefont {Jörg}\ \bibnamefont {Behler}}, \ and\ \bibinfo {author} {\bibfnamefont {Philipp}\ \bibnamefont {Marquetand}},\ }\bibfield  {title} {\enquote {\bibinfo {title} {Machine learning molecular dynamics for the simulation of infrared spectra},}\ }\href {\doibase 10.1039/C7SC02267K} {\bibfield  {journal} {\bibinfo  {journal} {Chem. Sci.}\ }\textbf {\bibinfo {volume} {8}},\ \bibinfo {pages} {6924--6935} (\bibinfo {year} {2017})}\BibitemShut {NoStop}%
\bibitem [{\citenamefont {Fischer}\ \emph {et~al.}(2016)\citenamefont {Fischer}, \citenamefont {Ueltschi}, \citenamefont {El-Khoury}, \citenamefont {Mifflin}, \citenamefont {Hess}, \citenamefont {Wang}, \citenamefont {Cramer},\ and\ \citenamefont {Govind}}]{Fischer2016}%
  \BibitemOpen
  \bibfield  {author} {\bibinfo {author} {\bibfnamefont {Sean~A.}\ \bibnamefont {Fischer}}, \bibinfo {author} {\bibfnamefont {Tyler~W.}\ \bibnamefont {Ueltschi}}, \bibinfo {author} {\bibfnamefont {Patrick~Z.}\ \bibnamefont {El-Khoury}}, \bibinfo {author} {\bibfnamefont {Amanda~L.}\ \bibnamefont {Mifflin}}, \bibinfo {author} {\bibfnamefont {Wayne~P.}\ \bibnamefont {Hess}}, \bibinfo {author} {\bibfnamefont {Hong-Fei}\ \bibnamefont {Wang}}, \bibinfo {author} {\bibfnamefont {Christopher~J.}\ \bibnamefont {Cramer}}, \ and\ \bibinfo {author} {\bibfnamefont {Niranjan}\ \bibnamefont {Govind}},\ }\bibfield  {title} {\enquote {\bibinfo {title} {Infrared and raman spectroscopy from ab initio molecular dynamics and static normal mode analysis: The c–h region of dmso as a case study},}\ }\href {\doibase 10.1021/acs.jpcb.5b03323} {\bibfield  {journal} {\bibinfo  {journal} {The Journal of Physical Chemistry B}\ }\textbf {\bibinfo {volume} {120}},\ \bibinfo {pages} {1429--1436} (\bibinfo {year} {2016})}\BibitemShut
  {NoStop}%
\bibitem [{\citenamefont {Gaigeot}\ and\ \citenamefont {Sprik}(2003)}]{Gaigeot2003}%
  \BibitemOpen
  \bibfield  {author} {\bibinfo {author} {\bibfnamefont {Marie-Pierre}\ \bibnamefont {Gaigeot}}\ and\ \bibinfo {author} {\bibfnamefont {Michiel}\ \bibnamefont {Sprik}},\ }\bibfield  {title} {\enquote {\bibinfo {title} {Ab initio molecular dynamics computation of the infrared spectrum of aqueous uracil},}\ }\href {\doibase 10.1021/jp034788u} {\bibfield  {journal} {\bibinfo  {journal} {The Journal of Physical Chemistry B}\ }\textbf {\bibinfo {volume} {107}},\ \bibinfo {pages} {10344--10358} (\bibinfo {year} {2003})}\BibitemShut {NoStop}%
\bibitem [{\citenamefont {Althorpe}(2024)}]{Althorpe2024}%
  \BibitemOpen
  \bibfield  {author} {\bibinfo {author} {\bibfnamefont {Stuart~C.}\ \bibnamefont {Althorpe}},\ }\bibfield  {title} {\enquote {\bibinfo {title} {Path integral simulations of condensed-phase vibrational spectroscopy},}\ }\href {\doibase https://doi.org/10.1146/annurev-physchem-090722-124705} {\bibfield  {journal} {\bibinfo  {journal} {Annual Review of Physical Chemistry}\ }\textbf {\bibinfo {volume} {75}},\ \bibinfo {pages} {397--420} (\bibinfo {year} {2024})}\BibitemShut {NoStop}%
\bibitem [{\citenamefont {Shepherd}\ \emph {et~al.}(2021)\citenamefont {Shepherd}, \citenamefont {Lan}, \citenamefont {Wilkins},\ and\ \citenamefont {Kapil}}]{Shepherd2021}%
  \BibitemOpen
  \bibfield  {author} {\bibinfo {author} {\bibfnamefont {Sam}\ \bibnamefont {Shepherd}}, \bibinfo {author} {\bibfnamefont {Jinggang}\ \bibnamefont {Lan}}, \bibinfo {author} {\bibfnamefont {David~M.}\ \bibnamefont {Wilkins}}, \ and\ \bibinfo {author} {\bibfnamefont {Venkat}\ \bibnamefont {Kapil}},\ }\bibfield  {title} {\enquote {\bibinfo {title} {Efficient quantum vibrational spectroscopy of water with high-order path integrals: From bulk to interfaces},}\ }\href {\doibase 10.1021/acs.jpclett.1c02574} {\bibfield  {journal} {\bibinfo  {journal} {The Journal of Physical Chemistry Letters}\ }\textbf {\bibinfo {volume} {12}},\ \bibinfo {pages} {9108--9114} (\bibinfo {year} {2021})}\BibitemShut {NoStop}%
\bibitem [{\citenamefont {Althorpe}(2021)}]{Althorpe2021}%
  \BibitemOpen
  \bibfield  {author} {\bibinfo {author} {\bibfnamefont {Stuart~C}\ \bibnamefont {Althorpe}},\ }\bibfield  {title} {\enquote {\bibinfo {title} {Path-integral approximations to quantum dynamics},}\ }\href {https://doi.org/10.1140/epjb/s10051-021-00155-2} {\bibfield  {journal} {\bibinfo  {journal} {Eur. Phys. J. B}\ }\textbf {\bibinfo {volume} {94}} (\bibinfo {year} {2021})}\BibitemShut {NoStop}%
\bibitem [{\citenamefont {Liu}\ and\ \citenamefont {Zhang}(2016)}]{Liu2016}%
  \BibitemOpen
  \bibfield  {author} {\bibinfo {author} {\bibfnamefont {Jian}\ \bibnamefont {Liu}}\ and\ \bibinfo {author} {\bibfnamefont {Zhijun}\ \bibnamefont {Zhang}},\ }\bibfield  {title} {\enquote {\bibinfo {title} {Path integral liouville dynamics: Applications to infrared spectra of oh, water, ammonia, and methane},}\ }\href {\doibase 10.1063/1.4939953} {\bibfield  {journal} {\bibinfo  {journal} {The Journal of Chemical Physics}\ }\textbf {\bibinfo {volume} {144}},\ \bibinfo {pages} {034307} (\bibinfo {year} {2016})}\BibitemShut {NoStop}%
\bibitem [{\citenamefont {Wu}\ \emph {et~al.}(2021)\citenamefont {Wu}, \citenamefont {Bao}, \citenamefont {Cao}, \citenamefont {Chen}, \citenamefont {Chen}, \citenamefont {Chen}, \citenamefont {Chung}, \citenamefont {Deng}, \citenamefont {Du}, \citenamefont {Fan}, \citenamefont {Gong}, \citenamefont {Guo}, \citenamefont {Guo}, \citenamefont {Guo}, \citenamefont {Han}, \citenamefont {Hong}, \citenamefont {Huang}, \citenamefont {Huo}, \citenamefont {Li}, \citenamefont {Li}, \citenamefont {Li}, \citenamefont {Li}, \citenamefont {Liang}, \citenamefont {Lin}, \citenamefont {Lin}, \citenamefont {Qian}, \citenamefont {Qiao}, \citenamefont {Rong}, \citenamefont {Su}, \citenamefont {Sun}, \citenamefont {Wang}, \citenamefont {Wang}, \citenamefont {Wu}, \citenamefont {Xu}, \citenamefont {Yan}, \citenamefont {Yang}, \citenamefont {Yang}, \citenamefont {Ye}, \citenamefont {Yin}, \citenamefont {Ying}, \citenamefont {Yu}, \citenamefont {Zha}, \citenamefont {Zhang}, \citenamefont {Zhang}, \citenamefont {Zhang}, \citenamefont
  {Zhang}, \citenamefont {Zhao}, \citenamefont {Zhao}, \citenamefont {Zhou}, \citenamefont {Zhu}, \citenamefont {Lu}, \citenamefont {Peng}, \citenamefont {Zhu},\ and\ \citenamefont {Pan}}]{Wu2021}%
  \BibitemOpen
  \bibfield  {author} {\bibinfo {author} {\bibfnamefont {Yulin}\ \bibnamefont {Wu}}, \bibinfo {author} {\bibfnamefont {Wan-Su}\ \bibnamefont {Bao}}, \bibinfo {author} {\bibfnamefont {Sirui}\ \bibnamefont {Cao}}, \bibinfo {author} {\bibfnamefont {Fusheng}\ \bibnamefont {Chen}}, \bibinfo {author} {\bibfnamefont {Ming-Cheng}\ \bibnamefont {Chen}}, \bibinfo {author} {\bibfnamefont {Xiawei}\ \bibnamefont {Chen}}, \bibinfo {author} {\bibfnamefont {Tung-Hsun}\ \bibnamefont {Chung}}, \bibinfo {author} {\bibfnamefont {Hui}\ \bibnamefont {Deng}}, \bibinfo {author} {\bibfnamefont {Yajie}\ \bibnamefont {Du}}, \bibinfo {author} {\bibfnamefont {Daojin}\ \bibnamefont {Fan}}, \bibinfo {author} {\bibfnamefont {Ming}\ \bibnamefont {Gong}}, \bibinfo {author} {\bibfnamefont {Cheng}\ \bibnamefont {Guo}}, \bibinfo {author} {\bibfnamefont {Chu}\ \bibnamefont {Guo}}, \bibinfo {author} {\bibfnamefont {Shaojun}\ \bibnamefont {Guo}}, \bibinfo {author} {\bibfnamefont {Lianchen}\ \bibnamefont {Han}}, \bibinfo {author} {\bibfnamefont
  {Linyin}\ \bibnamefont {Hong}}, \bibinfo {author} {\bibfnamefont {He-Liang}\ \bibnamefont {Huang}}, \bibinfo {author} {\bibfnamefont {Yong-Heng}\ \bibnamefont {Huo}}, \bibinfo {author} {\bibfnamefont {Liping}\ \bibnamefont {Li}}, \bibinfo {author} {\bibfnamefont {Na}~\bibnamefont {Li}}, \bibinfo {author} {\bibfnamefont {Shaowei}\ \bibnamefont {Li}}, \bibinfo {author} {\bibfnamefont {Yuan}\ \bibnamefont {Li}}, \bibinfo {author} {\bibfnamefont {Futian}\ \bibnamefont {Liang}}, \bibinfo {author} {\bibfnamefont {Chun}\ \bibnamefont {Lin}}, \bibinfo {author} {\bibfnamefont {Jin}\ \bibnamefont {Lin}}, \bibinfo {author} {\bibfnamefont {Haoran}\ \bibnamefont {Qian}}, \bibinfo {author} {\bibfnamefont {Dan}\ \bibnamefont {Qiao}}, \bibinfo {author} {\bibfnamefont {Hao}\ \bibnamefont {Rong}}, \bibinfo {author} {\bibfnamefont {Hong}\ \bibnamefont {Su}}, \bibinfo {author} {\bibfnamefont {Lihua}\ \bibnamefont {Sun}}, \bibinfo {author} {\bibfnamefont {Liangyuan}\ \bibnamefont {Wang}}, \bibinfo {author} {\bibfnamefont
  {Shiyu}\ \bibnamefont {Wang}}, \bibinfo {author} {\bibfnamefont {Dachao}\ \bibnamefont {Wu}}, \bibinfo {author} {\bibfnamefont {Yu}~\bibnamefont {Xu}}, \bibinfo {author} {\bibfnamefont {Kai}\ \bibnamefont {Yan}}, \bibinfo {author} {\bibfnamefont {Weifeng}\ \bibnamefont {Yang}}, \bibinfo {author} {\bibfnamefont {Yang}\ \bibnamefont {Yang}}, \bibinfo {author} {\bibfnamefont {Yangsen}\ \bibnamefont {Ye}}, \bibinfo {author} {\bibfnamefont {Jianghan}\ \bibnamefont {Yin}}, \bibinfo {author} {\bibfnamefont {Chong}\ \bibnamefont {Ying}}, \bibinfo {author} {\bibfnamefont {Jiale}\ \bibnamefont {Yu}}, \bibinfo {author} {\bibfnamefont {Chen}\ \bibnamefont {Zha}}, \bibinfo {author} {\bibfnamefont {Cha}\ \bibnamefont {Zhang}}, \bibinfo {author} {\bibfnamefont {Haibin}\ \bibnamefont {Zhang}}, \bibinfo {author} {\bibfnamefont {Kaili}\ \bibnamefont {Zhang}}, \bibinfo {author} {\bibfnamefont {Yiming}\ \bibnamefont {Zhang}}, \bibinfo {author} {\bibfnamefont {Han}\ \bibnamefont {Zhao}}, \bibinfo {author} {\bibfnamefont
  {Youwei}\ \bibnamefont {Zhao}}, \bibinfo {author} {\bibfnamefont {Liang}\ \bibnamefont {Zhou}}, \bibinfo {author} {\bibfnamefont {Qingling}\ \bibnamefont {Zhu}}, \bibinfo {author} {\bibfnamefont {Chao-Yang}\ \bibnamefont {Lu}}, \bibinfo {author} {\bibfnamefont {Cheng-Zhi}\ \bibnamefont {Peng}}, \bibinfo {author} {\bibfnamefont {Xiaobo}\ \bibnamefont {Zhu}}, \ and\ \bibinfo {author} {\bibfnamefont {Jian-Wei}\ \bibnamefont {Pan}},\ }\bibfield  {title} {\enquote {\bibinfo {title} {Strong quantum computational advantage using a superconducting quantum processor},}\ }\href {\doibase 10.1103/PhysRevLett.127.180501} {\bibfield  {journal} {\bibinfo  {journal} {Phys. Rev. Lett.}\ }\textbf {\bibinfo {volume} {127}},\ \bibinfo {pages} {180501} (\bibinfo {year} {2021})}\BibitemShut {NoStop}%
\bibitem [{\citenamefont {Madsen}\ \emph {et~al.}(2022)\citenamefont {Madsen}, \citenamefont {Laudenbach}, \citenamefont {Askarani}, \citenamefont {Rortais}, \citenamefont {Vincent}, \citenamefont {Bulmer}, \citenamefont {Miatto}, \citenamefont {Neuhaus}, \citenamefont {Helt}, \citenamefont {Collins}, \citenamefont {Lita}, \citenamefont {Gerrits}, \citenamefont {Nam}, \citenamefont {Vaidya}, \citenamefont {Menotti}, \citenamefont {Dhand}, \citenamefont {Vernon}, \citenamefont {Quesada},\ and\ \citenamefont {Lavoie}}]{Madsen2022}%
  \BibitemOpen
  \bibfield  {author} {\bibinfo {author} {\bibfnamefont {Lars~S.}\ \bibnamefont {Madsen}}, \bibinfo {author} {\bibfnamefont {Fabian}\ \bibnamefont {Laudenbach}}, \bibinfo {author} {\bibfnamefont {Mohsen~Falamarzi.}\ \bibnamefont {Askarani}}, \bibinfo {author} {\bibfnamefont {Fabien}\ \bibnamefont {Rortais}}, \bibinfo {author} {\bibfnamefont {Trevor}\ \bibnamefont {Vincent}}, \bibinfo {author} {\bibfnamefont {Jacob F.~F.}\ \bibnamefont {Bulmer}}, \bibinfo {author} {\bibfnamefont {Filippo~M.}\ \bibnamefont {Miatto}}, \bibinfo {author} {\bibfnamefont {Leonhard}\ \bibnamefont {Neuhaus}}, \bibinfo {author} {\bibfnamefont {Lukas~G.}\ \bibnamefont {Helt}}, \bibinfo {author} {\bibfnamefont {Matthew~J.}\ \bibnamefont {Collins}}, \bibinfo {author} {\bibfnamefont {Adriana~E.}\ \bibnamefont {Lita}}, \bibinfo {author} {\bibfnamefont {Thomas}\ \bibnamefont {Gerrits}}, \bibinfo {author} {\bibfnamefont {Sae~Woo}\ \bibnamefont {Nam}}, \bibinfo {author} {\bibfnamefont {Varun~D.}\ \bibnamefont {Vaidya}}, \bibinfo {author}
  {\bibfnamefont {Matteo}\ \bibnamefont {Menotti}}, \bibinfo {author} {\bibfnamefont {Ish}\ \bibnamefont {Dhand}}, \bibinfo {author} {\bibfnamefont {Zachary}\ \bibnamefont {Vernon}}, \bibinfo {author} {\bibfnamefont {Nicol{\'a}s}\ \bibnamefont {Quesada}}, \ and\ \bibinfo {author} {\bibfnamefont {Jonathan}\ \bibnamefont {Lavoie}},\ }\bibfield  {title} {\enquote {\bibinfo {title} {Quantum computational advantage with a programmable photonic processor},}\ }\href {\doibase 10.1038/s41586-022-04725-x} {\bibfield  {journal} {\bibinfo  {journal} {Nature}\ }\textbf {\bibinfo {volume} {606}},\ \bibinfo {pages} {75--81} (\bibinfo {year} {2022})}\BibitemShut {NoStop}%
\bibitem [{\citenamefont {Deng}\ \emph {et~al.}(2023)\citenamefont {Deng}, \citenamefont {Gu}, \citenamefont {Liu}, \citenamefont {Gong}, \citenamefont {Su}, \citenamefont {Zhang}, \citenamefont {Tang}, \citenamefont {Jia}, \citenamefont {Xu}, \citenamefont {Chen}, \citenamefont {Qin}, \citenamefont {Peng}, \citenamefont {Yan}, \citenamefont {Hu}, \citenamefont {Huang}, \citenamefont {Li}, \citenamefont {Li}, \citenamefont {Chen}, \citenamefont {Jiang}, \citenamefont {Gan}, \citenamefont {Yang}, \citenamefont {You}, \citenamefont {Li}, \citenamefont {Zhong}, \citenamefont {Wang}, \citenamefont {Liu}, \citenamefont {Renema}, \citenamefont {Lu},\ and\ \citenamefont {Pan}}]{Deng2023}%
  \BibitemOpen
  \bibfield  {author} {\bibinfo {author} {\bibfnamefont {Yu-Hao}\ \bibnamefont {Deng}}, \bibinfo {author} {\bibfnamefont {Yi-Chao}\ \bibnamefont {Gu}}, \bibinfo {author} {\bibfnamefont {Hua-Liang}\ \bibnamefont {Liu}}, \bibinfo {author} {\bibfnamefont {Si-Qiu}\ \bibnamefont {Gong}}, \bibinfo {author} {\bibfnamefont {Hao}\ \bibnamefont {Su}}, \bibinfo {author} {\bibfnamefont {Zhi-Jiong}\ \bibnamefont {Zhang}}, \bibinfo {author} {\bibfnamefont {Hao-Yang}\ \bibnamefont {Tang}}, \bibinfo {author} {\bibfnamefont {Meng-Hao}\ \bibnamefont {Jia}}, \bibinfo {author} {\bibfnamefont {Jia-Min}\ \bibnamefont {Xu}}, \bibinfo {author} {\bibfnamefont {Ming-Cheng}\ \bibnamefont {Chen}}, \bibinfo {author} {\bibfnamefont {Jian}\ \bibnamefont {Qin}}, \bibinfo {author} {\bibfnamefont {Li-Chao}\ \bibnamefont {Peng}}, \bibinfo {author} {\bibfnamefont {Jiarong}\ \bibnamefont {Yan}}, \bibinfo {author} {\bibfnamefont {Yi}~\bibnamefont {Hu}}, \bibinfo {author} {\bibfnamefont {Jia}\ \bibnamefont {Huang}}, \bibinfo {author}
  {\bibfnamefont {Hao}\ \bibnamefont {Li}}, \bibinfo {author} {\bibfnamefont {Yuxuan}\ \bibnamefont {Li}}, \bibinfo {author} {\bibfnamefont {Yaojian}\ \bibnamefont {Chen}}, \bibinfo {author} {\bibfnamefont {Xiao}\ \bibnamefont {Jiang}}, \bibinfo {author} {\bibfnamefont {Lin}\ \bibnamefont {Gan}}, \bibinfo {author} {\bibfnamefont {Guangwen}\ \bibnamefont {Yang}}, \bibinfo {author} {\bibfnamefont {Lixing}\ \bibnamefont {You}}, \bibinfo {author} {\bibfnamefont {Li}~\bibnamefont {Li}}, \bibinfo {author} {\bibfnamefont {Han-Sen}\ \bibnamefont {Zhong}}, \bibinfo {author} {\bibfnamefont {Hui}\ \bibnamefont {Wang}}, \bibinfo {author} {\bibfnamefont {Nai-Le}\ \bibnamefont {Liu}}, \bibinfo {author} {\bibfnamefont {Jelmer~J.}\ \bibnamefont {Renema}}, \bibinfo {author} {\bibfnamefont {Chao-Yang}\ \bibnamefont {Lu}}, \ and\ \bibinfo {author} {\bibfnamefont {Jian-Wei}\ \bibnamefont {Pan}},\ }\bibfield  {title} {\enquote {\bibinfo {title} {Gaussian boson sampling with pseudo-photon-number-resolving detectors and quantum
  computational advantage},}\ }\href {\doibase 10.1103/PhysRevLett.131.150601} {\bibfield  {journal} {\bibinfo  {journal} {Phys. Rev. Lett.}\ }\textbf {\bibinfo {volume} {131}},\ \bibinfo {pages} {150601} (\bibinfo {year} {2023})}\BibitemShut {NoStop}%
\bibitem [{\citenamefont {Daley}\ \emph {et~al.}(2022)\citenamefont {Daley}, \citenamefont {Bloch}, \citenamefont {Kokail}, \citenamefont {Flannigan}, \citenamefont {Pearson}, \citenamefont {Troyer},\ and\ \citenamefont {Zoller}}]{Daley2022}%
  \BibitemOpen
  \bibfield  {author} {\bibinfo {author} {\bibfnamefont {Andrew~J.}\ \bibnamefont {Daley}}, \bibinfo {author} {\bibfnamefont {Immanuel}\ \bibnamefont {Bloch}}, \bibinfo {author} {\bibfnamefont {Christian}\ \bibnamefont {Kokail}}, \bibinfo {author} {\bibfnamefont {Stuart}\ \bibnamefont {Flannigan}}, \bibinfo {author} {\bibfnamefont {Natalie}\ \bibnamefont {Pearson}}, \bibinfo {author} {\bibfnamefont {Matthias}\ \bibnamefont {Troyer}}, \ and\ \bibinfo {author} {\bibfnamefont {Peter}\ \bibnamefont {Zoller}},\ }\bibfield  {title} {\enquote {\bibinfo {title} {Practical quantum advantage in quantum simulation},}\ }\href {\doibase 10.1038/s41586-022-04940-6} {\bibfield  {journal} {\bibinfo  {journal} {Nature}\ }\textbf {\bibinfo {volume} {607}},\ \bibinfo {pages} {667--676} (\bibinfo {year} {2022})}\BibitemShut {NoStop}%
\bibitem [{\citenamefont {Sawaya}\ \emph {et~al.}(2021)\citenamefont {Sawaya}, \citenamefont {Paesani},\ and\ \citenamefont {Tabor}}]{Sawaya2021}%
  \BibitemOpen
  \bibfield  {author} {\bibinfo {author} {\bibfnamefont {Nicolas P.~D.}\ \bibnamefont {Sawaya}}, \bibinfo {author} {\bibfnamefont {Francesco}\ \bibnamefont {Paesani}}, \ and\ \bibinfo {author} {\bibfnamefont {Daniel~P.}\ \bibnamefont {Tabor}},\ }\bibfield  {title} {\enquote {\bibinfo {title} {Near- and long-term quantum algorithmic approaches for vibrational spectroscopy},}\ }\href {\doibase 10.1103/PhysRevA.104.062419} {\bibfield  {journal} {\bibinfo  {journal} {Phys. Rev. A}\ }\textbf {\bibinfo {volume} {104}},\ \bibinfo {pages} {062419} (\bibinfo {year} {2021})}\BibitemShut {NoStop}%
\bibitem [{\citenamefont {Ollitrault}\ \emph {et~al.}(2020)\citenamefont {Ollitrault}, \citenamefont {Baiardi}, \citenamefont {Reiher},\ and\ \citenamefont {Tavernelli}}]{Pauline2020}%
  \BibitemOpen
  \bibfield  {author} {\bibinfo {author} {\bibfnamefont {Pauline~J.}\ \bibnamefont {Ollitrault}}, \bibinfo {author} {\bibfnamefont {Alberto}\ \bibnamefont {Baiardi}}, \bibinfo {author} {\bibfnamefont {Markus}\ \bibnamefont {Reiher}}, \ and\ \bibinfo {author} {\bibfnamefont {Ivano}\ \bibnamefont {Tavernelli}},\ }\bibfield  {title} {\enquote {\bibinfo {title} {Hardware efficient quantum algorithms for vibrational structure calculations},}\ }\href {https://api.semanticscholar.org/CorpusID:214713906} {\bibfield  {journal} {\bibinfo  {journal} {Chemical Science}\ }\textbf {\bibinfo {volume} {11}},\ \bibinfo {pages} {6842 -- 6855} (\bibinfo {year} {2020})}\BibitemShut {NoStop}%
\bibitem [{\citenamefont {McArdle}\ \emph {et~al.}(2019{\natexlab{a}})\citenamefont {McArdle}, \citenamefont {Mayorov}, \citenamefont {Shan}, \citenamefont {Benjamin},\ and\ \citenamefont {Yuan}}]{McArdle2019}%
  \BibitemOpen
  \bibfield  {author} {\bibinfo {author} {\bibfnamefont {Sam}\ \bibnamefont {McArdle}}, \bibinfo {author} {\bibfnamefont {Alexander}\ \bibnamefont {Mayorov}}, \bibinfo {author} {\bibfnamefont {Xiao}\ \bibnamefont {Shan}}, \bibinfo {author} {\bibfnamefont {Simon}\ \bibnamefont {Benjamin}}, \ and\ \bibinfo {author} {\bibfnamefont {Xiao}\ \bibnamefont {Yuan}},\ }\bibfield  {title} {\enquote {\bibinfo {title} {Digital quantum simulation of molecular vibrations},}\ }\href {http://dx.doi.org/10.1039/c9sc01313j} {\bibfield  {journal} {\bibinfo  {journal} {Chemical Science}\ }\textbf {\bibinfo {volume} {10}},\ \bibinfo {pages} {5725–5735} (\bibinfo {year} {2019}{\natexlab{a}})}\BibitemShut {NoStop}%
\bibitem [{\citenamefont {Teplukhin}\ \emph {et~al.}(2019)\citenamefont {Teplukhin}, \citenamefont {Kendrick},\ and\ \citenamefont {Babikov}}]{Alexander2019}%
  \BibitemOpen
  \bibfield  {author} {\bibinfo {author} {\bibfnamefont {Alexander}\ \bibnamefont {Teplukhin}}, \bibinfo {author} {\bibfnamefont {Brian~K.}\ \bibnamefont {Kendrick}}, \ and\ \bibinfo {author} {\bibfnamefont {Dmitri}\ \bibnamefont {Babikov}},\ }\bibfield  {title} {\enquote {\bibinfo {title} {Calculation of molecular vibrational spectra on a quantum annealer},}\ }\href {\doibase 10.1021/acs.jctc.9b00402} {\bibfield  {journal} {\bibinfo  {journal} {Journal of Chemical Theory and Computation}\ }\textbf {\bibinfo {volume} {15}},\ \bibinfo {pages} {4555--4563} (\bibinfo {year} {2019})}\BibitemShut {NoStop}%
\bibitem [{\citenamefont {Hu}\ \emph {et~al.}(2018)\citenamefont {Hu}, \citenamefont {Ma}, \citenamefont {Xu}, \citenamefont {Wang}, \citenamefont {Ma}, \citenamefont {Liu}, \citenamefont {Wang}, \citenamefont {Song}, \citenamefont {Yung},\ and\ \citenamefont {Sun}}]{HU2018}%
  \BibitemOpen
  \bibfield  {author} {\bibinfo {author} {\bibfnamefont {Ling}\ \bibnamefont {Hu}}, \bibinfo {author} {\bibfnamefont {Yue-Chi}\ \bibnamefont {Ma}}, \bibinfo {author} {\bibfnamefont {Yuan}\ \bibnamefont {Xu}}, \bibinfo {author} {\bibfnamefont {Wei-Ting}\ \bibnamefont {Wang}}, \bibinfo {author} {\bibfnamefont {Yu-Wei}\ \bibnamefont {Ma}}, \bibinfo {author} {\bibfnamefont {Ke}~\bibnamefont {Liu}}, \bibinfo {author} {\bibfnamefont {Hai-Yan}\ \bibnamefont {Wang}}, \bibinfo {author} {\bibfnamefont {Yi-Pu}\ \bibnamefont {Song}}, \bibinfo {author} {\bibfnamefont {Man-Hong}\ \bibnamefont {Yung}}, \ and\ \bibinfo {author} {\bibfnamefont {Lu-Yan}\ \bibnamefont {Sun}},\ }\bibfield  {title} {\enquote {\bibinfo {title} {Simulation of molecular spectroscopy with circuit quantum electrodynamics},}\ }\href {\doibase https://doi.org/10.1016/j.scib.2018.02.001} {\bibfield  {journal} {\bibinfo  {journal} {Science Bulletin}\ }\textbf {\bibinfo {volume} {63}},\ \bibinfo {pages} {293--299} (\bibinfo {year} {2018})}\BibitemShut
  {NoStop}%
\bibitem [{\citenamefont {MacDonell}\ \emph {et~al.}(2023)\citenamefont {MacDonell}, \citenamefont {Navickas}, \citenamefont {Wohlers-Reichel}, \citenamefont {Valahu}, \citenamefont {Rao}, \citenamefont {Millican}, \citenamefont {Currington}, \citenamefont {Biercuk}, \citenamefont {Tan}, \citenamefont {Hempel},\ and\ \citenamefont {Kassal}}]{MacDonell2023}%
  \BibitemOpen
  \bibfield  {author} {\bibinfo {author} {\bibfnamefont {Ryan~J.}\ \bibnamefont {MacDonell}}, \bibinfo {author} {\bibfnamefont {Tomas}\ \bibnamefont {Navickas}}, \bibinfo {author} {\bibfnamefont {Tim~F.}\ \bibnamefont {Wohlers-Reichel}}, \bibinfo {author} {\bibfnamefont {Christophe~H.}\ \bibnamefont {Valahu}}, \bibinfo {author} {\bibfnamefont {Arjun~D.}\ \bibnamefont {Rao}}, \bibinfo {author} {\bibfnamefont {Maverick~J.}\ \bibnamefont {Millican}}, \bibinfo {author} {\bibfnamefont {Michael~A.}\ \bibnamefont {Currington}}, \bibinfo {author} {\bibfnamefont {Michael~J.}\ \bibnamefont {Biercuk}}, \bibinfo {author} {\bibfnamefont {Ting~Rei}\ \bibnamefont {Tan}}, \bibinfo {author} {\bibfnamefont {Cornelius}\ \bibnamefont {Hempel}}, \ and\ \bibinfo {author} {\bibfnamefont {Ivan}\ \bibnamefont {Kassal}},\ }\bibfield  {title} {\enquote {\bibinfo {title} {Predicting molecular vibronic spectra using time-domain analog quantum simulation},}\ }\href {\doibase 10.1039/d3sc02453a} {\bibfield  {journal} {\bibinfo  {journal}
  {Chemical Science}\ }\textbf {\bibinfo {volume} {14}},\ \bibinfo {pages} {9439–9451} (\bibinfo {year} {2023})}\BibitemShut {NoStop}%
\bibitem [{\citenamefont {Richerme}\ \emph {et~al.}(2023)\citenamefont {Richerme}, \citenamefont {Revelle}, \citenamefont {Yale}, \citenamefont {Lobser}, \citenamefont {Burch}, \citenamefont {Clark}, \citenamefont {Saha}, \citenamefont {Lopez-Ruiz}, \citenamefont {Dwivedi}, \citenamefont {Smith}, \citenamefont {Norrell}, \citenamefont {Sabry},\ and\ \citenamefont {Iyengar}}]{Richerme2023}%
  \BibitemOpen
  \bibfield  {author} {\bibinfo {author} {\bibfnamefont {Philip}\ \bibnamefont {Richerme}}, \bibinfo {author} {\bibfnamefont {Melissa~C.}\ \bibnamefont {Revelle}}, \bibinfo {author} {\bibfnamefont {Christopher~G.}\ \bibnamefont {Yale}}, \bibinfo {author} {\bibfnamefont {Daniel}\ \bibnamefont {Lobser}}, \bibinfo {author} {\bibfnamefont {Ashlyn~D.}\ \bibnamefont {Burch}}, \bibinfo {author} {\bibfnamefont {Susan~M.}\ \bibnamefont {Clark}}, \bibinfo {author} {\bibfnamefont {Debadrita}\ \bibnamefont {Saha}}, \bibinfo {author} {\bibfnamefont {Miguel~Angel}\ \bibnamefont {Lopez-Ruiz}}, \bibinfo {author} {\bibfnamefont {Anurag}\ \bibnamefont {Dwivedi}}, \bibinfo {author} {\bibfnamefont {Jeremy~M.}\ \bibnamefont {Smith}}, \bibinfo {author} {\bibfnamefont {Sam~A.}\ \bibnamefont {Norrell}}, \bibinfo {author} {\bibfnamefont {Amr}\ \bibnamefont {Sabry}}, \ and\ \bibinfo {author} {\bibfnamefont {Srinivasan~S.}\ \bibnamefont {Iyengar}},\ }\bibfield  {title} {\enquote {\bibinfo {title} {Quantum computation of hydrogen bond
  dynamics and vibrational spectra},}\ }\href {\doibase 10.1021/acs.jpclett.3c01601} {\bibfield  {journal} {\bibinfo  {journal} {The Journal of Physical Chemistry Letters}\ }\textbf {\bibinfo {volume} {14}},\ \bibinfo {pages} {7256–7263} (\bibinfo {year} {2023})}\BibitemShut {NoStop}%
\bibitem [{\citenamefont {{Komarova}}\ \emph {et~al.}(2020)\citenamefont {{Komarova}}, \citenamefont {{Gattuso}}, \citenamefont {{Levine}},\ and\ \citenamefont {{Remacle}}}]{Komarova2020}%
  \BibitemOpen
  \bibfield  {author} {\bibinfo {author} {\bibfnamefont {Ksenia}\ \bibnamefont {{Komarova}}}, \bibinfo {author} {\bibfnamefont {Hugo}\ \bibnamefont {{Gattuso}}}, \bibinfo {author} {\bibfnamefont {R.~D.}\ \bibnamefont {{Levine}}}, \ and\ \bibinfo {author} {\bibfnamefont {F.}~\bibnamefont {{Remacle}}},\ }\bibfield  {title} {\enquote {\bibinfo {title} {{Parallel Quantum Computation of Vibrational Dynamics}},}\ }\href {\doibase 10.3389/fphy.2020.590699} {\bibfield  {journal} {\bibinfo  {journal} {Frontiers in Physics}\ }\textbf {\bibinfo {volume} {8}},\ \bibinfo {eid} {486} (\bibinfo {year} {2020})}\BibitemShut {NoStop}%
\bibitem [{\citenamefont {Trenev}\ \emph {et~al.}(2023)\citenamefont {Trenev}, \citenamefont {Ollitrault}, \citenamefont {Harwood}, \citenamefont {Gujarati}, \citenamefont {Raman}, \citenamefont {Mezzacapo},\ and\ \citenamefont {Mostame}}]{trenev2023}%
  \BibitemOpen
  \bibfield  {author} {\bibinfo {author} {\bibfnamefont {Dimitar}\ \bibnamefont {Trenev}}, \bibinfo {author} {\bibfnamefont {Pauline~J}\ \bibnamefont {Ollitrault}}, \bibinfo {author} {\bibfnamefont {Stuart~M.}\ \bibnamefont {Harwood}}, \bibinfo {author} {\bibfnamefont {Tanvi~P.}\ \bibnamefont {Gujarati}}, \bibinfo {author} {\bibfnamefont {Sumathy}\ \bibnamefont {Raman}}, \bibinfo {author} {\bibfnamefont {Antonio}\ \bibnamefont {Mezzacapo}}, \ and\ \bibinfo {author} {\bibfnamefont {Sarah}\ \bibnamefont {Mostame}},\ }\href {https://arxiv.org/abs/2311.03719} {\enquote {\bibinfo {title} {Refining resource estimation for the quantum computation of vibrational molecular spectra through trotter error analysis},}\ } (\bibinfo {year} {2023})\BibitemShut {NoStop}%
\bibitem [{\citenamefont {Lauvergnat}\ and\ \citenamefont {Nauts}(2014)}]{LAUVERGNAT2014}%
  \BibitemOpen
  \bibfield  {author} {\bibinfo {author} {\bibfnamefont {David}\ \bibnamefont {Lauvergnat}}\ and\ \bibinfo {author} {\bibfnamefont {André}\ \bibnamefont {Nauts}},\ }\bibfield  {title} {\enquote {\bibinfo {title} {Quantum dynamics with sparse grids: A combination of smolyak scheme and cubature. application to methanol in full dimensionality},}\ }\href {\doibase https://doi.org/10.1016/j.saa.2013.05.068} {\bibfield  {journal} {\bibinfo  {journal} {Spectrochimica Acta Part A: Molecular and Biomolecular Spectroscopy}\ }\textbf {\bibinfo {volume} {119}},\ \bibinfo {pages} {18--25} (\bibinfo {year} {2014})},\ \bibinfo {note} {frontiers in molecular vibrational calculations and computational spectroscopy}\BibitemShut {NoStop}%
\bibitem [{\citenamefont {Avila}\ and\ \citenamefont {Carrington}(2011)}]{avila2011}%
  \BibitemOpen
  \bibfield  {author} {\bibinfo {author} {\bibfnamefont {Gustavo}\ \bibnamefont {Avila}}\ and\ \bibinfo {author} {\bibfnamefont {Jr.}\ \bibnamefont {Carrington}, \bibfnamefont {Tucker}},\ }\bibfield  {title} {\enquote {\bibinfo {title} {Using nonproduct quadrature grids to solve the vibrational schrödinger equation in 12d},}\ }\href {\doibase 10.1063/1.3549817} {\bibfield  {journal} {\bibinfo  {journal} {The Journal of Chemical Physics}\ }\textbf {\bibinfo {volume} {134}},\ \bibinfo {pages} {054126} (\bibinfo {year} {2011})}\BibitemShut {NoStop}%
\bibitem [{\citenamefont {Rakhuba}\ and\ \citenamefont {Oseledets}(2016)}]{Maxim2016}%
  \BibitemOpen
  \bibfield  {author} {\bibinfo {author} {\bibfnamefont {Maxim}\ \bibnamefont {Rakhuba}}\ and\ \bibinfo {author} {\bibfnamefont {Ivan}\ \bibnamefont {Oseledets}},\ }\bibfield  {title} {\enquote {\bibinfo {title} {Calculating vibrational spectra of molecules using tensor train decomposition},}\ }\href {\doibase 10.1063/1.4962420} {\bibfield  {journal} {\bibinfo  {journal} {The Journal of Chemical Physics}\ }\textbf {\bibinfo {volume} {145}},\ \bibinfo {pages} {124101} (\bibinfo {year} {2016})}\BibitemShut {NoStop}%
\bibitem [{\citenamefont {Chan}\ \emph {et~al.}(2023)\citenamefont {Chan}, \citenamefont {Meister}, \citenamefont {Jones}, \citenamefont {Tew},\ and\ \citenamefont {Benjamin}}]{Chan2023}%
  \BibitemOpen
  \bibfield  {author} {\bibinfo {author} {\bibfnamefont {Hans Hon~Sang}\ \bibnamefont {Chan}}, \bibinfo {author} {\bibfnamefont {Richard}\ \bibnamefont {Meister}}, \bibinfo {author} {\bibfnamefont {Tyson}\ \bibnamefont {Jones}}, \bibinfo {author} {\bibfnamefont {David~P.}\ \bibnamefont {Tew}}, \ and\ \bibinfo {author} {\bibfnamefont {Simon~C.}\ \bibnamefont {Benjamin}},\ }\bibfield  {title} {\enquote {\bibinfo {title} {Grid-based methods for chemistry simulations on a quantum computer},}\ }\href {http://dx.doi.org/10.1126/sciadv.abo7484} {\bibfield  {journal} {\bibinfo  {journal} {Science Advances}\ }\textbf {\bibinfo {volume} {9}} (\bibinfo {year} {2023})}\BibitemShut {NoStop}%
\bibitem [{\citenamefont {Childs}\ \emph {et~al.}(2022)\citenamefont {Childs}, \citenamefont {Leng}, \citenamefont {Li}, \citenamefont {Liu},\ and\ \citenamefont {Zhang}}]{Childs2022}%
  \BibitemOpen
  \bibfield  {author} {\bibinfo {author} {\bibfnamefont {Andrew~M.}\ \bibnamefont {Childs}}, \bibinfo {author} {\bibfnamefont {Jiaqi}\ \bibnamefont {Leng}}, \bibinfo {author} {\bibfnamefont {Tongyang}\ \bibnamefont {Li}}, \bibinfo {author} {\bibfnamefont {Jin-Peng}\ \bibnamefont {Liu}}, \ and\ \bibinfo {author} {\bibfnamefont {Chenyi}\ \bibnamefont {Zhang}},\ }\bibfield  {title} {\enquote {\bibinfo {title} {Quantum simulation of real-space dynamics},}\ }\href {\doibase 10.22331/q-2022-11-17-860} {\bibfield  {journal} {\bibinfo  {journal} {Quantum}\ }\textbf {\bibinfo {volume} {6}},\ \bibinfo {pages} {860} (\bibinfo {year} {2022})}\BibitemShut {NoStop}%
\bibitem [{\citenamefont {Ollitrault}\ \emph {et~al.}(2023)\citenamefont {Ollitrault}, \citenamefont {Jandura}, \citenamefont {Miessen}, \citenamefont {Burghardt}, \citenamefont {Martinazzo}, \citenamefont {Tacchino},\ and\ \citenamefont {Tavernelli}}]{Pauline2023}%
  \BibitemOpen
  \bibfield  {author} {\bibinfo {author} {\bibfnamefont {Pauline~J}\ \bibnamefont {Ollitrault}}, \bibinfo {author} {\bibfnamefont {Sven}\ \bibnamefont {Jandura}}, \bibinfo {author} {\bibfnamefont {Alexander}\ \bibnamefont {Miessen}}, \bibinfo {author} {\bibfnamefont {Irene}\ \bibnamefont {Burghardt}}, \bibinfo {author} {\bibfnamefont {Rocco}\ \bibnamefont {Martinazzo}}, \bibinfo {author} {\bibfnamefont {Francesco}\ \bibnamefont {Tacchino}}, \ and\ \bibinfo {author} {\bibfnamefont {Ivano}\ \bibnamefont {Tavernelli}},\ }\bibfield  {title} {\enquote {\bibinfo {title} {Quantum algorithms for grid-based variational time evolution},}\ }\href {\doibase 10.22331/q-2023-10-12-1139} {\bibfield  {journal} {\bibinfo  {journal} {{Quantum}}\ }\textbf {\bibinfo {volume} {7}},\ \bibinfo {pages} {1139} (\bibinfo {year} {2023})}\BibitemShut {NoStop}%
\bibitem [{\citenamefont {Lloyd}(1996)}]{Lloyd1996}%
  \BibitemOpen
  \bibfield  {author} {\bibinfo {author} {\bibfnamefont {Seth}\ \bibnamefont {Lloyd}},\ }\bibfield  {title} {\enquote {\bibinfo {title} {Universal quantum simulators},}\ }\href {https://api.semanticscholar.org/CorpusID:43496899} {\bibfield  {journal} {\bibinfo  {journal} {Science}\ }\textbf {\bibinfo {volume} {273}},\ \bibinfo {pages} {1073 -- 1078} (\bibinfo {year} {1996})}\BibitemShut {NoStop}%
\bibitem [{\citenamefont {Suzuki}(1990)}]{SUZUKI1990}%
  \BibitemOpen
  \bibfield  {author} {\bibinfo {author} {\bibfnamefont {Masuo}\ \bibnamefont {Suzuki}},\ }\bibfield  {title} {\enquote {\bibinfo {title} {Fractal decomposition of exponential operators with applications to many-body theories and monte carlo simulations},}\ }\href {\doibase https://doi.org/10.1016/0375-9601(90)90962-N} {\bibfield  {journal} {\bibinfo  {journal} {Physics Letters A}\ }\textbf {\bibinfo {volume} {146}},\ \bibinfo {pages} {319--323} (\bibinfo {year} {1990})}\BibitemShut {NoStop}%
\bibitem [{\citenamefont {Ostmeyer}(2023)}]{Ostmeyer2023}%
  \BibitemOpen
  \bibfield  {author} {\bibinfo {author} {\bibfnamefont {Johann}\ \bibnamefont {Ostmeyer}},\ }\bibfield  {title} {\enquote {\bibinfo {title} {Optimised trotter decompositions for classical and quantum computing},}\ }\href {\doibase 10.1088/1751-8121/acde7a} {\bibfield  {journal} {\bibinfo  {journal} {Journal of Physics A: Mathematical and Theoretical}\ }\textbf {\bibinfo {volume} {56}},\ \bibinfo {pages} {285303} (\bibinfo {year} {2023})}\BibitemShut {NoStop}%
\bibitem [{\citenamefont {Dhand}\ and\ \citenamefont {Sanders}(2014)}]{Dhand2014}%
  \BibitemOpen
  \bibfield  {author} {\bibinfo {author} {\bibfnamefont {Ish}\ \bibnamefont {Dhand}}\ and\ \bibinfo {author} {\bibfnamefont {Barry~C}\ \bibnamefont {Sanders}},\ }\bibfield  {title} {\enquote {\bibinfo {title} {Stability of the trotter–suzuki decomposition},}\ }\href {\doibase 10.1088/1751-8113/47/26/265206} {\bibfield  {journal} {\bibinfo  {journal} {Journal of Physics A: Mathematical and Theoretical}\ }\textbf {\bibinfo {volume} {47}},\ \bibinfo {pages} {265206} (\bibinfo {year} {2014})}\BibitemShut {NoStop}%
\bibitem [{\citenamefont {Zhu}\ \emph {et~al.}(2024)\citenamefont {Zhu}, \citenamefont {Wan},\ and\ \citenamefont {Zhang}}]{Zhu2024}%
  \BibitemOpen
  \bibfield  {author} {\bibinfo {author} {\bibfnamefont {Xiaogang}\ \bibnamefont {Zhu}}, \bibinfo {author} {\bibfnamefont {Haiyang}\ \bibnamefont {Wan}}, \ and\ \bibinfo {author} {\bibfnamefont {Yaping}\ \bibnamefont {Zhang}},\ }\bibfield  {title} {\enquote {\bibinfo {title} {A split-step finite element method for the space-fractional schr{\"o}dinger equation in two dimensions},}\ }\href {\doibase 10.1038/s41598-024-75547-2} {\bibfield  {journal} {\bibinfo  {journal} {Scientific Reports}\ }\textbf {\bibinfo {volume} {14}},\ \bibinfo {pages} {24257} (\bibinfo {year} {2024})}\BibitemShut {NoStop}%
\bibitem [{\citenamefont {Semenova}\ \emph {et~al.}(2021)\citenamefont {Semenova}, \citenamefont {Dyachenko}, \citenamefont {Korotkevich},\ and\ \citenamefont {Lushnikov}}]{Semenova2021}%
  \BibitemOpen
  \bibfield  {author} {\bibinfo {author} {\bibfnamefont {Anastassiya}\ \bibnamefont {Semenova}}, \bibinfo {author} {\bibfnamefont {Sergey~A.}\ \bibnamefont {Dyachenko}}, \bibinfo {author} {\bibfnamefont {Alexander~O.}\ \bibnamefont {Korotkevich}}, \ and\ \bibinfo {author} {\bibfnamefont {Pavel~M.}\ \bibnamefont {Lushnikov}},\ }\bibfield  {title} {\enquote {\bibinfo {title} {Comparison of split-step and hamiltonian integration methods for simulation of the nonlinear schrödinger type equations},}\ }\href {\doibase 10.1016/j.jcp.2020.110061} {\bibfield  {journal} {\bibinfo  {journal} {Journal of Computational Physics}\ }\textbf {\bibinfo {volume} {427}},\ \bibinfo {pages} {110061} (\bibinfo {year} {2021})}\BibitemShut {NoStop}%
\bibitem [{\citenamefont {Sathyamurthy}\ and\ \citenamefont {Mahapatra}(2021)}]{Sathya2021}%
  \BibitemOpen
  \bibfield  {author} {\bibinfo {author} {\bibfnamefont {Narayanasami}\ \bibnamefont {Sathyamurthy}}\ and\ \bibinfo {author} {\bibfnamefont {Susanta}\ \bibnamefont {Mahapatra}},\ }\bibfield  {title} {\enquote {\bibinfo {title} {Time-dependent quantum mechanical wave packet dynamics},}\ }\href {\doibase 10.1039/D0CP03929B} {\bibfield  {journal} {\bibinfo  {journal} {Phys. Chem. Chem. Phys.}\ }\textbf {\bibinfo {volume} {23}},\ \bibinfo {pages} {7586--7614} (\bibinfo {year} {2021})}\BibitemShut {NoStop}%
\bibitem [{\citenamefont {Roulet}\ and\ \citenamefont {Vaníček}(2021)}]{Roulet2021}%
  \BibitemOpen
  \bibfield  {author} {\bibinfo {author} {\bibfnamefont {Julien}\ \bibnamefont {Roulet}}\ and\ \bibinfo {author} {\bibfnamefont {Jiří}\ \bibnamefont {Vaníček}},\ }\bibfield  {title} {\enquote {\bibinfo {title} {An implicit split-operator algorithm for the nonlinear time-dependent schrödinger equation},}\ }\href {http://dx.doi.org/10.1063/5.0071153} {\bibfield  {journal} {\bibinfo  {journal} {The Journal of Chemical Physics}\ }\textbf {\bibinfo {volume} {155}} (\bibinfo {year} {2021})}\BibitemShut {NoStop}%
\bibitem [{\citenamefont {Bandrauk}\ and\ \citenamefont {Shen}(1993)}]{Bandrauk1993}%
  \BibitemOpen
  \bibfield  {author} {\bibinfo {author} {\bibfnamefont {André~D.}\ \bibnamefont {Bandrauk}}\ and\ \bibinfo {author} {\bibfnamefont {Hai}\ \bibnamefont {Shen}},\ }\bibfield  {title} {\enquote {\bibinfo {title} {Exponential split operator methods for solving coupled time‐dependent schrödinger equations},}\ }\href {\doibase 10.1063/1.465362} {\bibfield  {journal} {\bibinfo  {journal} {The Journal of Chemical Physics}\ }\textbf {\bibinfo {volume} {99}},\ \bibinfo {pages} {1185--1193} (\bibinfo {year} {1993})}\BibitemShut {NoStop}%
\bibitem [{\citenamefont {Bandrauk}\ and\ \citenamefont {Shen}(1991)}]{BANDRAUK1991}%
  \BibitemOpen
  \bibfield  {author} {\bibinfo {author} {\bibfnamefont {André~D.}\ \bibnamefont {Bandrauk}}\ and\ \bibinfo {author} {\bibfnamefont {Hai}\ \bibnamefont {Shen}},\ }\bibfield  {title} {\enquote {\bibinfo {title} {Improved exponential split operator method for solving the time-dependent schrödinger equation},}\ }\href {\doibase https://doi.org/10.1016/0009-2614(91)90232-X} {\bibfield  {journal} {\bibinfo  {journal} {Chemical Physics Letters}\ }\textbf {\bibinfo {volume} {176}},\ \bibinfo {pages} {428--432} (\bibinfo {year} {1991})}\BibitemShut {NoStop}%
\bibitem [{\citenamefont {Hermann}\ and\ \citenamefont {Fleck}(1988)}]{Hermann1988}%
  \BibitemOpen
  \bibfield  {author} {\bibinfo {author} {\bibfnamefont {Mark~R.}\ \bibnamefont {Hermann}}\ and\ \bibinfo {author} {\bibfnamefont {J.~A.}\ \bibnamefont {Fleck}},\ }\bibfield  {title} {\enquote {\bibinfo {title} {Split-operator spectral method for solving the time-dependent schr\"odinger equation in spherical coordinates},}\ }\href {\doibase 10.1103/PhysRevA.38.6000} {\bibfield  {journal} {\bibinfo  {journal} {Phys. Rev. A}\ }\textbf {\bibinfo {volume} {38}},\ \bibinfo {pages} {6000--6012} (\bibinfo {year} {1988})}\BibitemShut {NoStop}%
\bibitem [{\citenamefont {Chen}\ \emph {et~al.}(2023)\citenamefont {Chen}, \citenamefont {Stoudenmire},\ and\ \citenamefont {White}}]{Chen2023}%
  \BibitemOpen
  \bibfield  {author} {\bibinfo {author} {\bibfnamefont {Jielun}\ \bibnamefont {Chen}}, \bibinfo {author} {\bibfnamefont {E.M.}\ \bibnamefont {Stoudenmire}}, \ and\ \bibinfo {author} {\bibfnamefont {Steven~R.}\ \bibnamefont {White}},\ }\bibfield  {title} {\enquote {\bibinfo {title} {Quantum fourier transform has small entanglement},}\ }\href {http://dx.doi.org/10.1103/PRXQuantum.4.040318} {\bibfield  {journal} {\bibinfo  {journal} {PRX Quantum}\ }\textbf {\bibinfo {volume} {4}} (\bibinfo {year} {2023})}\BibitemShut {NoStop}%
\bibitem [{\citenamefont {Dixit}\ and\ \citenamefont {Jian}(2022)}]{Dixit2022}%
  \BibitemOpen
  \bibfield  {author} {\bibinfo {author} {\bibfnamefont {Vinayak}\ \bibnamefont {Dixit}}\ and\ \bibinfo {author} {\bibfnamefont {Sisi}\ \bibnamefont {Jian}},\ }\bibfield  {title} {\enquote {\bibinfo {title} {Quantum fourier transform to estimate drive cycles},}\ }\href {\doibase 10.1038/s41598-021-04639-0} {\bibfield  {journal} {\bibinfo  {journal} {Scientific Reports}\ }\textbf {\bibinfo {volume} {12}},\ \bibinfo {pages} {654} (\bibinfo {year} {2022})}\BibitemShut {NoStop}%
\bibitem [{\citenamefont {Camps}\ \emph {et~al.}(2020)\citenamefont {Camps}, \citenamefont {Van~Beeumen},\ and\ \citenamefont {Yang}}]{Camps2020}%
  \BibitemOpen
  \bibfield  {author} {\bibinfo {author} {\bibfnamefont {Daan}\ \bibnamefont {Camps}}, \bibinfo {author} {\bibfnamefont {Roel}\ \bibnamefont {Van~Beeumen}}, \ and\ \bibinfo {author} {\bibfnamefont {Chao}\ \bibnamefont {Yang}},\ }\bibfield  {title} {\enquote {\bibinfo {title} {Quantum fourier transform revisited},}\ }\href {http://dx.doi.org/10.1002/nla.2331} {\bibfield  {journal} {\bibinfo  {journal} {Numerical Linear Algebra with Applications}\ }\textbf {\bibinfo {volume} {28}} (\bibinfo {year} {2020})}\BibitemShut {NoStop}%
\bibitem [{\citenamefont {Wang}\ \emph {et~al.}(2020)\citenamefont {Wang}, \citenamefont {Curtis}, \citenamefont {Lester}, \citenamefont {Zhang}, \citenamefont {Gao}, \citenamefont {Freeze}, \citenamefont {Batista}, \citenamefont {Vaccaro}, \citenamefont {Chuang}, \citenamefont {Frunzio}, \citenamefont {Jiang}, \citenamefont {Girvin},\ and\ \citenamefont {Schoelkopf}}]{Wang_2020}%
  \BibitemOpen
  \bibfield  {author} {\bibinfo {author} {\bibfnamefont {Christopher~S.}\ \bibnamefont {Wang}}, \bibinfo {author} {\bibfnamefont {Jacob~C.}\ \bibnamefont {Curtis}}, \bibinfo {author} {\bibfnamefont {Brian~J.}\ \bibnamefont {Lester}}, \bibinfo {author} {\bibfnamefont {Yaxing}\ \bibnamefont {Zhang}}, \bibinfo {author} {\bibfnamefont {Yvonne~Y.}\ \bibnamefont {Gao}}, \bibinfo {author} {\bibfnamefont {Jessica}\ \bibnamefont {Freeze}}, \bibinfo {author} {\bibfnamefont {Victor~S.}\ \bibnamefont {Batista}}, \bibinfo {author} {\bibfnamefont {Patrick~H.}\ \bibnamefont {Vaccaro}}, \bibinfo {author} {\bibfnamefont {Isaac~L.}\ \bibnamefont {Chuang}}, \bibinfo {author} {\bibfnamefont {Luigi}\ \bibnamefont {Frunzio}}, \bibinfo {author} {\bibfnamefont {Liang}\ \bibnamefont {Jiang}}, \bibinfo {author} {\bibfnamefont {S.~M.}\ \bibnamefont {Girvin}}, \ and\ \bibinfo {author} {\bibfnamefont {Robert~J.}\ \bibnamefont {Schoelkopf}},\ }\bibfield  {title} {\enquote {\bibinfo {title} {Efficient multiphoton sampling of molecular
  vibronic spectra on a superconducting bosonic processor},}\ }\href {http://dx.doi.org/10.1103/PhysRevX.10.021060} {\bibfield  {journal} {\bibinfo  {journal} {Physical Review X}\ }\textbf {\bibinfo {volume} {10}} (\bibinfo {year} {2020})}\BibitemShut {NoStop}%
\bibitem [{\citenamefont {Chen}\ \emph {et~al.}(2021)\citenamefont {Chen}, \citenamefont {Gan}, \citenamefont {Zhang}, \citenamefont {Matuskevich},\ and\ \citenamefont {Kim}}]{Chen_2021}%
  \BibitemOpen
  \bibfield  {author} {\bibinfo {author} {\bibfnamefont {Wentao}\ \bibnamefont {Chen}}, \bibinfo {author} {\bibfnamefont {Jaren}\ \bibnamefont {Gan}}, \bibinfo {author} {\bibfnamefont {Jing-Ning}\ \bibnamefont {Zhang}}, \bibinfo {author} {\bibfnamefont {Dzmitry}\ \bibnamefont {Matuskevich}}, \ and\ \bibinfo {author} {\bibfnamefont {Kihwan}\ \bibnamefont {Kim}},\ }\bibfield  {title} {\enquote {\bibinfo {title} {Quantum computation and simulation with vibrational modes of trapped ions},}\ }\href {\doibase 10.1088/1674-1056/ac01e3} {\bibfield  {journal} {\bibinfo  {journal} {Chinese Physics B}\ }\textbf {\bibinfo {volume} {30}},\ \bibinfo {pages} {060311} (\bibinfo {year} {2021})}\BibitemShut {NoStop}%
\bibitem [{\citenamefont {Sawaya}\ \emph {et~al.}(2020)\citenamefont {Sawaya}, \citenamefont {Menke}, \citenamefont {Kyaw}, \citenamefont {Johri}, \citenamefont {Aspuru-Guzik},\ and\ \citenamefont {Guerreschi}}]{Sawaya_2020}%
  \BibitemOpen
  \bibfield  {author} {\bibinfo {author} {\bibfnamefont {Nicolas P.~D.}\ \bibnamefont {Sawaya}}, \bibinfo {author} {\bibfnamefont {Tim}\ \bibnamefont {Menke}}, \bibinfo {author} {\bibfnamefont {Thi~Ha}\ \bibnamefont {Kyaw}}, \bibinfo {author} {\bibfnamefont {Sonika}\ \bibnamefont {Johri}}, \bibinfo {author} {\bibfnamefont {Alán}\ \bibnamefont {Aspuru-Guzik}}, \ and\ \bibinfo {author} {\bibfnamefont {Gian~Giacomo}\ \bibnamefont {Guerreschi}},\ }\bibfield  {title} {\enquote {\bibinfo {title} {Resource-efficient digital quantum simulation of d-level systems for photonic, vibrational, and spin-s hamiltonians},}\ }\href {http://dx.doi.org/10.1038/s41534-020-0278-0} {\bibfield  {journal} {\bibinfo  {journal} {npj Quantum Information}\ }\textbf {\bibinfo {volume} {6}} (\bibinfo {year} {2020})}\BibitemShut {NoStop}%
\bibitem [{\citenamefont {Majland}\ \emph {et~al.}(2026)\citenamefont {Majland}, \citenamefont {Jensen}, \citenamefont {Ettenhuber}, \citenamefont {Shaik}, \citenamefont {Zinner},\ and\ \citenamefont {Christiansen}}]{Ove2026}%
  \BibitemOpen
  \bibfield  {author} {\bibinfo {author} {\bibfnamefont {Marco}\ \bibnamefont {Majland}}, \bibinfo {author} {\bibfnamefont {Rasmus~Berg}\ \bibnamefont {Jensen}}, \bibinfo {author} {\bibfnamefont {Patrick}\ \bibnamefont {Ettenhuber}}, \bibinfo {author} {\bibfnamefont {Irfansha}\ \bibnamefont {Shaik}}, \bibinfo {author} {\bibfnamefont {Nikolaj~Thomas}\ \bibnamefont {Zinner}}, \ and\ \bibinfo {author} {\bibfnamefont {Ove}\ \bibnamefont {Christiansen}},\ }\bibfield  {title} {\enquote {\bibinfo {title} {Fault-tolerant quantum computations of vibrational wave functions},}\ }\href {\doibase 10.1021/acs.jctc.5c01419} {\bibfield  {journal} {\bibinfo  {journal} {Journal of Chemical Theory and Computation}\ }\textbf {\bibinfo {volume} {22}},\ \bibinfo {pages} {30--51} (\bibinfo {year} {2026})},\ \bibinfo {note} {pMID: 41429007}\BibitemShut {NoStop}%
\bibitem [{\citenamefont {L\"otstedt}\ \emph {et~al.}(2021)\citenamefont {L\"otstedt}, \citenamefont {Yamanouchi}, \citenamefont {Tsuchiya},\ and\ \citenamefont {Tachikawa}}]{Erik2021}%
  \BibitemOpen
  \bibfield  {author} {\bibinfo {author} {\bibfnamefont {Erik}\ \bibnamefont {L\"otstedt}}, \bibinfo {author} {\bibfnamefont {Kaoru}\ \bibnamefont {Yamanouchi}}, \bibinfo {author} {\bibfnamefont {Takashi}\ \bibnamefont {Tsuchiya}}, \ and\ \bibinfo {author} {\bibfnamefont {Yutaka}\ \bibnamefont {Tachikawa}},\ }\bibfield  {title} {\enquote {\bibinfo {title} {Calculation of vibrational eigenenergies on a quantum computer: Application to the fermi resonance in ${\mathrm{co}}_{2}$},}\ }\href {\doibase 10.1103/PhysRevA.103.062609} {\bibfield  {journal} {\bibinfo  {journal} {Phys. Rev. A}\ }\textbf {\bibinfo {volume} {103}},\ \bibinfo {pages} {062609} (\bibinfo {year} {2021})}\BibitemShut {NoStop}%
\bibitem [{\citenamefont {Suzuki}(1991)}]{SUZUKI1991}%
  \BibitemOpen
  \bibfield  {author} {\bibinfo {author} {\bibfnamefont {Masuo}\ \bibnamefont {Suzuki}},\ }\bibfield  {title} {\enquote {\bibinfo {title} {General theory of fractal path integrals with applications to many‐body theories and statistical physics},}\ }\href {\doibase 10.1063/1.529425} {\bibfield  {journal} {\bibinfo  {journal} {Journal of Mathematical Physics}\ }\textbf {\bibinfo {volume} {32}},\ \bibinfo {pages} {400--407} (\bibinfo {year} {1991})}\BibitemShut {NoStop}%
\bibitem [{\citenamefont {Hatano}\ and\ \citenamefont {Suzuki}(2005)}]{Hatano2005}%
  \BibitemOpen
  \bibfield  {author} {\bibinfo {author} {\bibfnamefont {Naomichi}\ \bibnamefont {Hatano}}\ and\ \bibinfo {author} {\bibfnamefont {Masuo}\ \bibnamefont {Suzuki}},\ }\enquote {\bibinfo {title} {Finding exponential product formulas of higher orders},}\ in\ \href {\doibase 10.1007/11526216_2} {\emph {\bibinfo {booktitle} {Quantum Annealing and Other Optimization Methods}}}\ (\bibinfo  {publisher} {Springer Berlin Heidelberg},\ \bibinfo {year} {2005})\ p.\ \bibinfo {pages} {37–68}\BibitemShut {NoStop}%
\bibitem [{\citenamefont {Jones}\ \emph {et~al.}(2019)\citenamefont {Jones}, \citenamefont {O'Brien}, \citenamefont {White}, \citenamefont {Campbell},\ and\ \citenamefont {Clark}}]{jones2019}%
  \BibitemOpen
  \bibfield  {author} {\bibinfo {author} {\bibfnamefont {Benjamin D.~M.}\ \bibnamefont {Jones}}, \bibinfo {author} {\bibfnamefont {George~O.}\ \bibnamefont {O'Brien}}, \bibinfo {author} {\bibfnamefont {David~R.}\ \bibnamefont {White}}, \bibinfo {author} {\bibfnamefont {Earl~T.}\ \bibnamefont {Campbell}}, \ and\ \bibinfo {author} {\bibfnamefont {John~A.}\ \bibnamefont {Clark}},\ }\href {https://arxiv.org/abs/1904.01336} {\enquote {\bibinfo {title} {Optimising trotter-suzuki decompositions for quantum simulation using evolutionary strategies},}\ } (\bibinfo {year} {2019})\BibitemShut {NoStop}%
\bibitem [{\citenamefont {Balint-Kurti}\ \emph {et~al.}(1990)\citenamefont {Balint-Kurti}, \citenamefont {Dixon},\ and\ \citenamefont {Marston}}]{Balint1990}%
  \BibitemOpen
  \bibfield  {author} {\bibinfo {author} {\bibfnamefont {Gabriel~G.}\ \bibnamefont {Balint-Kurti}}, \bibinfo {author} {\bibfnamefont {Richard~N.}\ \bibnamefont {Dixon}}, \ and\ \bibinfo {author} {\bibfnamefont {C.~Clay}\ \bibnamefont {Marston}},\ }\bibfield  {title} {\enquote {\bibinfo {title} {Time-dependent quantum dynamics of molecular photofragmentation processes},}\ }\href {\doibase 10.1039/FT9908601741} {\bibfield  {journal} {\bibinfo  {journal} {J. Chem. Soc.{,} Faraday Trans.}\ }\textbf {\bibinfo {volume} {86}},\ \bibinfo {pages} {1741--1749} (\bibinfo {year} {1990})}\BibitemShut {NoStop}%
\bibitem [{\citenamefont {Vendrell}\ \emph {et~al.}(2007)\citenamefont {Vendrell}, \citenamefont {Gatti},\ and\ \citenamefont {Meyer}}]{Vendrell2007}%
  \BibitemOpen
  \bibfield  {author} {\bibinfo {author} {\bibfnamefont {Oriol}\ \bibnamefont {Vendrell}}, \bibinfo {author} {\bibfnamefont {Fabien}\ \bibnamefont {Gatti}}, \ and\ \bibinfo {author} {\bibfnamefont {Hans-Dieter}\ \bibnamefont {Meyer}},\ }\bibfield  {title} {\enquote {\bibinfo {title} {Full dimensional (15-dimensional) quantum-dynamical simulation of the protonated water dimer. ii. infrared spectrum and vibrational dynamics},}\ }\href {http://dx.doi.org/10.1063/1.2787596} {\bibfield  {journal} {\bibinfo  {journal} {The Journal of Chemical Physics}\ }\textbf {\bibinfo {volume} {127}} (\bibinfo {year} {2007})}\BibitemShut {NoStop}%
\bibitem [{\citenamefont {Beck}\ \emph {et~al.}(2000)\citenamefont {Beck}, \citenamefont {Jäckle}, \citenamefont {Worth},\ and\ \citenamefont {Meyer}}]{BECK2000}%
  \BibitemOpen
  \bibfield  {author} {\bibinfo {author} {\bibfnamefont {M.H.}\ \bibnamefont {Beck}}, \bibinfo {author} {\bibfnamefont {A.}~\bibnamefont {Jäckle}}, \bibinfo {author} {\bibfnamefont {G.A.}\ \bibnamefont {Worth}}, \ and\ \bibinfo {author} {\bibfnamefont {H.-D.}\ \bibnamefont {Meyer}},\ }\bibfield  {title} {\enquote {\bibinfo {title} {The multiconfiguration time-dependent hartree (mctdh) method: a highly efficient algorithm for propagating wavepackets},}\ }\href {\doibase https://doi.org/10.1016/S0370-1573(99)00047-2} {\bibfield  {journal} {\bibinfo  {journal} {Physics Reports}\ }\textbf {\bibinfo {volume} {324}},\ \bibinfo {pages} {1--105} (\bibinfo {year} {2000})}\BibitemShut {NoStop}%
\bibitem [{\citenamefont {Peláez}\ and\ \citenamefont {Meyer}(2017)}]{PELAEZ2017}%
  \BibitemOpen
  \bibfield  {author} {\bibinfo {author} {\bibfnamefont {Daniel}\ \bibnamefont {Peláez}}\ and\ \bibinfo {author} {\bibfnamefont {Hans-Dieter}\ \bibnamefont {Meyer}},\ }\bibfield  {title} {\enquote {\bibinfo {title} {On the infrared absorption spectrum of the hydrated hydroxide (h3o2-) cluster anion},}\ }\href {\doibase https://doi.org/10.1016/j.chemphys.2016.08.025} {\bibfield  {journal} {\bibinfo  {journal} {Chemical Physics}\ }\textbf {\bibinfo {volume} {482}},\ \bibinfo {pages} {100--105} (\bibinfo {year} {2017})}\BibitemShut {NoStop}%
\bibitem [{\citenamefont {Feng}\ \emph {et~al.}(2026)\citenamefont {Feng}, \citenamefont {Chan},\ and\ \citenamefont {Tew}}]{Feng_arxiv_2025}%
  \BibitemOpen
  \bibfield  {author} {\bibinfo {author} {\bibfnamefont {Xiaoning}\ \bibnamefont {Feng}}, \bibinfo {author} {\bibfnamefont {Hans Hon~Sang}\ \bibnamefont {Chan}}, \ and\ \bibinfo {author} {\bibfnamefont {David~P.}\ \bibnamefont {Tew}},\ }\href {https://arxiv.org/abs/2506.08609} {\enquote {\bibinfo {title} {Quantum resource assay for the grid-based simulation of the photodynamics of pyrazine},}\ } (\bibinfo {year} {2026}),\ \Eprint {http://arxiv.org/abs/2506.08609} {arXiv:2506.08609 [quant-ph]} \BibitemShut {NoStop}%
\bibitem [{\citenamefont {Lehtovaara}\ \emph {et~al.}(2007)\citenamefont {Lehtovaara}, \citenamefont {Toivanen},\ and\ \citenamefont {Eloranta}}]{LEHTOVAARA2007}%
  \BibitemOpen
  \bibfield  {author} {\bibinfo {author} {\bibfnamefont {L.}~\bibnamefont {Lehtovaara}}, \bibinfo {author} {\bibfnamefont {J.}~\bibnamefont {Toivanen}}, \ and\ \bibinfo {author} {\bibfnamefont {J.}~\bibnamefont {Eloranta}},\ }\bibfield  {title} {\enquote {\bibinfo {title} {Solution of time-independent schrödinger equation by the imaginary time propagation method},}\ }\href {\doibase https://doi.org/10.1016/j.jcp.2006.06.006} {\bibfield  {journal} {\bibinfo  {journal} {Journal of Computational Physics}\ }\textbf {\bibinfo {volume} {221}},\ \bibinfo {pages} {148--157} (\bibinfo {year} {2007})}\BibitemShut {NoStop}%
\bibitem [{\citenamefont {McClean}\ and\ \citenamefont {Aspuru-Guzik}(2015)}]{McClean2015}%
  \BibitemOpen
  \bibfield  {author} {\bibinfo {author} {\bibfnamefont {Jarrod~R.}\ \bibnamefont {McClean}}\ and\ \bibinfo {author} {\bibfnamefont {Alán}\ \bibnamefont {Aspuru-Guzik}},\ }\bibfield  {title} {\enquote {\bibinfo {title} {Compact wavefunctions from compressed imaginary time evolution},}\ }\href {\doibase 10.1039/c5ra23047k} {\bibfield  {journal} {\bibinfo  {journal} {RSC Advances}\ }\textbf {\bibinfo {volume} {5}},\ \bibinfo {pages} {102277–102283} (\bibinfo {year} {2015})}\BibitemShut {NoStop}%
\bibitem [{\citenamefont {McArdle}\ \emph {et~al.}(2019{\natexlab{b}})\citenamefont {McArdle}, \citenamefont {Jones}, \citenamefont {Endo}, \citenamefont {Li}, \citenamefont {Benjamin},\ and\ \citenamefont {Yuan}}]{McArdle20192}%
  \BibitemOpen
  \bibfield  {author} {\bibinfo {author} {\bibfnamefont {Sam}\ \bibnamefont {McArdle}}, \bibinfo {author} {\bibfnamefont {Tyson}\ \bibnamefont {Jones}}, \bibinfo {author} {\bibfnamefont {Suguru}\ \bibnamefont {Endo}}, \bibinfo {author} {\bibfnamefont {Ying}\ \bibnamefont {Li}}, \bibinfo {author} {\bibfnamefont {Simon~C.}\ \bibnamefont {Benjamin}}, \ and\ \bibinfo {author} {\bibfnamefont {Xiao}\ \bibnamefont {Yuan}},\ }\bibfield  {title} {\enquote {\bibinfo {title} {Variational ansatz-based quantum simulation of imaginary time evolution},}\ }\href {http://dx.doi.org/10.1038/s41534-019-0187-2} {\bibfield  {journal} {\bibinfo  {journal} {npj Quantum Information}\ }\textbf {\bibinfo {volume} {5}} (\bibinfo {year} {2019}{\natexlab{b}})}\BibitemShut {NoStop}%
\bibitem [{\citenamefont {Gomes}\ \emph {et~al.}(2020)\citenamefont {Gomes}, \citenamefont {Zhang}, \citenamefont {Berthusen}, \citenamefont {Wang}, \citenamefont {Ho}, \citenamefont {Orth},\ and\ \citenamefont {Yao}}]{Gomes2020}%
  \BibitemOpen
  \bibfield  {author} {\bibinfo {author} {\bibfnamefont {Niladri}\ \bibnamefont {Gomes}}, \bibinfo {author} {\bibfnamefont {Feng}\ \bibnamefont {Zhang}}, \bibinfo {author} {\bibfnamefont {Noah~F.}\ \bibnamefont {Berthusen}}, \bibinfo {author} {\bibfnamefont {Cai-Zhuang}\ \bibnamefont {Wang}}, \bibinfo {author} {\bibfnamefont {Kai-Ming}\ \bibnamefont {Ho}}, \bibinfo {author} {\bibfnamefont {Peter~P.}\ \bibnamefont {Orth}}, \ and\ \bibinfo {author} {\bibfnamefont {Yongxin}\ \bibnamefont {Yao}},\ }\bibfield  {title} {\enquote {\bibinfo {title} {Efficient step-merged quantum imaginary time evolution algorithm for quantum chemistry},}\ }\href {\doibase 10.1021/acs.jctc.0c00666} {\bibfield  {journal} {\bibinfo  {journal} {Journal of Chemical Theory and Computation}\ }\textbf {\bibinfo {volume} {16}},\ \bibinfo {pages} {6256–6266} (\bibinfo {year} {2020})}\BibitemShut {NoStop}%
\bibitem [{\citenamefont {Leadbeater}\ \emph {et~al.}(2024)\citenamefont {Leadbeater}, \citenamefont {Fitzpatrick}, \citenamefont {Ramo},\ and\ \citenamefont {Thom}}]{Leadbeater2024}%
  \BibitemOpen
  \bibfield  {author} {\bibinfo {author} {\bibfnamefont {Chiara}\ \bibnamefont {Leadbeater}}, \bibinfo {author} {\bibfnamefont {Nathan}\ \bibnamefont {Fitzpatrick}}, \bibinfo {author} {\bibfnamefont {David~Muñoz}\ \bibnamefont {Ramo}}, \ and\ \bibinfo {author} {\bibfnamefont {Alex J~W}\ \bibnamefont {Thom}},\ }\bibfield  {title} {\enquote {\bibinfo {title} {Non-unitary trotter circuits for imaginary time evolution},}\ }\href {\doibase 10.1088/2058-9565/ad53fb} {\bibfield  {journal} {\bibinfo  {journal} {Quantum Science and Technology}\ }\textbf {\bibinfo {volume} {9}},\ \bibinfo {pages} {045007} (\bibinfo {year} {2024})}\BibitemShut {NoStop}%
\bibitem [{\citenamefont {Xie}\ \emph {et~al.}(2024)\citenamefont {Xie}, \citenamefont {Wei}, \citenamefont {Yang}, \citenamefont {Wang}, \citenamefont {Chen}, \citenamefont {Fan},\ and\ \citenamefont {Long}}]{Xie2024}%
  \BibitemOpen
  \bibfield  {author} {\bibinfo {author} {\bibfnamefont {Hao-Nan}\ \bibnamefont {Xie}}, \bibinfo {author} {\bibfnamefont {Shi-Jie}\ \bibnamefont {Wei}}, \bibinfo {author} {\bibfnamefont {Fan}\ \bibnamefont {Yang}}, \bibinfo {author} {\bibfnamefont {Zheng-An}\ \bibnamefont {Wang}}, \bibinfo {author} {\bibfnamefont {Chi-Tong}\ \bibnamefont {Chen}}, \bibinfo {author} {\bibfnamefont {Heng}\ \bibnamefont {Fan}}, \ and\ \bibinfo {author} {\bibfnamefont {Gui-Lu}\ \bibnamefont {Long}},\ }\bibfield  {title} {\enquote {\bibinfo {title} {Probabilistic imaginary-time evolution algorithm based on nonunitary quantum circuits},}\ }\href {\doibase 10.1103/PhysRevA.109.052414} {\bibfield  {journal} {\bibinfo  {journal} {Phys. Rev. A}\ }\textbf {\bibinfo {volume} {109}},\ \bibinfo {pages} {052414} (\bibinfo {year} {2024})}\BibitemShut {NoStop}%
\bibitem [{\citenamefont {Motta}\ \emph {et~al.}(2019)\citenamefont {Motta}, \citenamefont {Sun}, \citenamefont {Tan}, \citenamefont {O’Rourke}, \citenamefont {Ye}, \citenamefont {Minnich}, \citenamefont {Brandão},\ and\ \citenamefont {Chan}}]{Motta2019}%
  \BibitemOpen
  \bibfield  {author} {\bibinfo {author} {\bibfnamefont {Mario}\ \bibnamefont {Motta}}, \bibinfo {author} {\bibfnamefont {Chong}\ \bibnamefont {Sun}}, \bibinfo {author} {\bibfnamefont {Adrian T.~K.}\ \bibnamefont {Tan}}, \bibinfo {author} {\bibfnamefont {Matthew~J.}\ \bibnamefont {O’Rourke}}, \bibinfo {author} {\bibfnamefont {Erika}\ \bibnamefont {Ye}}, \bibinfo {author} {\bibfnamefont {Austin~J.}\ \bibnamefont {Minnich}}, \bibinfo {author} {\bibfnamefont {Fernando G. S.~L.}\ \bibnamefont {Brandão}}, \ and\ \bibinfo {author} {\bibfnamefont {Garnet Kin-Lic}\ \bibnamefont {Chan}},\ }\bibfield  {title} {\enquote {\bibinfo {title} {Determining eigenstates and thermal states on a quantum computer using quantum imaginary time evolution},}\ }\href {\doibase 10.1038/s41567-019-0704-4} {\bibfield  {journal} {\bibinfo  {journal} {Nature Physics}\ }\textbf {\bibinfo {volume} {16}},\ \bibinfo {pages} {205–210} (\bibinfo {year} {2019})}\BibitemShut {NoStop}%
\bibitem [{\citenamefont {Yeter-Aydeniz}\ \emph {et~al.}(2020)\citenamefont {Yeter-Aydeniz}, \citenamefont {Pooser},\ and\ \citenamefont {Siopsis}}]{Yeter2020}%
  \BibitemOpen
  \bibfield  {author} {\bibinfo {author} {\bibfnamefont {K{\"u}bra}\ \bibnamefont {Yeter-Aydeniz}}, \bibinfo {author} {\bibfnamefont {Raphael~C.}\ \bibnamefont {Pooser}}, \ and\ \bibinfo {author} {\bibfnamefont {George}\ \bibnamefont {Siopsis}},\ }\bibfield  {title} {\enquote {\bibinfo {title} {Practical quantum computation of chemical and nuclear energy levels using quantum imaginary time evolution and lanczos algorithms},}\ }\href {\doibase 10.1038/s41534-020-00290-1} {\bibfield  {journal} {\bibinfo  {journal} {npj Quantum Information}\ }\textbf {\bibinfo {volume} {6}},\ \bibinfo {pages} {63} (\bibinfo {year} {2020})}\BibitemShut {NoStop}%
\bibitem [{\citenamefont {M\"{o}tt\"{o}nen}\ \emph {et~al.}(2005)\citenamefont {M\"{o}tt\"{o}nen}, \citenamefont {Vartiainen}, \citenamefont {Bergholm},\ and\ \citenamefont {Salomaa}}]{Mottenen2005}%
  \BibitemOpen
  \bibfield  {author} {\bibinfo {author} {\bibfnamefont {Mikko}\ \bibnamefont {M\"{o}tt\"{o}nen}}, \bibinfo {author} {\bibfnamefont {Juha~J.}\ \bibnamefont {Vartiainen}}, \bibinfo {author} {\bibfnamefont {Ville}\ \bibnamefont {Bergholm}}, \ and\ \bibinfo {author} {\bibfnamefont {Martti~M.}\ \bibnamefont {Salomaa}},\ }\bibfield  {title} {\enquote {\bibinfo {title} {Transformation of quantum states using uniformly controlled rotations},}\ }\href {https://doi.org/10.26421/QIC5.6-5} {\bibfield  {journal} {\bibinfo  {journal} {Quantum Info. Comput.}\ }\textbf {\bibinfo {volume} {5}},\ \bibinfo {pages} {467–473} (\bibinfo {year} {2005})}\BibitemShut {NoStop}%
\bibitem [{\citenamefont {Moosa}\ \emph {et~al.}(2023)\citenamefont {Moosa}, \citenamefont {Watts}, \citenamefont {Chen}, \citenamefont {Sarma},\ and\ \citenamefont {McMahon}}]{Moosa2023}%
  \BibitemOpen
  \bibfield  {author} {\bibinfo {author} {\bibfnamefont {Mudassir}\ \bibnamefont {Moosa}}, \bibinfo {author} {\bibfnamefont {Thomas~W}\ \bibnamefont {Watts}}, \bibinfo {author} {\bibfnamefont {Yiyou}\ \bibnamefont {Chen}}, \bibinfo {author} {\bibfnamefont {Abhijat}\ \bibnamefont {Sarma}}, \ and\ \bibinfo {author} {\bibfnamefont {Peter~L}\ \bibnamefont {McMahon}},\ }\bibfield  {title} {\enquote {\bibinfo {title} {Linear-depth quantum circuits for loading fourier approximations of arbitrary functions},}\ }\href {\doibase 10.1088/2058-9565/acfc62} {\bibfield  {journal} {\bibinfo  {journal} {Quantum Science and Technology}\ }\textbf {\bibinfo {volume} {9}},\ \bibinfo {pages} {015002} (\bibinfo {year} {2023})}\BibitemShut {NoStop}%
\bibitem [{\citenamefont {Iaconis}\ \emph {et~al.}(2024)\citenamefont {Iaconis}, \citenamefont {Johri},\ and\ \citenamefont {Zhu}}]{Iaconis2024}%
  \BibitemOpen
  \bibfield  {author} {\bibinfo {author} {\bibfnamefont {Jason}\ \bibnamefont {Iaconis}}, \bibinfo {author} {\bibfnamefont {Sonika}\ \bibnamefont {Johri}}, \ and\ \bibinfo {author} {\bibfnamefont {Elton~Yechao}\ \bibnamefont {Zhu}},\ }\bibfield  {title} {\enquote {\bibinfo {title} {Quantum state preparation of normal distributions using matrix product states},}\ }\href {http://dx.doi.org/10.1038/s41534-024-00805-0} {\bibfield  {journal} {\bibinfo  {journal} {npj Quantum Information}\ }\textbf {\bibinfo {volume} {10}} (\bibinfo {year} {2024})}\BibitemShut {NoStop}%
\bibitem [{\citenamefont {Rattew}\ \emph {et~al.}(2021)\citenamefont {Rattew}, \citenamefont {Sun}, \citenamefont {Minssen},\ and\ \citenamefont {Pistoia}}]{Rattew2021}%
  \BibitemOpen
  \bibfield  {author} {\bibinfo {author} {\bibfnamefont {Arthur~G.}\ \bibnamefont {Rattew}}, \bibinfo {author} {\bibfnamefont {Yue}\ \bibnamefont {Sun}}, \bibinfo {author} {\bibfnamefont {Pierre}\ \bibnamefont {Minssen}}, \ and\ \bibinfo {author} {\bibfnamefont {Marco}\ \bibnamefont {Pistoia}},\ }\bibfield  {title} {\enquote {\bibinfo {title} {The efficient preparation of normal distributions in quantum registers},}\ }\href {\doibase 10.22331/q-2021-12-23-609} {\bibfield  {journal} {\bibinfo  {journal} {Quantum}\ }\textbf {\bibinfo {volume} {5}},\ \bibinfo {pages} {609} (\bibinfo {year} {2021})}\BibitemShut {NoStop}%
\bibitem [{\citenamefont {Bauer}\ \emph {et~al.}(2021)\citenamefont {Bauer}, \citenamefont {Deliyannis}, \citenamefont {Freytsis},\ and\ \citenamefont {Nachman}}]{bauer2021}%
  \BibitemOpen
  \bibfield  {author} {\bibinfo {author} {\bibfnamefont {Christian~W.}\ \bibnamefont {Bauer}}, \bibinfo {author} {\bibfnamefont {Plato}\ \bibnamefont {Deliyannis}}, \bibinfo {author} {\bibfnamefont {Marat}\ \bibnamefont {Freytsis}}, \ and\ \bibinfo {author} {\bibfnamefont {Benjamin}\ \bibnamefont {Nachman}},\ }\href {https://arxiv.org/abs/2109.10918} {\enquote {\bibinfo {title} {Practical considerations for the preparation of multivariate gaussian states on quantum computers},}\ } (\bibinfo {year} {2021})\BibitemShut {NoStop}%
\bibitem [{\citenamefont {Nielsen}\ and\ \citenamefont {Chuang}(2010)}]{nielsen2010}%
  \BibitemOpen
  \bibfield  {author} {\bibinfo {author} {\bibfnamefont {Michael~A.}\ \bibnamefont {Nielsen}}\ and\ \bibinfo {author} {\bibfnamefont {Isaac~L.}\ \bibnamefont {Chuang}},\ }\href {\doibase 10.1017/CBO9780511976667} {\emph {\bibinfo {title} {Quantum Computation and Quantum Information: 10th Anniversary Edition}}}\ (\bibinfo  {publisher} {Cambridge University Press},\ \bibinfo {year} {2010})\BibitemShut {NoStop}%
\bibitem [{\citenamefont {Kosugi}\ \emph {et~al.}(2022)\citenamefont {Kosugi}, \citenamefont {Nishiya}, \citenamefont {Nishi},\ and\ \citenamefont {Matsushita}}]{Kosugi2022}%
  \BibitemOpen
  \bibfield  {author} {\bibinfo {author} {\bibfnamefont {Taichi}\ \bibnamefont {Kosugi}}, \bibinfo {author} {\bibfnamefont {Yusuke}\ \bibnamefont {Nishiya}}, \bibinfo {author} {\bibfnamefont {Hirofumi}\ \bibnamefont {Nishi}}, \ and\ \bibinfo {author} {\bibfnamefont {Yu-ichiro}\ \bibnamefont {Matsushita}},\ }\bibfield  {title} {\enquote {\bibinfo {title} {Imaginary-time evolution using forward and backward real-time evolution with a single ancilla: First-quantized eigensolver algorithm for quantum chemistry},}\ }\href {\doibase 10.1103/PhysRevResearch.4.033121} {\bibfield  {journal} {\bibinfo  {journal} {Phys. Rev. Res.}\ }\textbf {\bibinfo {volume} {4}},\ \bibinfo {pages} {033121} (\bibinfo {year} {2022})}\BibitemShut {NoStop}%
\bibitem [{\citenamefont {Aharonov}\ \emph {et~al.}(2009)\citenamefont {Aharonov}, \citenamefont {Jones},\ and\ \citenamefont {Landau}}]{Aharonov2009}%
  \BibitemOpen
  \bibfield  {author} {\bibinfo {author} {\bibfnamefont {Dorit}\ \bibnamefont {Aharonov}}, \bibinfo {author} {\bibfnamefont {Vaughan}\ \bibnamefont {Jones}}, \ and\ \bibinfo {author} {\bibfnamefont {Zeph}\ \bibnamefont {Landau}},\ }\bibfield  {title} {\enquote {\bibinfo {title} {A polynomial quantum algorithm for approximating the jones polynomial},}\ }\href {\doibase 10.1007/s00453-008-9168-0} {\bibfield  {journal} {\bibinfo  {journal} {Algorithmica}\ }\textbf {\bibinfo {volume} {55}},\ \bibinfo {pages} {395--421} (\bibinfo {year} {2009})}\BibitemShut {NoStop}%
\bibitem [{\citenamefont {Mitarai}\ and\ \citenamefont {Fujii}(2019)}]{Mitarai2019}%
  \BibitemOpen
  \bibfield  {author} {\bibinfo {author} {\bibfnamefont {Kosuke}\ \bibnamefont {Mitarai}}\ and\ \bibinfo {author} {\bibfnamefont {Keisuke}\ \bibnamefont {Fujii}},\ }\bibfield  {title} {\enquote {\bibinfo {title} {Methodology for replacing indirect measurements with direct measurements},}\ }\href {http://dx.doi.org/10.1103/PhysRevResearch.1.013006} {\bibfield  {journal} {\bibinfo  {journal} {Physical Review Research}\ }\textbf {\bibinfo {volume} {1}} (\bibinfo {year} {2019})}\BibitemShut {NoStop}%
\bibitem [{\citenamefont {J.}\ \emph {et~al.}(2022)\citenamefont {J.}, \citenamefont {Adedoyin}, \citenamefont {Ambrosiano}, \citenamefont {Anisimov}, \citenamefont {Casper}, \citenamefont {Chennupati}, \citenamefont {Coffrin}, \citenamefont {Djidjev}, \citenamefont {Gunter}, \citenamefont {Karra}, \citenamefont {Lemons}, \citenamefont {Lin}, \citenamefont {Malyzhenkov}, \citenamefont {Mascarenas}, \citenamefont {Mniszewski}, \citenamefont {Nadiga}, \citenamefont {O’malley}, \citenamefont {Oyen}, \citenamefont {Pakin}, \citenamefont {Prasad}, \citenamefont {Roberts}, \citenamefont {Romero}, \citenamefont {Santhi}, \citenamefont {Sinitsyn}, \citenamefont {Swart}, \citenamefont {Wendelberger}, \citenamefont {Yoon}, \citenamefont {Zamora}, \citenamefont {Zhu}, \citenamefont {Eidenbenz}, \citenamefont {Bärtschi}, \citenamefont {Coles}, \citenamefont {Vuffray},\ and\ \citenamefont {Lokhov}}]{Abhijith2022}%
  \BibitemOpen
  \bibfield  {author} {\bibinfo {author} {\bibfnamefont {Abhijith}\ \bibnamefont {J.}}, \bibinfo {author} {\bibfnamefont {Adetokunbo}\ \bibnamefont {Adedoyin}}, \bibinfo {author} {\bibfnamefont {John}\ \bibnamefont {Ambrosiano}}, \bibinfo {author} {\bibfnamefont {Petr}\ \bibnamefont {Anisimov}}, \bibinfo {author} {\bibfnamefont {William}\ \bibnamefont {Casper}}, \bibinfo {author} {\bibfnamefont {Gopinath}\ \bibnamefont {Chennupati}}, \bibinfo {author} {\bibfnamefont {Carleton}\ \bibnamefont {Coffrin}}, \bibinfo {author} {\bibfnamefont {Hristo}\ \bibnamefont {Djidjev}}, \bibinfo {author} {\bibfnamefont {David}\ \bibnamefont {Gunter}}, \bibinfo {author} {\bibfnamefont {Satish}\ \bibnamefont {Karra}}, \bibinfo {author} {\bibfnamefont {Nathan}\ \bibnamefont {Lemons}}, \bibinfo {author} {\bibfnamefont {Shizeng}\ \bibnamefont {Lin}}, \bibinfo {author} {\bibfnamefont {Alexander}\ \bibnamefont {Malyzhenkov}}, \bibinfo {author} {\bibfnamefont {David}\ \bibnamefont {Mascarenas}}, \bibinfo {author} {\bibfnamefont
  {Susan}\ \bibnamefont {Mniszewski}}, \bibinfo {author} {\bibfnamefont {Balu}\ \bibnamefont {Nadiga}}, \bibinfo {author} {\bibfnamefont {Daniel}\ \bibnamefont {O’malley}}, \bibinfo {author} {\bibfnamefont {Diane}\ \bibnamefont {Oyen}}, \bibinfo {author} {\bibfnamefont {Scott}\ \bibnamefont {Pakin}}, \bibinfo {author} {\bibfnamefont {Lakshman}\ \bibnamefont {Prasad}}, \bibinfo {author} {\bibfnamefont {Randy}\ \bibnamefont {Roberts}}, \bibinfo {author} {\bibfnamefont {Phillip}\ \bibnamefont {Romero}}, \bibinfo {author} {\bibfnamefont {Nandakishore}\ \bibnamefont {Santhi}}, \bibinfo {author} {\bibfnamefont {Nikolai}\ \bibnamefont {Sinitsyn}}, \bibinfo {author} {\bibfnamefont {Pieter~J.}\ \bibnamefont {Swart}}, \bibinfo {author} {\bibfnamefont {James~G.}\ \bibnamefont {Wendelberger}}, \bibinfo {author} {\bibfnamefont {Boram}\ \bibnamefont {Yoon}}, \bibinfo {author} {\bibfnamefont {Richard}\ \bibnamefont {Zamora}}, \bibinfo {author} {\bibfnamefont {Wei}\ \bibnamefont {Zhu}}, \bibinfo {author} {\bibfnamefont
  {Stephan}\ \bibnamefont {Eidenbenz}}, \bibinfo {author} {\bibfnamefont {Andreas}\ \bibnamefont {Bärtschi}}, \bibinfo {author} {\bibfnamefont {Patrick~J.}\ \bibnamefont {Coles}}, \bibinfo {author} {\bibfnamefont {Marc}\ \bibnamefont {Vuffray}}, \ and\ \bibinfo {author} {\bibfnamefont {Andrey~Y.}\ \bibnamefont {Lokhov}},\ }\bibfield  {title} {\enquote {\bibinfo {title} {Quantum algorithm implementations for beginners},}\ }\href {\doibase 10.1145/3517340} {\bibfield  {journal} {\bibinfo  {journal} {ACM Transactions on Quantum Computing}\ }\textbf {\bibinfo {volume} {3}},\ \bibinfo {pages} {1–92} (\bibinfo {year} {2022})}\BibitemShut {NoStop}%
\bibitem [{\citenamefont {Culot}\ and\ \citenamefont {Li{\'e}vin}(1992)}]{Culot1992}%
  \BibitemOpen
  \bibfield  {author} {\bibinfo {author} {\bibfnamefont {F}~\bibnamefont {Culot}}\ and\ \bibinfo {author} {\bibfnamefont {J}~\bibnamefont {Li{\'e}vin}},\ }\bibfield  {title} {\enquote {\bibinfo {title} {\textit{Ab initio} calculation of vibrational dipole moment matrix elements. {II}. the water molecule as a polyatomic test case},}\ }\href {\doibase 10.1088/0031-8949/46/6/004} {\bibfield  {journal} {\bibinfo  {journal} {Phys. Scr.}\ }\textbf {\bibinfo {volume} {46}},\ \bibinfo {pages} {502--517} (\bibinfo {year} {1992})}\BibitemShut {NoStop}%
\bibitem [{\citenamefont {Rauhut}(2004)}]{Guntram2004}%
  \BibitemOpen
  \bibfield  {author} {\bibinfo {author} {\bibfnamefont {Guntram}\ \bibnamefont {Rauhut}},\ }\bibfield  {title} {\enquote {\bibinfo {title} {Efficient calculation of potential energy surfaces for the generation of vibrational wave functions},}\ }\href {\doibase 10.1063/1.1804174} {\bibfield  {journal} {\bibinfo  {journal} {The Journal of Chemical Physics}\ }\textbf {\bibinfo {volume} {121}},\ \bibinfo {pages} {9313--9322} (\bibinfo {year} {2004})}\BibitemShut {NoStop}%
\bibitem [{\citenamefont {Braams}\ and\ \citenamefont {Bowman}(2009)}]{Bowman2009}%
  \BibitemOpen
  \bibfield  {author} {\bibinfo {author} {\bibfnamefont {Bastiaan~J.}\ \bibnamefont {Braams}}\ and\ \bibinfo {author} {\bibfnamefont {Joel~M.}\ \bibnamefont {Bowman}},\ }\bibfield  {title} {\enquote {\bibinfo {title} {Permutationally invariant potential energy surfaces in high dimensionality},}\ }\href {\doibase 10.1080/01442350903234923} {\bibfield  {journal} {\bibinfo  {journal} {International Reviews in Physical Chemistry}\ }\textbf {\bibinfo {volume} {28}},\ \bibinfo {pages} {577--606} (\bibinfo {year} {2009})}\BibitemShut {NoStop}%
\bibitem [{\citenamefont {König}\ and\ \citenamefont {Christiansen}(2016)}]{Ove2016}%
  \BibitemOpen
  \bibfield  {author} {\bibinfo {author} {\bibfnamefont {Carolin}\ \bibnamefont {König}}\ and\ \bibinfo {author} {\bibfnamefont {Ove}\ \bibnamefont {Christiansen}},\ }\bibfield  {title} {\enquote {\bibinfo {title} {Linear-scaling generation of potential energy surfaces using a double incremental expansion},}\ }\href {\doibase 10.1063/1.4960189} {\bibfield  {journal} {\bibinfo  {journal} {The Journal of Chemical Physics}\ }\textbf {\bibinfo {volume} {145}},\ \bibinfo {pages} {064105} (\bibinfo {year} {2016})}\BibitemShut {NoStop}%
\bibitem [{\citenamefont {Mizukami}\ \emph {et~al.}(2014)\citenamefont {Mizukami}, \citenamefont {Habershon},\ and\ \citenamefont {Tew}}]{Tew2014}%
  \BibitemOpen
  \bibfield  {author} {\bibinfo {author} {\bibfnamefont {Wataru}\ \bibnamefont {Mizukami}}, \bibinfo {author} {\bibfnamefont {Scott}\ \bibnamefont {Habershon}}, \ and\ \bibinfo {author} {\bibfnamefont {David~P.}\ \bibnamefont {Tew}},\ }\bibfield  {title} {\enquote {\bibinfo {title} {A compact and accurate semi-global potential energy surface for malonaldehyde from constrained least squares regression},}\ }\href {\doibase 10.1063/1.4897486} {\bibfield  {journal} {\bibinfo  {journal} {The Journal of Chemical Physics}\ }\textbf {\bibinfo {volume} {141}},\ \bibinfo {pages} {144310} (\bibinfo {year} {2014})}\BibitemShut {NoStop}%
\bibitem [{\citenamefont {Tew}\ and\ \citenamefont {Mizukami}(2016)}]{Tew2016}%
  \BibitemOpen
  \bibfield  {author} {\bibinfo {author} {\bibfnamefont {David~P.}\ \bibnamefont {Tew}}\ and\ \bibinfo {author} {\bibfnamefont {Wataru}\ \bibnamefont {Mizukami}},\ }\bibfield  {title} {\enquote {\bibinfo {title} {Ab initio vibrational spectroscopy of cis- and trans-formic acid from a global potential energy surface},}\ }\href {\doibase 10.1021/acs.jpca.6b09952} {\bibfield  {journal} {\bibinfo  {journal} {The Journal of Physical Chemistry A}\ }\textbf {\bibinfo {volume} {120}},\ \bibinfo {pages} {9815--9828} (\bibinfo {year} {2016})},\ \bibinfo {note} {pMID: 27973803}\BibitemShut {NoStop}%
\bibitem [{\citenamefont {Schr{\"o}der}\ and\ \citenamefont {Rauhut}(2024{\natexlab{b}})}]{Guntram2024}%
  \BibitemOpen
  \bibfield  {author} {\bibinfo {author} {\bibfnamefont {Benjamin}\ \bibnamefont {Schr{\"o}der}}\ and\ \bibinfo {author} {\bibfnamefont {Guntram}\ \bibnamefont {Rauhut}},\ }\bibfield  {title} {\enquote {\bibinfo {title} {From the automated calculation of potential energy surfaces to accurate infrared spectra},}\ }\href {\doibase 10.1021/acs.jpclett.4c00186} {\bibfield  {journal} {\bibinfo  {journal} {The Journal of Physical Chemistry Letters}\ }\textbf {\bibinfo {volume} {15}},\ \bibinfo {pages} {3159--3169} (\bibinfo {year} {2024}{\natexlab{b}})},\ \bibinfo {note} {pMID: 38478898}\BibitemShut {NoStop}%
\bibitem [{\citenamefont {Hättig}\ \emph {et~al.}(2012)\citenamefont {Hättig}, \citenamefont {Klopper}, \citenamefont {K{\"o}hn},\ and\ \citenamefont {Tew}}]{Tew2012}%
  \BibitemOpen
  \bibfield  {author} {\bibinfo {author} {\bibfnamefont {Christof}\ \bibnamefont {Hättig}}, \bibinfo {author} {\bibfnamefont {Wim}\ \bibnamefont {Klopper}}, \bibinfo {author} {\bibfnamefont {Andreas}\ \bibnamefont {K{\"o}hn}}, \ and\ \bibinfo {author} {\bibfnamefont {David~P.}\ \bibnamefont {Tew}},\ }\bibfield  {title} {\enquote {\bibinfo {title} {Explicitly correlated electrons in molecules},}\ }\href {\doibase 10.1021/cr200168z} {\bibfield  {journal} {\bibinfo  {journal} {Chemical Reviews}\ }\textbf {\bibinfo {volume} {112}},\ \bibinfo {pages} {4--74} (\bibinfo {year} {2012})},\ \bibinfo {note} {pMID: 22206503}\BibitemShut {NoStop}%
\bibitem [{\citenamefont {Tennyson}\ \emph {et~al.}(2016)\citenamefont {Tennyson}, \citenamefont {Yurchenko}, \citenamefont {Al-Refaie}, \citenamefont {Barton}, \citenamefont {Chubb}, \citenamefont {Coles}, \citenamefont {Diamantopoulou}, \citenamefont {Gorman}, \citenamefont {Hill}, \citenamefont {Lam}, \citenamefont {Lodi}, \citenamefont {McKemmish}, \citenamefont {Na}, \citenamefont {Owens}, \citenamefont {Polyansky}, \citenamefont {Rivlin}, \citenamefont {Sousa-Silva}, \citenamefont {Underwood}, \citenamefont {Yachmenev},\ and\ \citenamefont {Zak}}]{Tennyson2016}%
  \BibitemOpen
  \bibfield  {author} {\bibinfo {author} {\bibfnamefont {Jonathan}\ \bibnamefont {Tennyson}}, \bibinfo {author} {\bibfnamefont {Sergei~N.}\ \bibnamefont {Yurchenko}}, \bibinfo {author} {\bibfnamefont {Ahmed~F.}\ \bibnamefont {Al-Refaie}}, \bibinfo {author} {\bibfnamefont {Emma~J.}\ \bibnamefont {Barton}}, \bibinfo {author} {\bibfnamefont {Katy~L.}\ \bibnamefont {Chubb}}, \bibinfo {author} {\bibfnamefont {Phillip~A.}\ \bibnamefont {Coles}}, \bibinfo {author} {\bibfnamefont {S.}~\bibnamefont {Diamantopoulou}}, \bibinfo {author} {\bibfnamefont {Maire~N.}\ \bibnamefont {Gorman}}, \bibinfo {author} {\bibfnamefont {Christian}\ \bibnamefont {Hill}}, \bibinfo {author} {\bibfnamefont {Aden~Z.}\ \bibnamefont {Lam}}, \bibinfo {author} {\bibfnamefont {Lorenzo}\ \bibnamefont {Lodi}}, \bibinfo {author} {\bibfnamefont {Laura~K.}\ \bibnamefont {McKemmish}}, \bibinfo {author} {\bibfnamefont {Yueqi}\ \bibnamefont {Na}}, \bibinfo {author} {\bibfnamefont {Alec}\ \bibnamefont {Owens}}, \bibinfo {author} {\bibfnamefont {Oleg~L.}\
  \bibnamefont {Polyansky}}, \bibinfo {author} {\bibfnamefont {Tom}\ \bibnamefont {Rivlin}}, \bibinfo {author} {\bibfnamefont {Clara}\ \bibnamefont {Sousa-Silva}}, \bibinfo {author} {\bibfnamefont {Daniel~S.}\ \bibnamefont {Underwood}}, \bibinfo {author} {\bibfnamefont {Andrey}\ \bibnamefont {Yachmenev}}, \ and\ \bibinfo {author} {\bibfnamefont {Emil}\ \bibnamefont {Zak}},\ }\bibfield  {title} {\enquote {\bibinfo {title} {The exomol database: Molecular line lists for exoplanet and other hot atmospheres},}\ }\href {\doibase 10.1016/j.jms.2016.05.002} {\bibfield  {journal} {\bibinfo  {journal} {Journal of Molecular Spectroscopy}\ }\textbf {\bibinfo {volume} {327}},\ \bibinfo {pages} {73–94} (\bibinfo {year} {2016})}\BibitemShut {NoStop}%
\bibitem [{\citenamefont {Huang}\ and\ \citenamefont {Lee}(2008)}]{Huang_JCP_2008_044312}%
  \BibitemOpen
  \bibfield  {author} {\bibinfo {author} {\bibfnamefont {Xinchuan}\ \bibnamefont {Huang}}\ and\ \bibinfo {author} {\bibfnamefont {Timothy~J.}\ \bibnamefont {Lee}},\ }\bibfield  {title} {\enquote {\bibinfo {title} {A procedure for computing accurate ab initio quartic force fields: Application to {HO$_2^+$} and {H$_2$O}},}\ }\href {\doibase 10.1063/1.2957488} {\bibfield  {journal} {\bibinfo  {journal} {J. Chem. Phys.}\ }\textbf {\bibinfo {volume} {129}},\ \bibinfo {pages} {044312} (\bibinfo {year} {2008})}\BibitemShut {NoStop}%
\bibitem [{\citenamefont {Yang}\ \emph {et~al.}(2024)\citenamefont {Yang}, \citenamefont {Christianen}, \citenamefont {Ba\~nuls}, \citenamefont {Wild},\ and\ \citenamefont {Cirac}}]{Yang2024}%
  \BibitemOpen
  \bibfield  {author} {\bibinfo {author} {\bibfnamefont {Yilun}\ \bibnamefont {Yang}}, \bibinfo {author} {\bibfnamefont {Arthur}\ \bibnamefont {Christianen}}, \bibinfo {author} {\bibfnamefont {Mari~Carmen}\ \bibnamefont {Ba\~nuls}}, \bibinfo {author} {\bibfnamefont {Dominik~S.}\ \bibnamefont {Wild}}, \ and\ \bibinfo {author} {\bibfnamefont {J.~Ignacio}\ \bibnamefont {Cirac}},\ }\bibfield  {title} {\enquote {\bibinfo {title} {Phase-sensitive quantum measurement without controlled operations},}\ }\href {\doibase 10.1103/PhysRevLett.132.220601} {\bibfield  {journal} {\bibinfo  {journal} {Phys. Rev. Lett.}\ }\textbf {\bibinfo {volume} {132}},\ \bibinfo {pages} {220601} (\bibinfo {year} {2024})}\BibitemShut {NoStop}%
\bibitem [{\citenamefont {Lu}\ \emph {et~al.}(2021)\citenamefont {Lu}, \citenamefont {Ba\~nuls},\ and\ \citenamefont {Cirac}}]{Lu2021}%
  \BibitemOpen
  \bibfield  {author} {\bibinfo {author} {\bibfnamefont {Sirui}\ \bibnamefont {Lu}}, \bibinfo {author} {\bibfnamefont {Mari~Carmen}\ \bibnamefont {Ba\~nuls}}, \ and\ \bibinfo {author} {\bibfnamefont {J.~Ignacio}\ \bibnamefont {Cirac}},\ }\bibfield  {title} {\enquote {\bibinfo {title} {Algorithms for quantum simulation at finite energies},}\ }\href {\doibase 10.1103/PRXQuantum.2.020321} {\bibfield  {journal} {\bibinfo  {journal} {PRX Quantum}\ }\textbf {\bibinfo {volume} {2}},\ \bibinfo {pages} {020321} (\bibinfo {year} {2021})}\BibitemShut {NoStop}%
\bibitem [{\citenamefont {O'Brien}\ \emph {et~al.}(2021)\citenamefont {O'Brien}, \citenamefont {Polla}, \citenamefont {Rubin}, \citenamefont {Huggins}, \citenamefont {McArdle}, \citenamefont {Boixo}, \citenamefont {McClean},\ and\ \citenamefont {Babbush}}]{Brien2021}%
  \BibitemOpen
  \bibfield  {author} {\bibinfo {author} {\bibfnamefont {Thomas~E.}\ \bibnamefont {O'Brien}}, \bibinfo {author} {\bibfnamefont {Stefano}\ \bibnamefont {Polla}}, \bibinfo {author} {\bibfnamefont {Nicholas~C.}\ \bibnamefont {Rubin}}, \bibinfo {author} {\bibfnamefont {William~J.}\ \bibnamefont {Huggins}}, \bibinfo {author} {\bibfnamefont {Sam}\ \bibnamefont {McArdle}}, \bibinfo {author} {\bibfnamefont {Sergio}\ \bibnamefont {Boixo}}, \bibinfo {author} {\bibfnamefont {Jarrod~R.}\ \bibnamefont {McClean}}, \ and\ \bibinfo {author} {\bibfnamefont {Ryan}\ \bibnamefont {Babbush}},\ }\bibfield  {title} {\enquote {\bibinfo {title} {Error mitigation via verified phase estimation},}\ }\href {\doibase 10.1103/PRXQuantum.2.020317} {\bibfield  {journal} {\bibinfo  {journal} {PRX Quantum}\ }\textbf {\bibinfo {volume} {2}},\ \bibinfo {pages} {020317} (\bibinfo {year} {2021})}\BibitemShut {NoStop}%
\end{thebibliography}%

\clearpage
\newpage
\onecolumngrid
\appendix

\section{Supplementary Circuits for Time Evolution Operators}
\label{App evolve}

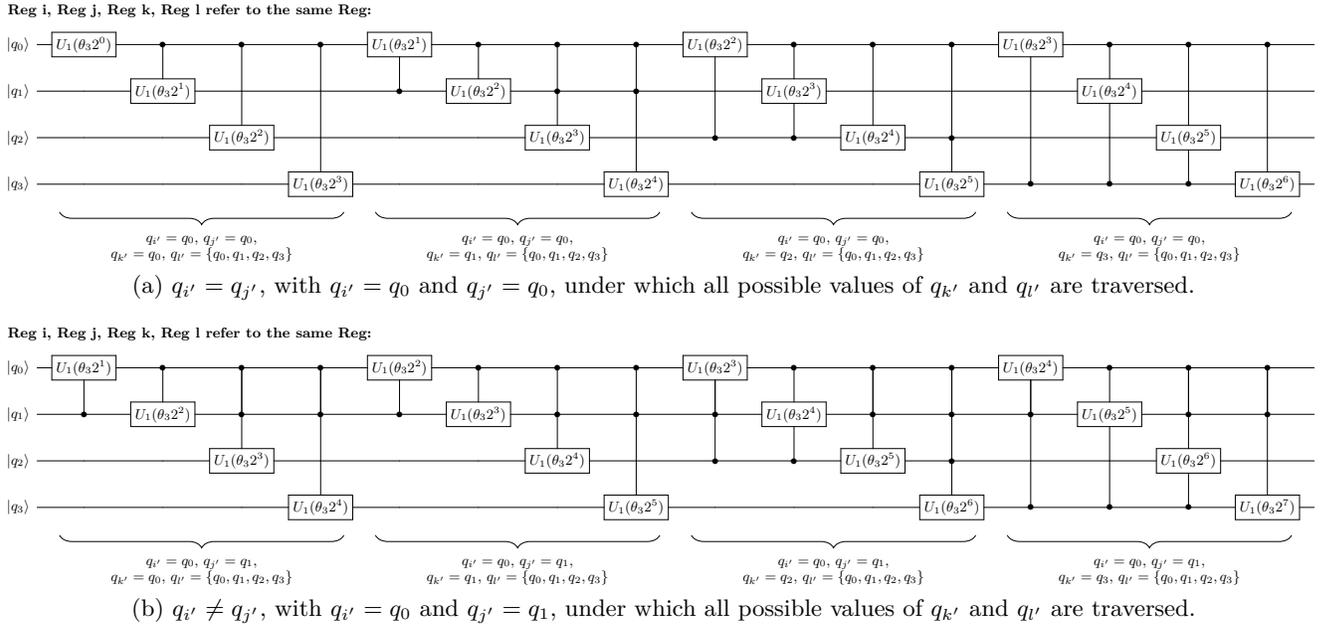
\begin{figure}[htbp]
\flushright
\subfloat[$q_{i^\prime} = q_{j^\prime}$, with $q_{i^\prime} = q_0$ and $q_{j^\prime} = q_0$, under which all possible values of $q_{k^\prime}$ and $q_{l^\prime}$ are traversed. \label{fourthexam1cir1}]{
\vspace{2.5em} 
\scalebox{0.6}{
\Qcircuit @C=1em @R=1.5em {
\makebox[4em][l]{\textbf{Reg i, Reg j, Reg k, Reg l refer to the same Reg:}}\\
\lstick{\ket{q_0}}&\gate{U_1(\theta_32^0)} & \ctrl{1} & \ctrl{2} & \ctrl{3} 
&\gate{U_1(\theta_32^1)} &\ctrl{1} &\ctrl{1} &\ctrl{1} 
&\gate{U_1(\theta_32^2)} & \ctrl{1} & \ctrl{2} & \ctrl{2} 
&\gate{U_1(\theta_32^3)} & \ctrl{1} & \ctrl{2} & \ctrl{3}&\qw \\
\lstick{\ket{q_1}}&\qw & \gate{U_1(\theta_32^1)}&\qw &\qw 
&\ctrl{-1}& \gate{U_1(\theta_32^2)}&\ctrl{1} & \ctrl{2} 
&\qw & \gate{U_1(\theta_32^3)}&\qw &\qw
&\qw & \gate{U_1(\theta_32^4)}&\qw &\qw&\qw\\
\lstick{\ket{q_2}}&\qw & \qw & \gate{U_1(\theta_32^2)} &\qw 
& \qw &\qw &\gate{U_1(\theta_32^3)} &\qw
&\ctrl{-2} & \ctrl{-1}&\gate{U_1(\theta_32^4)}&\ctrl{1}
& \qw &\qw &\gate{U_1(\theta_32^5)} &\qw&\qw\\
\lstick{\ket{q_3}}&\qw & \qw &\qw & \gate{U_1(\theta_32^3)} 
& \qw &\qw&\qw &\gate{U_1(\theta_32^4)}
&\qw & \qw &\qw &\gate{U_1(\theta_32^5)}
&\ctrl{-3} & \ctrl{-2} &\ctrl{-1} &\gate{U_1(\theta_32^6)}&\qw\\
}
\begin{tikzpicture}[overlay, remember picture]
    \draw[decorate, decoration={brace, amplitude=8pt, mirror}, yshift=-2ex] 
        (-27.8,-4.2) -- (-21.5,-4.2) node[midway, yshift=-2.5em, align=center]{$q_{i^\prime} = q_0$, $q_{j^\prime} = q_0$, \\ $q_{k^\prime} = q_0$, $q_{l^\prime} = \{q_0,q_1,q_2,q_3\}$};
    \draw[decorate, decoration={brace, amplitude=8pt, mirror}, yshift=-2ex] 
        (-20.8,-4.2) -- (-14.5,-4.2) node[midway, yshift=-2.5em, align=center]{$q_{i^\prime} = q_0$, $q_{j^\prime} = q_0$, \\ $q_{k^\prime} = q_1$, $q_{l^\prime} = \{q_0,q_1,q_2,q_3\}$};
    \draw[decorate, decoration={brace, amplitude=8pt, mirror}, yshift=-2ex] 
        (-13.8,-4.2) -- (-7.5,-4.2) node[midway, yshift=-2.5em, align=center]{$q_{i^\prime} = q_0$, $q_{j^\prime} = q_0$, \\ $q_{k^\prime} = q_2$, $q_{l^\prime} = \{q_0,q_1,q_2,q_3\}$};
    \draw[decorate, decoration={brace, amplitude=8pt, mirror}, yshift=-2ex] 
        (-6.8,-4.2) -- (-0.5,-4.2) node[midway, yshift=-2.5em, align=center]{$q_{i^\prime} = q_0$, $q_{j^\prime} = q_0$, \\ $q_{k^\prime} = q_3$, $q_{l^\prime} = \{q_0,q_1,q_2,q_3\}$};
\end{tikzpicture}}}
\\
\vspace{1em}
\subfloat[$q_{i^\prime} \neq q_{j^\prime}$, with $q_{i^\prime} = q_0$ and $q_{j^\prime} = q_1$, under which all possible values of $q_{k^\prime}$ and $q_{l^\prime}$ are traversed. \label{fourthexam1cir2}]{
\vspace{2.5em} 
\scalebox{0.6}{
\Qcircuit @C=1em @R=1.5em {
\makebox[4em][l]{\textbf{Reg i, Reg j, Reg k, Reg l refer to the same Reg:}}\\
\lstick{\ket{q_0}}&\gate{U_1(\theta_32^1)} & \ctrl{1} & \ctrl{2} & \ctrl{3} 
&\gate{U_1(\theta_32^2)} &\ctrl{1} &\ctrl{1} &\ctrl{1} 
&\gate{U_1(\theta_32^3)} & \ctrl{1} & \ctrl{2} & \ctrl{2} 
&\gate{U_1(\theta_32^4)} & \ctrl{1} & \ctrl{2} & \ctrl{3}&\qw \\
\lstick{\ket{q_1}} & \ctrl{-1} & \gate{U_1(\theta_32^2)}&\ctrl{-1} &\ctrl{-1}
&\ctrl{-1}& \gate{U_1(\theta_32^3)}&\ctrl{1} & \ctrl{2} 
&\ctrl{-1} & \gate{U_1(\theta_32^4)}&\ctrl{-1} &\ctrl{-1}
&\ctrl{-1} & \gate{U_1(\theta_32^5)}&\ctrl{-1} &\ctrl{-1}&\qw\\
\lstick{\ket{q_2}}&\qw & \qw & \gate{U_1(\theta_32^3)} &\qw 
& \qw &\qw &\gate{U_1(\theta_32^4)} &\qw
&\ctrl{-2} & \ctrl{-1}&\gate{U_1(\theta_32^5)}&\ctrl{1}
& \qw &\qw &\gate{U_1(\theta_32^6)} &\qw&\qw\\
\lstick{\ket{q_3}}&\qw & \qw &\qw & \gate{U_1(\theta_32^4)} 
& \qw &\qw&\qw &\gate{U_1(\theta_32^5)}
&\qw & \qw &\qw &\gate{U_1(\theta_32^6)}
&\ctrl{-3} & \ctrl{-2} &\ctrl{-1} &\gate{U_1(\theta_32^7)}&\qw\\
}
\begin{tikzpicture}[overlay, remember picture]
    \draw[decorate, decoration={brace, amplitude=8pt, mirror}, yshift=-2ex] 
        (-27.8,-4.2) -- (-21.5,-4.2) node[midway, yshift=-2.5em, align=center]{$q_{i^\prime} = q_0$, $q_{j^\prime} = q_1$, \\ $q_{k^\prime} = q_0$, $q_{l^\prime} = \{q_0,q_1,q_2,q_3\}$};
    \draw[decorate, decoration={brace, amplitude=8pt, mirror}, yshift=-2ex] 
        (-20.8,-4.2) -- (-14.5,-4.2) node[midway, yshift=-2.5em, align=center]{$q_{i^\prime} = q_0$, $q_{j^\prime} = q_1$, \\ $q_{k^\prime} = q_1$, $q_{l^\prime} = \{q_0,q_1,q_2,q_3\}$};
    \draw[decorate, decoration={brace, amplitude=8pt, mirror}, yshift=-2ex] 
        (-13.8,-4.2) -- (-7.5,-4.2) node[midway, yshift=-2.5em, align=center]{$q_{i^\prime} = q_0$, $q_{j^\prime} = q_1$, \\ $q_{k^\prime} = q_2$, $q_{l^\prime} = \{q_0,q_1,q_2,q_3\}$};
    \draw[decorate, decoration={brace, amplitude=8pt, mirror}, yshift=-2ex] 
        (-6.8,-4.2) -- (-0.5,-4.2) node[midway, yshift=-2.5em, align=center]{$q_{i^\prime} = q_0$, $q_{j^\prime} = q_1$, \\ $q_{k^\prime} = q_3$, $q_{l^\prime} = \{q_0,q_1,q_2,q_3\}$};
\end{tikzpicture}}}
\caption{\justifying Example fragment circuit of $U_{ijkl}$ when $i=j=k=l$ (i.e. $x_i$, $x_j$, $x_k$ and $x_l$ indicate the same 4-qubit register). }
\label{fig:4th examples 1}
\end{figure}

\begin{figure}[htbp]
\flushright
\subfloat[$q_{i^\prime} = q_{j^\prime}$, with $q_{i^\prime} = q_0$ and $q_{j^\prime} = q_0$, under which all possible values of $q_{k^\prime}$ and $q_{l^\prime}$ are traversed. \label{fourthexam3cir1}]{
\vspace{2.5em} 
\scalebox{0.6}{
\Qcircuit @C=1em @R=1.5em {
\makebox[4em][l]{\textbf{Reg i \& j:}}\\
\lstick{\ket{q_0}}&\ctrl{5}&\ctrl{6}&\ctrl{7}&\ctrl{8}   
&\ctrl{5}&\ctrl{6}&\ctrl{7}&\ctrl{8}
&\ctrl{5}&\ctrl{6}&\ctrl{7}&\ctrl{8}
&\ctrl{5}&\ctrl{6}&\ctrl{7}&\ctrl{8}&\qw\\
\lstick{\ket{q_1}}&\qw&\qw&\qw&\qw   
&\qw&\qw&\qw&\qw
&\qw&\qw&\qw&\qw 
&\qw&\qw&\qw&\qw &\qw\\
\lstick{\ket{q_2}}&\qw&\qw&\qw&\qw   
&\qw&\qw&\qw&\qw 
&\qw&\qw&\qw&\qw 
&\qw&\qw&\qw&\qw &\qw \\
\lstick{\ket{q_3}}&\qw&\qw&\qw&\qw   
&\qw&\qw&\qw&\qw 
&\qw&\qw&\qw&\qw 
&\qw&\qw&\qw&\qw &\qw \\
\makebox[4em][l]{\textbf{Reg k \& l:}}\\
\lstick{\ket{q_0}}&\gate{U_1(\theta_32^0)} &\ctrl{1} & \ctrl{2} &\ctrl{3}   
&\gate{U_1(\theta_32^1)} &\qw & \qw &\qw
&\gate{U_1(\theta_32^2)} &\qw & \qw &\qw
&\gate{U_1(\theta_32^3)} &\qw & \qw &\qw&  \qw \\
\lstick{\ket{q_1}}&\qw &\gate{U_1(\theta_32^1)} & \qw &\qw   
&\ctrl{-1} &\gate{U_1(\theta_32^2)} & \ctrl{1} &\ctrl{2}   
&\qw &\gate{U_1(\theta_32^3)} & \qw &\qw   
&\qw &\gate{U_1(\theta_32^4)} & \qw &\qw   &\qw \\
\lstick{\ket{q_2}}&\qw & \qw &\gate{U_1(\theta_32^2)} &\qw   
&\qw & \qw &\gate{U_1(\theta_32^3)} &\qw   
&\ctrl{-2} & \ctrl{-1} &\gate{U_1(\theta_32^4)} &\ctrl{1}   
&\qw & \qw &\gate{U_1(\theta_32^5)} &\qw   &\qw \\
\lstick{\ket{q_3}}&\qw & \qw &\qw &\gate{U_1(\theta_32^3)}   
&\qw & \qw &\qw &\gate{U_1(\theta_32^4)}   
&\qw & \qw &\qw &\gate{U_1(\theta_32^5)}  
&\ctrl{-3} & \ctrl{-2} &\ctrl{-1} &\gate{U_1(\theta_32^6)}  &\qw \\
}
\begin{tikzpicture}[overlay, remember picture]
    \draw[decorate, decoration={brace, amplitude=8pt, mirror}, yshift=-2ex] 
        (-27.8,-7) -- (-21.5,-7) node[midway, yshift=-2em, align=center]{$q_{i^\prime} = q_0$, $q_{j^\prime} = q_0$, \\ $q_{k^\prime} = q_0$, $q_{l^\prime} = \{q_0,q_1,q_2,q_3\}$};
    \draw[decorate, decoration={brace, amplitude=8pt, mirror}, yshift=-2ex] 
        (-20.8,-7) -- (-14.5,-7) node[midway, yshift=-2em, align=center]{$q_{i^\prime} = q_0$, $q_{j^\prime} = q_0$, \\ $q_{k^\prime} = q_1$, $q_{l^\prime} = \{q_0,q_1,q_2,q_3\}$};
    \draw[decorate, decoration={brace, amplitude=8pt, mirror}, yshift=-2ex] 
        (-13.8,-7) -- (-7.5,-7) node[midway, yshift=-2em, align=center]{$q_{i^\prime} = q_0$, $q_{j^\prime} = q_0$, \\ $q_{k^\prime} = q_2$, $q_{l^\prime} = \{q_0,q_1,q_2,q_3\}$};
    \draw[decorate, decoration={brace, amplitude=8pt, mirror}, yshift=-2ex] 
        (-6.8,-7) -- (-0.5,-7) node[midway, yshift=-2em, align=center]{$q_{i^\prime} = q_0$, $q_{j^\prime} = q_0$, \\ $q_{k^\prime} = q_3$, $q_{l^\prime} = \{q_0,q_1,q_2,q_3\}$};
\end{tikzpicture}}
}
\\
\vspace{1em}
\subfloat[$q_{i^\prime} \neq q_{j^\prime}$, with $q_{i^\prime} = q_0$ and $q_{j^\prime} = q_1$, under which all possible values of $q_{k^\prime}$ and $q_{l^\prime}$ are traversed. \label{fourthexam3cir2}]{
\vspace{2.5em} 
\scalebox{0.6}{
\Qcircuit @C=1em @R=1.5em {
\makebox[4em][l]{\textbf{Reg i \& j:}}\\
\lstick{\ket{q_0}}&\ctrl{5}&\ctrl{6}&\ctrl{7}&\ctrl{8}   
&\ctrl{5}&\ctrl{6}&\ctrl{7}&\ctrl{8}
&\ctrl{5}&\ctrl{6}&\ctrl{7}&\ctrl{8}
&\ctrl{5}&\ctrl{6}&\ctrl{7}&\ctrl{8}&\qw\\
\lstick{\ket{q_1}}&\ctrl{-1}&\ctrl{-1}&\ctrl{-1}&\ctrl{-1}   
&\ctrl{4}&\ctrl{5}&\ctrl{6}&\ctrl{7}
&\ctrl{-1}&\ctrl{-1}&\ctrl{-1}&\ctrl{-1} 
&\ctrl{-1}&\ctrl{-1}&\ctrl{-1}&\ctrl{-1} &\qw\\
\lstick{\ket{q_2}}&\qw&\qw&\qw&\qw   
&\qw&\qw&\qw&\qw 
&\qw&\qw&\qw&\qw 
&\qw&\qw&\qw&\qw &\qw \\
\lstick{\ket{q_3}}&\qw&\qw&\qw&\qw   
&\qw&\qw&\qw&\qw 
&\qw&\qw&\qw&\qw 
&\qw&\qw&\qw&\qw &\qw \\
\makebox[4em][l]{\textbf{Reg k \& l:}}\\
\lstick{\ket{q_0}}&\gate{U_1(\theta_32^1)} &\ctrl{1} & \ctrl{2} &\ctrl{3}   
&\gate{U_1(\theta_32^2)} &\qw & \qw &\qw
&\gate{U_1(\theta_32^3)} &\qw & \qw &\qw
&\gate{U_1(\theta_32^4)} &\qw & \qw &\qw&  \qw \\
\lstick{\ket{q_1}}&\qw &\gate{U_1(\theta_32^2)} & \qw &\qw   
&\ctrl{-1} &\gate{U_1(\theta_32^3)} & \ctrl{1} &\ctrl{2}   
&\qw &\gate{U_1(\theta_32^4)} & \qw &\qw   
&\qw &\gate{U_1(\theta_32^5)} & \qw &\qw   &\qw \\
\lstick{\ket{q_2}}&\qw & \qw &\gate{U_1(\theta_32^3)} &\qw   
&\qw & \qw &\gate{U_1(\theta_32^4)} &\qw   
&\ctrl{-2} & \ctrl{-1} &\gate{U_1(\theta_32^5)} &\ctrl{1}   
&\qw & \qw &\gate{U_1(\theta_32^6)} &\qw   &\qw \\
\lstick{\ket{q_3}}&\qw & \qw &\qw &\gate{U_1(\theta_32^4)}   
&\qw & \qw &\qw &\gate{U_1(\theta_32^5)}   
&\qw & \qw &\qw &\gate{U_1(\theta_32^6)}  
&\ctrl{-3} & \ctrl{-2} &\ctrl{-1} &\gate{U_1(\theta_32^7)}  &\qw \\
}
\begin{tikzpicture}[overlay, remember picture]
    \draw[decorate, decoration={brace, amplitude=8pt, mirror}, yshift=-2ex] 
        (-27.8,-7) -- (-21.5,-7) node[midway, yshift=-2em, align=center]{$q_{i^\prime} = q_0$, $q_{j^\prime} = q_1$, \\ $q_{k^\prime} = q_0$, $q_{l^\prime} = \{q_0,q_1,q_2,q_3\}$};
    \draw[decorate, decoration={brace, amplitude=8pt, mirror}, yshift=-2ex] 
        (-20.8,-7) -- (-14.5,-7) node[midway, yshift=-2em, align=center]{$q_{i^\prime} = q_0$, $q_{j^\prime} = q_1$, \\ $q_{k^\prime} = q_1$, $q_{l^\prime} = \{q_0,q_1,q_2,q_3\}$};
    \draw[decorate, decoration={brace, amplitude=8pt, mirror}, yshift=-2ex] 
        (-13.8,-7) -- (-7.5,-7) node[midway, yshift=-2em, align=center]{$q_{i^\prime} = q_0$, $q_{j^\prime} = q_1$, \\ $q_{k^\prime} = q_2$, $q_{l^\prime} = \{q_0,q_1,q_2,q_3\}$};
    \draw[decorate, decoration={brace, amplitude=8pt, mirror}, yshift=-2ex] 
        (-6.8,-7) -- (-0.5,-7) node[midway, yshift=-2em, align=center]{$q_{i^\prime} = q_0$, $q_{j^\prime} = q_1$, \\ $q_{k^\prime} = q_3$, $q_{l^\prime} = \{q_0,q_1,q_2,q_3\}$};
\end{tikzpicture}}
}
\caption{\justifying Example fragment circuit of $U_{ijkl}$ when $i=j\neq k=l$ ($x_i$ and $x_j$ indicate the same register, $x_k$ and $x_l$ indicate the same register) with 4 qubits in each register.}
\label{fig:4th examples 3}
\end{figure}
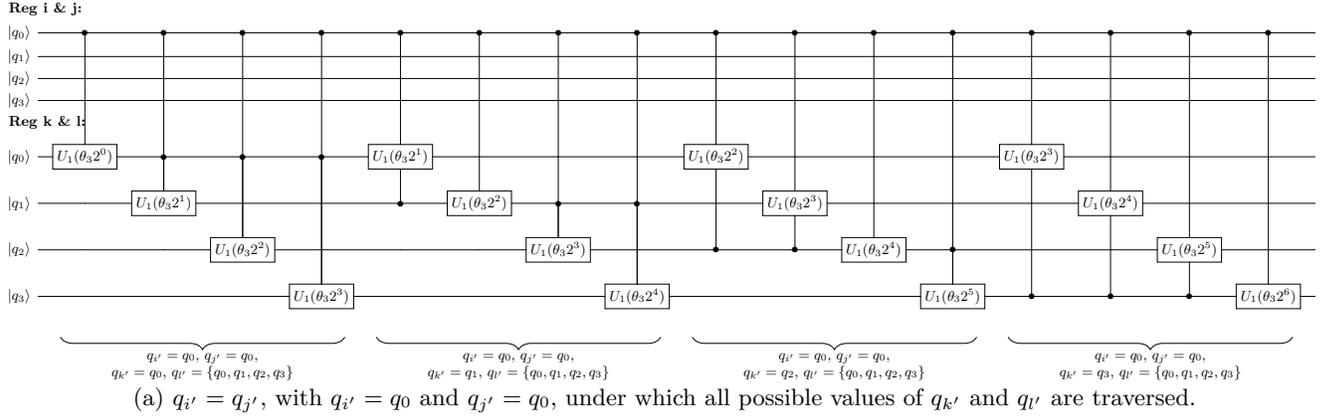
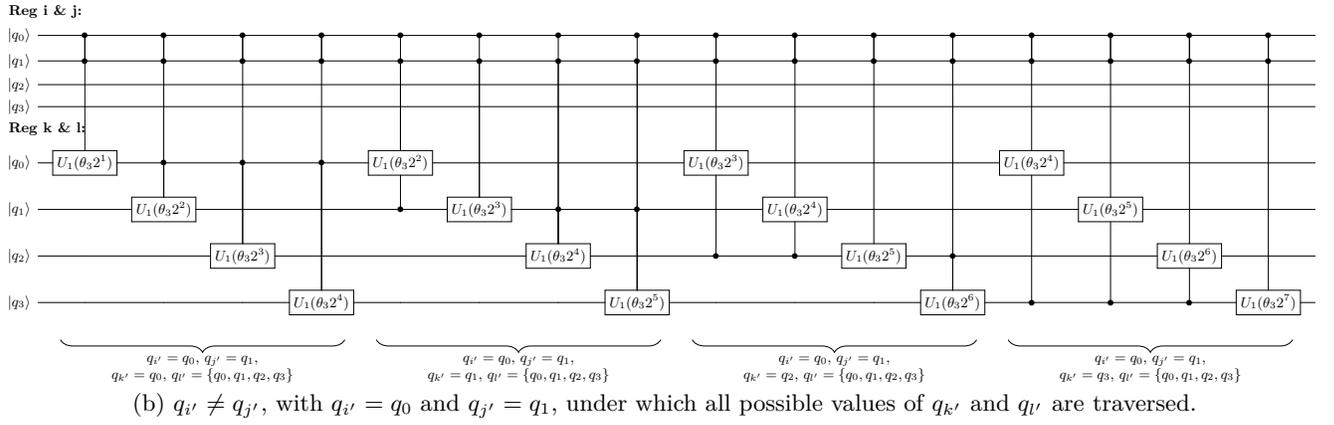

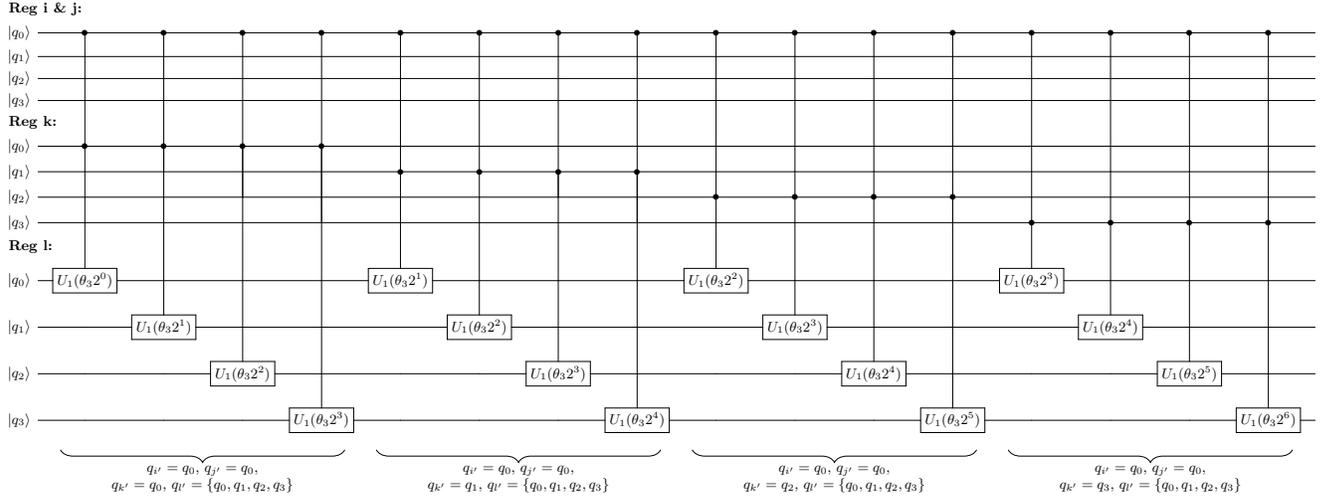
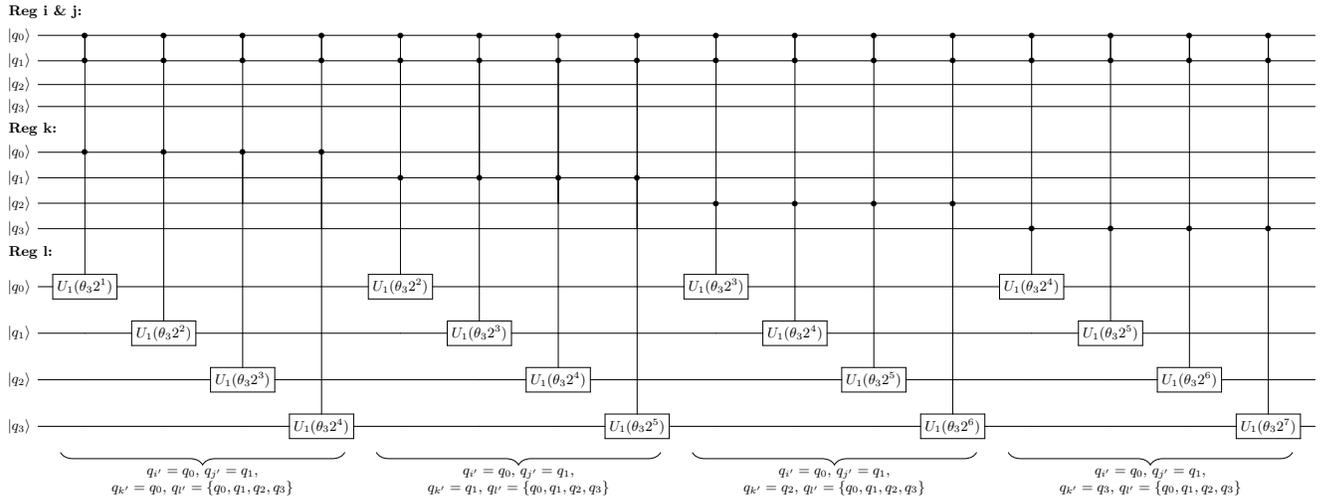
\begin{figure}[htbp]
\flushright
\subfloat[$q_{i^\prime} = q_{j^\prime}$, with $q_{i^\prime} = q_0$ and $q_{j^\prime} = q_0$, under which all possible values of $q_{k^\prime}$ and $q_{l^\prime}$ are traversed. \label{fourthexam4cir1}]{
\vspace{2.5em} 
\scalebox{0.6}{
\Qcircuit @C=1em @R=1.5em {
\makebox[4em][l]{\textbf{Reg i \& j:}}\\
\lstick{\ket{q_0}}&\ctrl{5}&\ctrl{6}&\ctrl{7}&\ctrl{8}   
&\ctrl{6}&\ctrl{6}&\ctrl{7}&\ctrl{8}
&\ctrl{7}&\ctrl{7}&\ctrl{7}&\ctrl{8}
&\ctrl{8}&\ctrl{8}&\ctrl{8}&\ctrl{8}&\qw\\
\lstick{\ket{q_1}}&\qw&\qw&\qw&\qw   
&\qw&\qw&\qw&\qw
&\qw&\qw&\qw&\qw 
&\qw&\qw&\qw&\qw &\qw\\
\lstick{\ket{q_2}}&\qw&\qw&\qw&\qw   
&\qw&\qw&\qw&\qw 
&\qw&\qw&\qw&\qw 
&\qw&\qw&\qw&\qw &\qw \\
\lstick{\ket{q_3}}&\qw&\qw&\qw&\qw   
&\qw&\qw&\qw&\qw 
&\qw&\qw&\qw&\qw 
&\qw&\qw&\qw&\qw &\qw \\
\makebox[4em][l]{\textbf{Reg k:}}\\
\lstick{\ket{q_0}}&\ctrl{5}&\ctrl{6}&\ctrl{7}&\ctrl{8}   
&\qw&\qw&\qw&\qw
&\qw&\qw&\qw&\qw 
&\qw&\qw&\qw&\qw &\qw\\
\lstick{\ket{q_1}}&\qw&\qw&\qw&\qw   
&\ctrl{4}&\ctrl{5}&\ctrl{6}&\ctrl{7}
&\qw&\qw&\qw&\qw 
&\qw&\qw&\qw&\qw &\qw\\
\lstick{\ket{q_2}}&\qw&\qw&\qw&\qw   
&\qw&\qw&\qw&\qw 
&\ctrl{3}&\ctrl{4}&\ctrl{5}&\ctrl{6}
&\qw&\qw&\qw&\qw &\qw \\
\lstick{\ket{q_3}}&\qw&\qw&\qw&\qw   
&\qw&\qw&\qw&\qw 
&\qw&\qw&\qw&\qw 
&\ctrl{2}&\ctrl{3}&\ctrl{4}&\ctrl{5}&\qw \\
\makebox[4em][l]{\textbf{Reg l:}}\\
\lstick{\ket{q_0}}&\gate{U_1(\theta_32^0)} &\qw & \qw &\qw
&\gate{U_1(\theta_32^1)} &\qw & \qw &\qw
&\gate{U_1(\theta_32^2)} &\qw & \qw &\qw
&\gate{U_1(\theta_32^3)} &\qw & \qw &\qw&  \qw \\
\lstick{\ket{q_1}}&\qw &\gate{U_1(\theta_32^1)} & \qw &\qw   
&\qw &\gate{U_1(\theta_32^2)} & \qw &\qw   
&\qw &\gate{U_1(\theta_32^3)} & \qw &\qw   
&\qw &\gate{U_1(\theta_32^4)} & \qw &\qw   &\qw \\
\lstick{\ket{q_2}}&\qw & \qw &\gate{U_1(\theta_32^2)} &\qw   
&\qw & \qw &\gate{U_1(\theta_32^3)} &\qw   
&\qw & \qw &\gate{U_1(\theta_32^4)} &\qw   
&\qw & \qw &\gate{U_1(\theta_32^5)} &\qw   &\qw \\
\lstick{\ket{q_3}}&\qw & \qw &\qw &\gate{U_1(\theta_32^3)}   
&\qw & \qw &\qw &\gate{U_1(\theta_32^4)}   
&\qw & \qw &\qw &\gate{U_1(\theta_32^5)}  
&\qw & \qw &\qw &\gate{U_1(\theta_32^6)}  &\qw \\
}
\begin{tikzpicture}[overlay, remember picture]
    \draw[decorate, decoration={brace, amplitude=8pt, mirror}, yshift=-2ex] 
        (-27.8,-9.5) -- (-21.5,-9.5) node[midway, yshift=-2em, align=center]{$q_{i^\prime} = q_0$, $q_{j^\prime} = q_0$, \\ $q_{k^\prime} = q_0$, $q_{l^\prime} = \{q_0,q_1,q_2,q_3\}$};
    \draw[decorate, decoration={brace, amplitude=8pt, mirror}, yshift=-2ex] 
        (-20.8,-9.5) -- (-14.5,-9.5) node[midway, yshift=-2em, align=center]{$q_{i^\prime} = q_0$, $q_{j^\prime} = q_0$, \\ $q_{k^\prime} = q_1$, $q_{l^\prime} = \{q_0,q_1,q_2,q_3\}$};
    \draw[decorate, decoration={brace, amplitude=8pt, mirror}, yshift=-2ex] 
        (-13.8,-9.5) -- (-7.5,-9.5) node[midway, yshift=-2em, align=center]{$q_{i^\prime} = q_0$, $q_{j^\prime} = q_0$, \\ $q_{k^\prime} = q_2$, $q_{l^\prime} = \{q_0,q_1,q_2,q_3\}$};
    \draw[decorate, decoration={brace, amplitude=8pt, mirror}, yshift=-2ex] 
        (-6.8,-9.5) -- (-0.5,-9.5) node[midway, yshift=-2em, align=center]{$q_{i^\prime} = q_0$, $q_{j^\prime} = q_0$, \\ $q_{k^\prime} = q_3$, $q_{l^\prime} = \{q_0,q_1,q_2,q_3\}$};
\end{tikzpicture}}
}
\\
\vspace{1em}
\subfloat[$q_{i^\prime} \neq q_{j^\prime}$, with $q_{i^\prime} = q_0$ and $q_{j^\prime} = q_1$, under which all possible values of $q_{k^\prime}$ and $q_{l^\prime}$ are traversed. \label{fourthexam4cir2}]{
\vspace{2.5em} 
\scalebox{0.6}{
\Qcircuit @C=1em @R=1.5em {
\makebox[4em][l]{\textbf{Reg i \& j:}}\\
\lstick{\ket{q_0}}&\ctrl{5}&\ctrl{6}&\ctrl{7}&\ctrl{8}   
&\ctrl{6}&\ctrl{6}&\ctrl{7}&\ctrl{8}
&\ctrl{7}&\ctrl{7}&\ctrl{7}&\ctrl{8}
&\ctrl{8}&\ctrl{8}&\ctrl{8}&\ctrl{8}&\qw\\
\lstick{\ket{q_1}}&\ctrl{-1}&\ctrl{-1}&\ctrl{-1}&\ctrl{-1}   
&\ctrl{4}&\ctrl{5}&\ctrl{6}&\ctrl{7}
&\ctrl{-1}&\ctrl{-1}&\ctrl{-1}&\ctrl{-1} 
&\ctrl{-1}&\ctrl{-1}&\ctrl{-1}&\ctrl{-1} &\qw\\
\lstick{\ket{q_2}}&\qw&\qw&\qw&\qw   
&\qw&\qw&\qw&\qw 
&\qw&\qw&\qw&\qw 
&\qw&\qw&\qw&\qw &\qw \\
\lstick{\ket{q_3}}&\qw&\qw&\qw&\qw   
&\qw&\qw&\qw&\qw 
&\qw&\qw&\qw&\qw 
&\qw&\qw&\qw&\qw &\qw \\
\makebox[4em][l]{\textbf{Reg k:}}\\
\lstick{\ket{q_0}}&\ctrl{5}&\ctrl{6}&\ctrl{7}&\ctrl{8}   
&\qw&\qw&\qw&\qw
&\qw&\qw&\qw&\qw 
&\qw&\qw&\qw&\qw &\qw\\
\lstick{\ket{q_1}}&\qw&\qw&\qw&\qw   
&\ctrl{4}&\ctrl{5}&\ctrl{6}&\ctrl{7}
&\qw&\qw&\qw&\qw 
&\qw&\qw&\qw&\qw &\qw\\
\lstick{\ket{q_2}}&\qw&\qw&\qw&\qw   
&\qw&\qw&\qw&\qw 
&\ctrl{3}&\ctrl{4}&\ctrl{5}&\ctrl{6}
&\qw&\qw&\qw&\qw &\qw \\
\lstick{\ket{q_3}}&\qw&\qw&\qw&\qw   
&\qw&\qw&\qw&\qw 
&\qw&\qw&\qw&\qw 
&\ctrl{2}&\ctrl{3}&\ctrl{4}&\ctrl{5}&\qw \\
\makebox[4em][l]{\textbf{Reg l:}}\\
\lstick{\ket{q_0}}&\gate{U_1(\theta_32^1)} &\qw & \qw &\qw   
&\gate{U_1(\theta_32^2)} &\qw & \qw &\qw
&\gate{U_1(\theta_32^3)} &\qw & \qw &\qw
&\gate{U_1(\theta_32^4)} &\qw & \qw &\qw&  \qw \\
\lstick{\ket{q_1}}&\qw &\gate{U_1(\theta_32^2)} & \qw &\qw   
&\qw &\gate{U_1(\theta_32^3)} & \qw &\qw    
&\qw &\gate{U_1(\theta_32^4)} & \qw &\qw   
&\qw &\gate{U_1(\theta_32^5)} & \qw &\qw   &\qw \\
\lstick{\ket{q_2}}&\qw & \qw &\gate{U_1(\theta_32^3)} &\qw   
&\qw & \qw &\gate{U_1(\theta_32^4)} &\qw   
& \qw &\qw &\gate{U_1(\theta_32^5)} &\qw   
&\qw & \qw &\gate{U_1(\theta_32^6)} &\qw   &\qw \\
\lstick{\ket{q_3}}&\qw & \qw &\qw &\gate{U_1(\theta_32^4)}   
&\qw & \qw &\qw &\gate{U_1(\theta_32^5)}   
&\qw & \qw &\qw &\gate{U_1(\theta_32^6)}  
&\qw & \qw &\qw &\gate{U_1(\theta_32^7)}  &\qw \\
}
\begin{tikzpicture}[overlay, remember picture]
    \draw[decorate, decoration={brace, amplitude=8pt, mirror}, yshift=-2ex] 
        (-27.8,-9.5) -- (-21.5,-9.5) node[midway, yshift=-2em, align=center]{$q_{i^\prime} = q_0$, $q_{j^\prime} = q_1$, \\ $q_{k^\prime} = q_0$, $q_{l^\prime} = \{q_0,q_1,q_2,q_3\}$};
    \draw[decorate, decoration={brace, amplitude=8pt, mirror}, yshift=-2ex] 
        (-20.8,-9.5) -- (-14.5,-9.5) node[midway, yshift=-2em, align=center]{$q_{i^\prime} = q_0$, $q_{j^\prime} = q_1$, \\ $q_{k^\prime} = q_1$, $q_{l^\prime} = \{q_0,q_1,q_2,q_3\}$};
    \draw[decorate, decoration={brace, amplitude=8pt, mirror}, yshift=-2ex] 
        (-13.8,-9.5) -- (-7.5,-9.5) node[midway, yshift=-2em, align=center]{$q_{i^\prime} = q_0$, $q_{j^\prime} = q_1$, \\ $q_{k^\prime} = q_2$, $q_{l^\prime} = \{q_0,q_1,q_2,q_3\}$};
    \draw[decorate, decoration={brace, amplitude=8pt, mirror}, yshift=-2ex] 
        (-6.8,-9.5) -- (-0.5,-9.5) node[midway, yshift=-2em, align=center]{$q_{i^\prime} = q_0$, $q_{j^\prime} = q_1$, \\ $q_{k^\prime} = q_3$, $q_{l^\prime} = \{q_0,q_1,q_2,q_3\}$};
\end{tikzpicture}}
}
\caption{\justifying Example fragment circuit of $U_{ijkl}$ when $i=j\neq k\neq l$ ($x_i$ and $x_j$ indicate the same register) with 4 qubits in each register. }
\label{fig:4th examples 4}
\end{figure}

Figure~\ref{fig:4th examples 1}-\ref{fig:4th examples 4} display representative circuit fragments for the remaining index patterns, drawn with $n=4$ qubits per normal mode register for illustration. Each figure consists of two subplots: panel (a) depicts the case $q_{i^\prime}=q_0$ and $q_{j^\prime}=q_0$, while panel (b) exemplifies $q_{i^\prime}=q_0$ and $q_{j^\prime}=q_1$. To evaluate the full gate depth, we systematically traverse all qubit assignments. Starting with fixed $q_{i^\prime}=q_0$, one must account for all possible values of $q_{j^\prime}$. 
The circuit in panel (a) covers one case, while three additional circuits identical in depth to panel (b) cover the remaining $q_{j^\prime} \in \{q_1,q_2,q_3\}$.
Thus, the total contribution under $q_{i^\prime}=q_0$ is $N_{(a)}+3N_{(b)}$, where $N_{(a)}$ and $N_{(b)}$ denote the depths of the circuits in panels (a) and (b), respectively. Repeating the same procedure for $q_{i^\prime}\in \{q_1,q_2,q_3\}$ yields three further groups of equal contribution, leading to a total of $4(N_{(a)}+3N_{(b)})$ gates. The generalisation to arbitrary $n$ follows straightforwardly, and the resulting classifications of gate depths across different ${i,j,k,l}$ scenarios are summarised in Table~\ref{Fragment Gate}.

\begin{figure}[htbp]
\flushright
\subfloat[$q_{i^\prime} = q_{j^\prime}$, with $q_{i^\prime} = q_0$ and $q_{j^\prime} = q_0$, under which all possible values of $q_{k^\prime}$ and $q_{l^\prime}$ are traversed. \label{fourthexam2cir1}]{
\vspace{2.5em} 
\scalebox{0.6}{
\Qcircuit @C=1em @R=1.5em {
\makebox[4em][l]{\textbf{Reg i, Reg j, Reg k refer to the same Reg:}}\\
\lstick{\ket{q_0}}&\ctrl{5}&\ctrl{6}&\ctrl{7}&\ctrl{8}   
&\ctrl{5}&\ctrl{6}&\ctrl{7}&\ctrl{8}
&\ctrl{5}&\ctrl{6}&\ctrl{7}&\ctrl{8}
&\ctrl{5}&\ctrl{6}&\ctrl{7}&\ctrl{8}&\qw\\
\lstick{\ket{q_1}}&\qw&\qw&\qw&\qw   
&\ctrl{4}&\ctrl{5}&\ctrl{6}&\ctrl{7}
&\qw&\qw&\qw&\qw 
&\qw&\qw&\qw&\qw &\qw\\
\lstick{\ket{q_2}}&\qw&\qw&\qw&\qw   
&\qw&\qw&\qw&\qw 
&\ctrl{3}&\ctrl{4}&\ctrl{5}&\ctrl{6}
&\qw&\qw&\qw&\qw &\qw \\
\lstick{\ket{q_3}}&\qw&\qw&\qw&\qw   
&\qw&\qw&\qw&\qw 
&\qw&\qw&\qw&\qw 
&\ctrl{2}&\ctrl{3}&\ctrl{4}&\ctrl{5}&\qw \\
\makebox[4em][l]{\textbf{Reg l:}}\\
\lstick{\ket{q_0}}&\gate{U_1(\theta_32^0)} &\qw & \qw &\qw   
&\gate{U_1(\theta_32^1)} &\qw & \qw &\qw
&\gate{U_1(\theta_32^2)} &\qw & \qw &\qw
&\gate{U_1(\theta_32^3)} &\qw & \qw &\qw&  \qw \\
\lstick{\ket{q_1}}&\qw &\gate{U_1(\theta_32^1)} & \qw &\qw   
&\qw &\gate{U_1(\theta_32^2)} & \qw &\qw   
&\qw &\gate{U_1(\theta_32^3)} & \qw &\qw   
&\qw &\gate{U_1(\theta_32^4)} & \qw &\qw   &\qw \\
\lstick{\ket{q_2}}&\qw & \qw &\gate{U_1(\theta_32^2)} &\qw   
&\qw & \qw &\gate{U_1(\theta_32^3)} &\qw   
&\qw & \qw &\gate{U_1(\theta_32^4)} &\qw   
&\qw & \qw &\gate{U_1(\theta_32^5)} &\qw   &\qw \\
\lstick{\ket{q_3}}&\qw & \qw &\qw &\gate{U_1(\theta_32^3)}   
&\qw & \qw &\qw &\gate{U_1(\theta_32^4)}   
&\qw & \qw &\qw &\gate{U_1(\theta_32^5)}  
&\qw & \qw &\qw &\gate{U_1(\theta_32^6)}  &\qw \\
}
\begin{tikzpicture}[overlay, remember picture]
    \draw[decorate, decoration={brace, amplitude=8pt, mirror}, yshift=-2ex] 
        (-27.8,-7) -- (-21.5,-7) node[midway, yshift=-2em, align=center]{$q_{i^\prime} = q_0$, $q_{j^\prime} = q_0$, \\ $q_{k^\prime} = q_0$, $q_{l^\prime} = \{q_0,q_1,q_2,q_3\}$};
    \draw[decorate, decoration={brace, amplitude=8pt, mirror}, yshift=-2ex] 
        (-20.8,-7) -- (-14.5,-7) node[midway, yshift=-2em, align=center]{$q_{i^\prime} = q_0$, $q_{j^\prime} = q_0$, \\ $q_{k^\prime} = q_1$, $q_{l^\prime} = \{q_0,q_1,q_2,q_3\}$};
    \draw[decorate, decoration={brace, amplitude=8pt, mirror}, yshift=-2ex] 
        (-13.8,-7) -- (-7.5,-7) node[midway, yshift=-2em, align=center]{$q_{i^\prime} = q_0$, $q_{j^\prime} = q_0$, \\ $q_{k^\prime} = q_2$, $q_{l^\prime} = \{q_0,q_1,q_2,q_3\}$};
    \draw[decorate, decoration={brace, amplitude=8pt, mirror}, yshift=-2ex] 
        (-6.8,-7) -- (-0.5,-7) node[midway, yshift=-2em, align=center]{$q_{i^\prime} = q_0$, $q_{j^\prime} = q_0$, \\ $q_{k^\prime} = q_3$, $q_{l^\prime} = \{q_0,q_1,q_2,q_3\}$};
\end{tikzpicture}}
}
\\
\vspace{1em}
\subfloat[$q_{i^\prime} \neq q_{j^\prime}$, with $q_{i^\prime} = q_0$ and $q_{j^\prime} = q_1$, under which all possible values of $q_{k^\prime}$ and $q_{l^\prime}$ are traversed. \label{fourthexam2cir2}]{
\vspace{2.5em} 
\scalebox{0.6}{
\Qcircuit @C=1em @R=1.5em {
\makebox[4em][l]{\textbf{Reg i, Reg j, Reg k refer to the same Reg:}}\\
\lstick{\ket{q_0}}&\ctrl{5}&\ctrl{6}&\ctrl{7}&\ctrl{8}   
&\ctrl{5}&\ctrl{6}&\ctrl{7}&\ctrl{8}
&\ctrl{5}&\ctrl{6}&\ctrl{7}&\ctrl{8}
&\ctrl{5}&\ctrl{6}&\ctrl{7}&\ctrl{8}&\qw\\
\lstick{\ket{q_1}}&\ctrl{-1}&\ctrl{-1}&\ctrl{-1}&\ctrl{-1}   
&\ctrl{4}&\ctrl{5}&\ctrl{6}&\ctrl{7}
&\ctrl{-1}&\ctrl{-1}&\ctrl{-1}&\ctrl{-1} 
&\ctrl{-1}&\ctrl{-1}&\ctrl{-1}&\ctrl{-1} &\qw\\
\lstick{\ket{q_2}}&\qw&\qw&\qw&\qw   
&\qw&\qw&\qw&\qw 
&\ctrl{3}&\ctrl{4}&\ctrl{5}&\ctrl{6}
&\qw&\qw&\qw&\qw &\qw \\
\lstick{\ket{q_3}}&\qw&\qw&\qw&\qw   
&\qw&\qw&\qw&\qw 
&\qw&\qw&\qw&\qw 
&\ctrl{2}&\ctrl{3}&\ctrl{4}&\ctrl{5}&\qw \\
\makebox[4em][l]{\textbf{Reg l:}}\\
\lstick{\ket{q_0}}&\gate{U_1(\theta_32^1)} &\qw & \qw &\qw   
&\gate{U_1(\theta_32^2)} &\qw & \qw &\qw
&\gate{U_1(\theta_32^3)} &\qw & \qw &\qw
&\gate{U_1(\theta_32^4)} &\qw & \qw &\qw&  \qw \\
\lstick{\ket{q_1}}&\qw &\gate{U_1(\theta_32^2)} & \qw &\qw   
&\qw &\gate{U_1(\theta_32^3)} & \qw &\qw   
&\qw &\gate{U_1(\theta_32^4)} & \qw &\qw   
&\qw &\gate{U_1(\theta_32^5)} & \qw &\qw   &\qw \\
\lstick{\ket{q_2}}&\qw & \qw &\gate{U_1(\theta_32^3)} &\qw   
&\qw & \qw &\gate{U_1(\theta_32^4)} &\qw   
&\qw & \qw &\gate{U_1(\theta_32^5)} &\qw   
&\qw & \qw &\gate{U_1(\theta_32^6)} &\qw   &\qw \\
\lstick{\ket{q_3}}&\qw & \qw &\qw &\gate{U_1(\theta_32^4)}   
&\qw & \qw &\qw &\gate{U_1(\theta_32^5)}   
&\qw & \qw &\qw &\gate{U_1(\theta_32^6)}  
&\qw & \qw &\qw &\gate{U_1(\theta_32^7)}  &\qw \\
}
\begin{tikzpicture}[overlay, remember picture]
    \draw[decorate, decoration={brace, amplitude=8pt, mirror}, yshift=-2ex] 
        (-27.8,-7) -- (-21.5,-7) node[midway, yshift=-2em, align=center]{$q_{i^\prime} = q_0$, $q_{j^\prime} = q_1$, \\ $q_{k^\prime} = q_0$, $q_{l^\prime} = \{q_0,q_1,q_2,q_3\}$};
    \draw[decorate, decoration={brace, amplitude=8pt, mirror}, yshift=-2ex] 
        (-20.8,-7) -- (-14.5,-7) node[midway, yshift=-2em, align=center]{$q_{i^\prime} = q_0$, $q_{j^\prime} = q_1$, \\ $q_{k^\prime} = q_1$, $q_{l^\prime} = \{q_0,q_1,q_2,q_3\}$};
    \draw[decorate, decoration={brace, amplitude=8pt, mirror}, yshift=-2ex] 
        (-13.8,-7) -- (-7.5,-7) node[midway, yshift=-2em, align=center]{$q_{i^\prime} = q_0$, $q_{j^\prime} = q_1$, \\ $q_{k^\prime} = q_2$, $q_{l^\prime} = \{q_0,q_1,q_2,q_3\}$};
    \draw[decorate, decoration={brace, amplitude=8pt, mirror}, yshift=-2ex] 
        (-6.8,-7) -- (-0.5,-7) node[midway, yshift=-2em, align=center]{$q_{i^\prime} = q_0$, $q_{j^\prime} = q_1$, \\ $q_{k^\prime} = q_3$, $q_{l^\prime} = \{q_0,q_1,q_2,q_3\}$};
\end{tikzpicture}}
}
\caption{\justifying Example fragment circuit of $U_{ijkl}$ when $i=j=k\neq l$ ($x_i$, $x_j$ and $x_k$ indicate the same register) with 4 qubits in each register.}
\label{fig:4th examples 2}
\end{figure}

\clearpage
\newpage

\section{Supplementary Datasets for Fundamental Bands of Water}
\label{app:watertable} 
This appendix provides the supplementary numerical datasets for fundamental bands of water. Table~\ref{approxdipoledata} reports fundamentals obtained from simulations using an approximate dipole operator: the first-order dipole operator is constructed via the circuit in Figure~\ref{fig:non-unitary dipole} with gate phases specified by the approximation in Eq.~\ref{taylordipole}, and then applied to the approximate harmonic ground state. We tabulate the band positions and intensities under systematic variations of both total evolution time and time resolution. 
Tables~\ref{datasetvaryT} and~\ref{datasetvaryRes} present large datasets of fundamental bands for all four $\mu\Psi$ approximation schemes, with the former covering cases of varying total evolution time at fixed resolution, and the latter covering varying resolution at fixed total time.

\begin{table*}[htbp]
    \centering 
    \renewcommand{\arraystretch}{1.3}
    \begin{tabular}{|c<{\centering\arraybackslash}|c<{\centering\arraybackslash}|c<{\centering\arraybackslash}|c<{\centering\arraybackslash}|c<{\centering\arraybackslash}|c<{\centering\arraybackslash}|c<{\centering\arraybackslash}|c<{\centering\arraybackslash}|c<{\centering\arraybackslash}|} 
        \hline
        \multirow{2}{*}{$T$ (fs)}  &\multirow{2}{*}{$n_t$} &\multirow{2}{*}{$dt$ (fs)}
        & \multicolumn{2}{c|}{Mode 1} & \multicolumn{2}{c|}{Mode 2} & \multicolumn{2}{c|}{Mode 3} \\
        \cline{4-9}
        & & &$\widetilde{\upsilon}_c$ (cm$^{-1}$) &$I_\mathrm{IR}$ (km/mol) &$\widetilde{\upsilon}_c$ (cm$^{-1}$) &$I_\mathrm{IR}$ (km/mol) &$\widetilde{\upsilon}_c$ (cm$^{-1}$) &$I_\mathrm{IR}$ (km/mol) \\
        \hline
        \multicolumn{9}{l}{Various total evolution time $T$, with a fixed time resolution $dt$:} \\
        \hline
        1975& 30,000 & 0.066 & 3808.1 $\pm$ 14.5 & 4.5 & 1624.5 $\pm$ 1.9 & 69.6 & 3875.8 $\pm$ 7.4 & 36.5  \\
        \hline
        3950& 60,000 & 0.066 & 3809.1 $\pm$ 4.0 &4.6 & 1624.4 $\pm$ 1.2 &69.6 & 3876.3 $\pm$ 3.1&36.4\\
        \hline
        7900& 120,000& 0.066 & 3809.1 $\pm$ 0.8  &4.6 & 1624.3 $\pm$ 0.7 &69.6 & 3876.4 $\pm$ 0.5  &36.4\\
        \hline
        13165& 200,000 & 0.066 & 3809.1 $\pm$ 0.7 & 4.6 & 1624.3 $\pm$ 0.2 &69.6 & 3876.4 $\pm$ 0.4&36.4\\
        \hline
        \multicolumn{9}{l}{Various time resolution $dt$, with a fixed total evolution time $T$:} \\
        \hline
        3950& 20,000 &0.198 & 3815.2 $\pm$ 3.8 & 4.6 & 1627.8 $\pm$ 1.8 & 69.8 & 3882.7  $\pm$ 2.7 & 36.6 \\
        \hline
        3950& 40,000 & 0.099 & 3810.2 $\pm$ 4.0 &4.7 & 1624.9 $\pm$ 0.8 &69.7 & 3877.5 $\pm$ 2.9&36.5 \\
        \hline
        3950& 60,000 & 0.066 & 3809.1 $\pm$ 4.0 &4.6 & 1624.4 $\pm$ 1.2 &69.6 & 3876.3 $\pm$ 3.1 &36.4 \\
        \hline
        3950& 80,000 & 0.049 & 3808.7 $\pm$ 3.9&4.6 & 1624.2 $\pm$ 1.3&69.6 & 3875.9 $\pm$ 3.0&36.4 \\
        \hline
    \end{tabular}
    \caption{\justifying Fundamental bands simulated from various time parameters. Here $\widetilde{\upsilon}_c$ and $I_\mathrm{IR}$ represent the band position and absolute band intensity, respectively. All simulations start from a harmonic ground state and utilise approximated first-order dipole operators realised by the quantum circuit shown in Figure~\ref{fig:non-unitary dipole} with gate angles defined as Eq.~\ref{taylordipole}.}
    \label{approxdipoledata} 
\end{table*}

\begin{table*}[htbp]
    \centering 
    \fontsize{8.5pt}{10.2pt}\selectfont
    \renewcommand{\arraystretch}{1.3}
    \begin{tabular}{|c<{\centering\arraybackslash}|c<{\centering\arraybackslash}|c<{\centering\arraybackslash}|c<{\centering\arraybackslash}|c<{\centering\arraybackslash}|c<{\centering\arraybackslash}|c<{\centering\arraybackslash}|c<{\centering\arraybackslash}|c<{\centering\arraybackslash}|c<{\centering\arraybackslash}|c<{\centering\arraybackslash}|} 
        \hline 
        \multicolumn{3}{|c|}{Time Parameters} &\multicolumn{4}{c|}{$\widetilde{\upsilon}_c$ ($cm^{-1}$)} & \multicolumn{4}{c|}{$I_\mathrm{IR}$ (km/mol)} \\ 
        \hline 
        $T$~fs & $n_t$ &$dt$~fs &$\mu_\text{3rd}\Psi_\text{ITE}$&$\mu_\text{1st}\Psi_\text{ITE}$&$\mu_\text{3rd}\Psi_\text{Gauss}$ &$\mu_\text{1st}\Psi_\text{Gauss}$& $\mu_\text{3rd}\Psi_\text{ITE}$& $\mu_\text{1st}\Psi_\text{ITE}$& $\mu_\text{3rd}\Psi_\text{Gauss}$& $\mu_\text{1st}\Psi_\text{Gauss}$ \\ 
        \hline
        \multicolumn{1}{c}{Mode 1} \\
        \hline
        1975& 30,000& 0.066 & 3808.1 $\pm$ 14.5 &  3808.1 $\pm$ 14.5  & 3808.1 $\pm$ 14.5  &  3808.1 $\pm$ 14.5 &4.5 & 5.7  & 4.3 & 4.6\\ 
        \hline
        3950& 60,000& 0.066 & 3809.1 $\pm$ 4.1 &  3809.1 $\pm$ 4.0   & 3809.1 $\pm$ 4.1  & 3809.1 $\pm$ 4.0  & 4.6& 5.8  & 4.4 & 4.7\\
        \hline
        7900& 120,000& 0.066 & 3809.1 $\pm$ 0.8&  3809.1 $\pm$ 0.7  & 3809.1 $\pm$ 0.8  &  3809.1 $\pm$ 0.8  & 4.6& 5.8  & 4.4 & 4.7\\   
        \hline
        13165& 200,000& 0.066 & 3809.1 $\pm$ 0.7& 3809.1 $\pm$ 0.7  & 3809.1 $\pm$ 0.7 &  3809.1 $\pm$ 0.7  & 4.6& 5.8  & 4.4 & 4.7\\ 
        \hline
        \multicolumn{1}{c}{Mode 2}  \\
        \hline
        1975& 30,000& 0.066 & 1624.5 $\pm$ 1.9 & 1624.5 $\pm$ 1.9  & 1624.5 $\pm$ 1.9 & 1624.5 $\pm$ 1.9 &76.7 & 75.7  & 72.5 & 70.9 \\ 
        \hline
        3950 & 60,000& 0.066 & 1624.4 $\pm$ 1.2 &  1624.4 $\pm$ 1.2  & 1624.4 $\pm$ 1.2 & 1624.4 $\pm$ 1.2  &76.7 & 75.7  & 72.5 & 70.9 \\
        \hline
        7900& 120,000& 0.066 & 1624.3 $\pm$ 0.7 & 1624.3 $\pm$ 0.7  &  1624.3 $\pm$ 0.7 & 1624.3 $\pm$ 0.7 &76.7 & 75.7  & 72.5 & 70.9 \\   
        \hline
        13165& 200,000& 0.066 & 1624.3 $\pm$ 0.2 &  1624.3 $\pm$ 0.2  &  1624.3 $\pm$ 0.2 &  1624.3 $\pm$ 0.2 &76.7 & 75.7  & 72.5  & 70.9  \\ 
        \hline
        \multicolumn{1}{c}{Mode 3}   \\
        \hline
        1975& 30,000& 0.066 & 3875.8 $\pm$ 7.4 &  3875.8 $\pm$ 7.4   & 3875.8 $\pm$ 7.4  &   3875.8 $\pm$ 7.4  & 37.0& 40.8  & 34.5 & 36.6 \\ 
        \hline
        3950& 60,000& 0.066 & 3876.3 $\pm$ 3.1 & 3876.3 $\pm$ 3.1   &  3876.3 $\pm$ 3.1  &  3876.3 $\pm$ 3.1 &37.0 & 40.8  & 34.5 & 36.6  \\
        \hline
        7900& 120,000& 0.066 &3876.4 $\pm$ 0.5 &  3876.4 $\pm$ 0.5   & 3876.4 $\pm$ 0.5  &  3876.4 $\pm$ 0.5 & 37.0& 40.8  & 34.5 & 36.6  \\   
        \hline
        13165& 200,000& 0.066 & 3876.4 $\pm$ 0.4 &  3876.4 $\pm$ 0.4 &  3876.4 $\pm$ 0.4  &  3876.4 $\pm$ 0.4  &37.0 & 40.8  &  34.5  & 36.6  \\ 
        \hline
    \end{tabular}
    \caption{\justifying Fundamental bands obtained from simulations with four different approximation schemes in $\mu\Psi$, under varying total evolution time at fixed time resolution. Here $\widetilde{\upsilon}_c$ and $I_\mathrm{IR}$ represent the band position and absolute band intensity, respectively.}
    \label{datasetvaryT} 
\end{table*}

\begin{table*}[htbp]
    \centering 
    \fontsize{8.5pt}{10.2pt}\selectfont
    \renewcommand{\arraystretch}{1.3}
    \begin{tabular}{|c<{\centering\arraybackslash}|c<{\centering\arraybackslash}|c<{\centering\arraybackslash}|c<{\centering\arraybackslash}|c<{\centering\arraybackslash}|c<{\centering\arraybackslash}|c<{\centering\arraybackslash}|c<{\centering\arraybackslash}|c<{\centering\arraybackslash}|c<{\centering\arraybackslash}|c<{\centering\arraybackslash}|} 
        \hline 
        \multicolumn{3}{|c|}{Time Parameters} &\multicolumn{4}{c|}{$\widetilde{\upsilon}_c$ ($cm^{-1}$)} & \multicolumn{4}{c|}{$I_\mathrm{IR}$ (km/mol)} \\ 
        \hline 
        $T$~fs & $n_t$ &$dt$~fs &$\mu_\text{3rd}\Psi_\text{ITE}$&$\mu_\text{1st}\Psi_\text{ITE}$&$\mu_\text{3rd}\Psi_\text{Gauss}$ &$\mu_\text{1st}\Psi_\text{Gauss}$& $\mu_\text{3rd}\Psi_\text{ITE}$& $\mu_\text{1st}\Psi_\text{ITE}$& $\mu_\text{3rd}\Psi_\text{Gauss}$& $\mu_\text{1st}\Psi_\text{Gauss}$ \\ 
        \hline
        \multicolumn{1}{c}{Mode 1}  \\
        \hline
        3950 &20,000& 0.198 &3815.2 $\pm$ 3.9 & 3815.2 $\pm$ 3.7  & 3815.2 $\pm$ 3.9 &   3815.2 $\pm$ 3.8 &4.6 & 5.8   & 4.5 & 4.7 \\
        \hline
        3950&40,000& 0.099 &3810.2 $\pm$ 4.0 & 3810.3 $\pm$ 3.9 & 3810.2 $\pm$ 4.0 &  3810.2 $\pm$ 4.0 &4.6 & 5.8 & 4.5 & 4.7 \\
        \hline
        3950&60,000& 0.066 & 3809.1 $\pm$ 4.1 & 3809.1 $\pm$ 4.0  &  3809.1 $\pm$ 4.1  &   3809.1 $\pm$ 4.0 &4.6 &  5.8  & 4.4 & 4.7 \\   
        \hline
        3950&80,000& 0.049 &3808.7 $\pm$ 3.9 &  3808.7 $\pm$ 3.9 & 3808.7 $\pm$ 3.9 &  3808.7 $\pm$ 3.9 &4.5 & 5.8   & 4.4 & 4.7 \\ 
        \hline
        \multicolumn{1}{c}{Mode 2} \\
        \hline
        3950 &20,000& 0.198 & 1627.8 $\pm$ 1.8 &  1627.8 $\pm$ 1.8   & 1627.8 $\pm$ 1.8   &  1627.8 $\pm$ 1.8   &76.9 & 75.9   & 72.6 & 71.1 \\
        \hline
        3950&40,000& 0.099 & 1624.9 $\pm$ 0.8 & 1624.9 $\pm$ 0.8   & 1624.9 $\pm$ 0.8 &  1624.9 $\pm$ 0.8 &76.7 & 75.7   & 72.5 & 71.0 \\
        \hline
        3950&60,000& 0.066 & 1624.4 $\pm$ 1.2 & 1624.4 $\pm$ 1.2   & 1624.4 $\pm$ 1.2 &  1624.4 $\pm$ 1.2  &76.7 & 75.7   & 72.5 & 70.9 \\   
        \hline
        3950&80,000& 0.049 & 1624.2 $\pm$ 1.3 & 1624.2 $\pm$ 1.3  & 1624.2 $\pm$ 1.3  &  1624.2 $\pm$ 1.3 &76.7&  75.7  & 72.4 & 70.9 \\ 
        \hline
        \multicolumn{1}{c}{Mode 3}  \\
        \hline
        3950 &20,000& 0.198 & 3882.7 $\pm$ 2.7  & 3882.7 $\pm$ 2.7  &  3882.7 $\pm$ 2.7 &  3882.7 $\pm$ 2.7  &37.1 & 40.9   & 34.6 & 36.7 \\
        \hline
        3950&40,000& 0.099 & 3877.5 $\pm$ 2.9 &  3877.5 $\pm$ 2.9  &  3877.5 $\pm$ 2.9 &  3877.5 $\pm$ 2.9  &37.0 & 40.8   & 34.5 & 36.6 \\
        \hline
        3950&60,000& 0.066 & 3876.3 $\pm$ 3.1 & 3876.3 $\pm$ 3.1  & 3876.3 $\pm$ 3.1 &   3876.3 $\pm$ 3.1 &37.0 & 40.8   & 34.5 & 36.6  \\   
        \hline
        3950&80,000& 0.049 & 3875.9 $\pm$ 3.0 & 3875.9 $\pm$ 3.0  &  3875.9 $\pm$ 3.0   & 3875.9 $\pm$ 3.0  & 37.0& 40.8   & 34.5 & 36.6  \\ 
        \hline
    \end{tabular}
    \caption{\justifying Fundamental bands obtained from simulations with four different approximation schemes in $\mu\Psi$, under varying time resolution at fixed total evolution time. Here $\widetilde{\upsilon}_c$ and $I_\mathrm{IR}$ represent the band position and absolute band intensity, respectively.}
    \label{datasetvaryRes} 
\end{table*}

\clearpage
\newpage

\section{Supplementary Datasets for Overtone Bands of Water}
\label{App overtone}

In this appendix, overtone bands of water simulated with four different $\mu\Psi$ schemes are reported in Table~\ref{Overtone 1and4} (for $\mu_\text{3rd}\Psi_\text{ITE}$ and $\mu_\text{1st}\Psi_\text{Gauss}$) and Table~\ref{Overtone 2and3} (for $\mu_\text{1st}\Psi_\text{ITE}$ and $\mu_\text{3rd}\Psi_\text{Gauss}$).


\begin{table*}[htbp]
    \centering 
    \renewcommand{\arraystretch}{1.3}
    \begin{tabular}{|c<{\centering\arraybackslash}|c<{\centering\arraybackslash}|c<{\centering\arraybackslash}|c<{\centering\arraybackslash}|c<{\centering\arraybackslash}|c<{\centering\arraybackslash}|c<{\centering\arraybackslash}|c<{\centering\arraybackslash}|c<{\centering\arraybackslash}|} 
        \hline
        \multicolumn{4}{|c|}{$\mu_\text{3rd}\Psi_\text{ITE}$}&&\multicolumn{4}{c|}{$\mu_\text{1st}\Psi_\text{Gauss}$} \\
        \cline{1-4} \cline{6-9}
        \multicolumn{2}{|c|}{Band Position (cm$^{-1}$)}& \multicolumn{2}{c|}{Intensity (km/mol)}&&\multicolumn{2}{c|}{Band Position (cm$^{-1}$)}& \multicolumn{2}{c|}{Intensity (km/mol)}\\
        \cline{1-4} \cline{6-9}      
        T=3950 fs&T=13165 fs&T=3950 fs&T=13165 fs && T=3950 fs&T=13165 fs&T=3950 fs&T=13165 fs\\
        \cline{1-4} \cline{6-9}  
        3207.0 $\pm$ 6.5 & 3207.2 $\pm$ 3.4&0.66&  0.66&&3200.8 $\pm$ 17.6 & 3201.4 $\pm$ 9.3 &0.076&0.078 \\
        4733.9 $\pm$ 16.2 &  4738.0 $\pm$ 8.1&0.046 &0.047&&4733.0 $\pm$ 10.6 & 4734.6 $\pm$ 5.4 &0.10 &0.10  \\
        5383.8 $\pm$ 6.0 &  5384.5 $\pm$ 2.2&0.12 & 0.13&&5384.6 $\pm$ 1.4 & 5384.8 $\pm$ 0.5 &3.8&3.8  \\
        5428.0 $\pm$ 2.5 & 5428.1 $\pm$ 0.7&7.8 &  7.8&&5427.2 $\pm$ 8.2 & 5428.9 $\pm$ 3.5 &0.26&0.25\\
        6131.9 $\pm$ 14.9 &  6133.4 $\pm$ 6.7&0.038 &0.038&&6134.6 $\pm$ 9.0 & 6135.2 $\pm$ 4.0 &0.10&0.10  \\
        6944.4 $\pm$ 14.4 & 6948.4 $\pm$ 7.3&0.14 & 0.14&&6941.7 $\pm$ 15.4 & 6946.0 $\pm$ 7.9 &0.12&0.12  \\
        7005.8$\pm$17.1 & 7008.6$\pm$4.7&0.021 & 0.015&&7006.0$\pm$16.8 & 7008.7$\pm$4.5 &0.021 &0.015  \\
        7594.9 $\pm$6.0 &  7595.0 $\pm$ 2.7&0.47 &0.48&&7595.4 $\pm$ 4.1 & 7595.3 $\pm$ 1.7 &1.2&1.2 \\
        7643.8 $\pm$ 4.7 &  7643.9 $\pm$ 1.9&2.1 & 2.1&&7643.5 $\pm$ 3.8  & 7643.6 $\pm$ 1.2 &6.0&6.0 \\
        7726.7 $\pm$ 8.9 & 7727.7 $\pm$ 4.4 & 0.27 & 0.27 && 7727.8 $\pm$ 11.0 & 7729.2 $\pm$ 5.5 & 0.16 & 0.16 \\
        7812.4 $\pm$ 37.9 &  7813.1 $\pm$ 4.8&0.016 & 0.019&&7813.0 $\pm$ 37.5 & 7813.3 $\pm$ 4.1 &0.022&0.025  \\
        8413.0 $\pm$ 15.3 & 8414.4 $\pm$ 5.8&0.037 & 0.038&&8415.3 $\pm$ 15.2 & 8416.0 $\pm$ 5.5 &0.046 &0.045 \\
        8437.6 $\pm$ 37.9 & 8441.4 $\pm$ 4.4&0.018 &  0.016&&8439.9 $\pm$ 30.6 & 8439.7 $\pm$ 1.8 & 0.16&0.15  \\
        / &    9082.7 $\pm$ 4.4&/ & 0.012&&/ & 9083.1 $\pm$ 4.1 &/ &0.014  \\
        / &  9091.4 $\pm$ 4.9&/ &  0.0097 &&/ & 9091.4 $\pm$ 5.1 &/ &0.0078\\
        9111.3 $\pm$ 6.8 &   9112.0 $\pm$ 2.5&0.37 & 0.36&&9112.2 $\pm$ 12.0 & 9114.4 $\pm$ 5.0 &0.087 &0.079  \\
        9826.8 $\pm$ 15.6 &   9828.8 $\pm$ 7.2&0.069 & 0.068&&9826.8 $\pm$ 15.5 & 9828.8 $\pm$ 7.2 &0.065 &0.064 \\
        9869.0 $\pm$ 17.1 &  9870.6 $\pm$ 5.2&0.030 & 0.030&&9869.6 $\pm$ 16.9 & 9871.0 $\pm$ 4.9 &0.032& 0.032 \\
        / &  10532.0 $\pm$ 11.4&/ & 0.0067&&/ & 10532.4 $\pm$ 10.7 &/ &0.0088 \\
        /  &  10539.5 $\pm$ 4.5&/ & 0.013&&/  &  10539.8 $\pm$ 3.2 &/ &0.035   \\
        /  & 10663.7 $\pm$ 4.5&/ & 0.013&&/  & 10664.3 $\pm$ 3.2&/ &0.026  \\
        / &  11143.9 $\pm$ 5.1&/ & 0.010&&/ &  11143.9 $\pm$ 5.0  &/ &0.0099 \\
        11385.8 $\pm$ 14.6 &   11391.0 $\pm$ 5.6&0.051 & 0.054&&11388.6 $\pm$ 11.8&11391.5 $\pm$ 4.0& 0.097 & 0.099 \\
        11446.0 $\pm$ 11.6 &   11447.4 $\pm$ 5.2& 0.12 &0.12&&11447.7 $\pm$ 5.9  &11448.0 $\pm$ 2.6 &0.45 &0.45 \\
        11503.8 $\pm$ 13.7 &   11506.1 $\pm$ 5.1&0.057&  0.056&&11503.8 $\pm$ 13.8 &11506.1 $\pm$ 5.2 &0.053 & 0.052  \\
        11547.7 $\pm$ 12.4 &   11550.2 $\pm$ 5.6&0.11 & 0.11&&11545.6 $\pm$ 15.1 &11549.3 $\pm$ 7.0 &0.065 &0.068  \\
        / &  11857.3 $\pm$ 11.3&/ & 0.0078&&/ &11857.6 $\pm$ 10.7 &/ &0.0099  \\
        / &  11912.9 $\pm$ 11.4&/ & 0.0073&&/ &11914.0 $\pm$ 5.6 &/ &0.045  \\
        11972.1 $\pm$ 17.0 & 11971.2 $\pm$ 5.2&0.037 &  0.037&& 11972.0 $\pm$ 16.2 &11971.3 $\pm$ 4.9 &0.038 &0.038  \\
        \hline
    \end{tabular}
    \caption{\justifying Overtones simulated using different total propagation time lengths and approximation schemes in $\mu\ket{\Psi}$ ($\mu_\text{3rd}\Psi_\text{ITE}$ and $\mu_\text{1st}\Psi_\text{Gauss}$), with a fixed time step size $dt\approx 0.066$~fs.}
    \label{Overtone 1and4} 
\end{table*}

\begin{table*}[htbp]
    \centering 
    \renewcommand{\arraystretch}{1.3}
    \begin{tabular}{|c<{\centering\arraybackslash}|c<{\centering\arraybackslash}|c<{\centering\arraybackslash}|c<{\centering\arraybackslash}|c<{\centering\arraybackslash}|c<{\centering\arraybackslash}|c<{\centering\arraybackslash}|c<{\centering\arraybackslash}|c<{\centering\arraybackslash}|} 
        \hline
        \multicolumn{4}{|c|}{$\mu_\text{1st}\Psi_\text{ITE}$}&&\multicolumn{4}{c|}{$\mu_\text{3rd}\Psi_\text{Gauss}$} \\
        \cline{1-4} \cline{6-9}
        \multicolumn{2}{|c|}{Band Position (cm$^{-1}$)}& \multicolumn{2}{c|}{Intensity (km/mol)}&&\multicolumn{2}{c|}{Band Position (cm$^{-1}$)}& \multicolumn{2}{c|}{Intensity (km/mol)}\\
        \cline{1-4} \cline{6-9}      
        T=3950 fs&T=13165 fs&T=3950 fs&T=13165 fs && T=3950 fs&T=13165 fs&T=3950 fs&T=13165 fs\\
        \cline{1-4} \cline{6-9}  
        3198.7 $\pm$ 19.9 & 3199.6 $\pm$ 10.5&0.062&  0.064 &&3206.7 $\pm$ 7.5 & 3206.9 $\pm$ 3.9 &0.49&0.49\\
        4733.7 $\pm$ 15.4 &  4737.4 $\pm$ 7.7&0.050 &0.051 &&4733.3 $\pm$ 13.0 & 4736.0$\pm$ 6.6 &0.070 &0.072\\
        5381.9 $\pm$ 12.2 &  5383.9 $\pm$ 4.2&0.035 & 0.033 &&5384.7 $\pm$ 1.1 & 5384.8 $\pm$ 0.5 &5.6&5.6\\
        5428.7 $\pm$ 8.5 & 5429.7 $\pm$ 4.2&0.17 &  0.17&& 5427.9 $\pm$ 2.7 & 5428.1 $\pm$ 0.8 &7.4&7.4\\
        6131.7 $\pm$ 15.1 &  6133.2 $\pm$ 6.8&0.036 &0.036 && 6135.0 $\pm$ 7.6 & 6135.4 $\pm$ 3.4 &0.16&0.16\\
        6942.4 $\pm$ 17.9 & 6948.5 $\pm$ 9.1&0.089 & 0.089 && 6943.0 $\pm$ 16.3 & 6948.0 $\pm$ 8.3 &0.11&0.11 \\
        7005.8 $\pm$ 17.1 & 7008.6 $\pm$ 4.7&0.021 & 0.015 && 7006.3 $\pm$ 16.2 & 7008.7 $\pm$ 4.1 &0.025 &0.019  \\
        7593.0 $\pm$ 10.4 &  7593.5 $\pm$ 4.8&0.13 &0.14 && 7595.5 $\pm$ 3.6 & 7595.4 $\pm$ 1.4 &1.8&1.8 \\
        7644.4 $\pm$ 6.3 &  7644.8 $\pm$ 3.0&0.83 & 0.83 && 7643.5 $\pm$ 3.7  & 7643.5 $\pm$ 1.0 &8.4&8.4  \\
        7728.8 $\pm$ 12.7 & 7730.8 $\pm$ 6.2 & 0.13 & 0.12 && 7726.5 $\pm$ 8.4 & 7727.3 $\pm$ 4.2 & 0.30 & 0.30 \\
        7812.4 $\pm$ 37.9 &  7813.1 $\pm$ 4.9&0.015 & 0.018 && 7813.3 $\pm$ 37.0 & 7813.4 $\pm$ 3.6 &0.032&0.035  \\
        8412.1 $\pm$ 15.0 & 8413.7 $\pm$ 5.9&0.033 & 0.035 && 8418.7 $\pm$ 13.4 & 8418.9 $\pm$ 4.5 &0.087 &0.086 \\
        8437.4 $\pm$ 37.9 & 8441.5 $\pm$ 4.5&0.017 &  0.015 && 8440.0 $\pm$ 30.2 & 8439.7 $\pm$ 1.7 & 0.18&0.17  \\
        / &    9082.8 $\pm$ 4.4&/ & 0.012 && / & 9084.0 $\pm$ 2.6 &/ &0.052  \\
        / &  9091.3 $\pm$ 4.9&/ &  0.0097 && / & 9090.9 $\pm$ 3.4 &/ &0.039\\
        9112.4 $\pm$ 14.7 &   9116.2 $\pm$ 6.1&0.061 & 0.051 && 9110.8 $\pm$ 5.4 & 9111.5 $\pm$ 1.3 &1.5 &1.4  \\
        9826.8 $\pm$ 15.6 &   9828.8 $\pm$ 7.3&0.068 & 0.067 && 9827.3 $\pm$ 14.1 & 9828.9 $\pm$ 6.5 &0.084 &0.083 \\
        9869.0 $\pm$ 17.1 &  9870.6 $\pm$ 5.2&0.029 & 0.029 && 9870.7 $\pm$ 16.0 & 9871.7 $\pm$ 4.3 &0.046& 0.046 \\
        / &  10532.0 $\pm$ 11.4&/ & 0.0063 && / & 10532.2 $\pm$ 11.2 &/ &0.0076 \\
        /  &  10539.6 $\pm$ 4.5&/ & 0.013 && /  &  10539.6 $\pm$ 4.0 &/ &0.018   \\
        /  & 10663.7 $\pm$ 4.5&/ & 0.013 && /  & 10664.1 $\pm$ 3.7&/ &0.020  \\
        / &  11143.9 $\pm$ 5.1&/ & 0.0099 && / &  11143.9 $\pm$ 5.1  &/ &0.010 \\
        11385.4 $\pm$ 14.9 &   11391.0 $\pm$ 5.8&0.047 & 0.049 && 11390.1 $\pm$ 9.3&11391.7 $\pm$ 2.8& 0.21 & 0.21 \\
        11445.5 $\pm$ 12.6 &   11447.2 $\pm$ 5.6& 0.10 &0.10 && 11447.9 $\pm$ 4.6  &11448.1 $\pm$ 2.0 &0.79 &0.79 \\
        11503.6 $\pm$ 14.2 &   11506.2 $\pm$ 5.4&0.052&  0.050 && 11504.9 $\pm$ 10.6 &11506.0 $\pm$ 3.7 &0.11 & 0.11  \\
        11546.2 $\pm$ 14.4 &   11549.6 $\pm$ 6.6&0.076 & 0.080 && 11547.3 $\pm$ 13.1 &11550.0 $\pm$ 5.9 &0.096 &0.10  \\
        / &  11857.2 $\pm$ 11.4&/ & 0.0071 && / &11858.1 $\pm$ 7.9 &/ &0.024  \\
        / &  11912.9 $\pm$ 11.4&/ & 0.0070 && / &11914.0 $\pm$ 5.0 &/ &0.057  \\
        11972.1 $\pm$ 17.1 & 11971.2 $\pm$ 5.2&0.035 &  0.035 && 11972.0 $\pm$ 12.7 &11971.6 $\pm$ 3.9 &0.064 &0.064  \\
        \hline
    \end{tabular}
    \caption{\justifying Overtones simulated using different total propagation time lengths and approximation schemes in $\mu\ket{\Psi}$ ($\mu_\text{1st}\Psi_\text{ITE}$ and $\mu_\text{3rd}\Psi_\text{Gauss}$), with a fixed time step size $dt\approx 0.066$~fs.}
    \label{Overtone 2and3} 
\end{table*}

\end{document}